\documentclass[aps,PRX,reprint,superscriptaddress,longbibliography,floatfix]{revtex4-1}
\usepackage{graphicx}
\usepackage{extpfeil}
\usepackage{float}
\usepackage{caption}
\usepackage{setspace}
\usepackage{multirow}
\usepackage{arydshln}
\usepackage{appendix}
\usepackage{hyperref}
\usepackage{color,colortbl}

\newcommand{\romanNum}[1]{\MakeUppercase{\romannumeral #1}}
\begin{document}
	\title{Where to find lossless metals?}
\author{Xiaolei Hu}
\author{Zhengran Wu}
\affiliation{Institute of Physics, Chinese Academy of Sciences/Beijing National Laboratory for Condensed Matter Physics, Beijing 100190, China.}
\affiliation{School of Physical Sciences, University of Chinese Academy of Sciences, Beijing 100049, China.}
\author{Zhilin Li}
\email{lizhilin@iphy.ac.cn}
\affiliation{Institute of Physics, Chinese Academy of Sciences/Beijing National Laboratory for Condensed Matter Physics, Beijing 100190, China.}
\affiliation{Songshan Lake
	Materials Laboratory, Dongguan, Guangdong 523808, China.}
\author{Qiunan Xu}
\affiliation{Qingdao Institute for Theoretical and Computational Sciences, Shandong University, Qingdao 266237, China.}
\author{Kun Chen}
\affiliation{Institute of Physics, Chinese Academy of Sciences/Beijing National
	Laboratory for Condensed Matter Physics, Beijing 100190, China.}
\affiliation{School of Physical Sciences, University of Chinese Academy of Sciences, Beijing 100049, China.}
\author{Kui Jin}
\author{Hongming Weng}
\email{hmweng@iphy.ac.cn}
\author{Ling Lu}
\email{linglu@iphy.ac.cn}
\affiliation{Institute of Physics, Chinese Academy of Sciences/Beijing National
	Laboratory for Condensed Matter Physics, Beijing 100190, China.}
\affiliation{Songshan Lake
	Materials Laboratory, Dongguan, Guangdong 523808, China.}


\begin{abstract}
	Hypothetical metals having optical absorption losses as low as those of the transparent insulators, if found, could revolutionize optoelectronics.
	We perform the first high-throughput search for lossless metals among all known inorganic materials in the databases of over 100,000 entries.
	The 381 candidates are identified --- having well-isolated partially-filled bands --- and are analyzed by defining the figures of merit and classifying their real-space conductive connectivity.
	The existing experimental evidence of most candidates being insulating, instead of conducting, is due to the limitation of current density functional theory in predicting narrow-band metals that are unstable against magnetism, structural distortion, or electron-electron interactions.
	We propose future research directions including conductive oxides, intercalating layered materials, and compressing these false-metal candidates under high pressures into eventual lossless metals.
\end{abstract}
\maketitle%
	
	\section{Introduction}
\begin{figure*}[t] 
	\centering
	\includegraphics{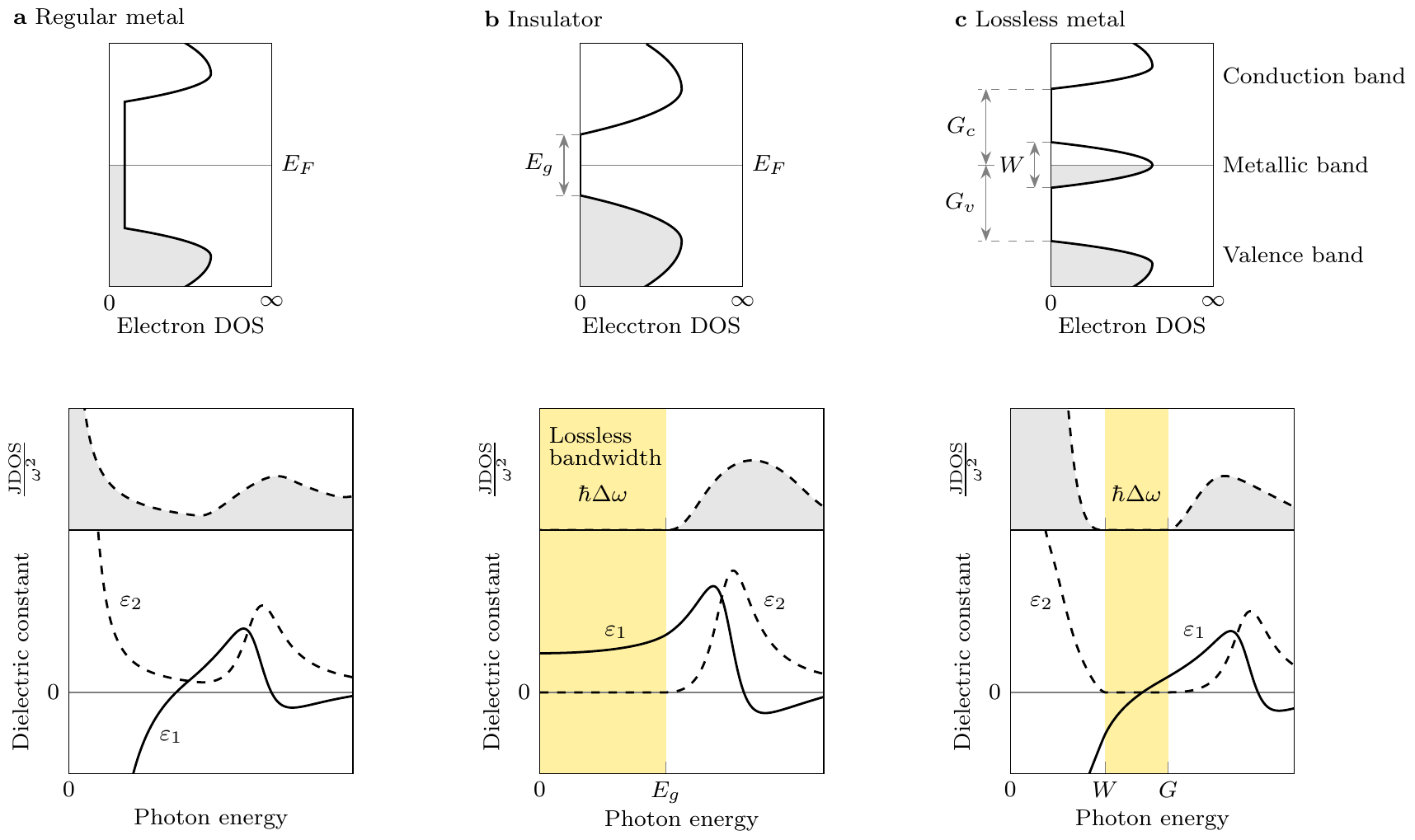} 
	\caption{\textbf{Comparison between lossless metal, regular metal and insulator}.\label{fig:banddos}
		\textbf{a}, Regular metal is lossy across the whole bandwidth.
		\textbf{b}, Insulator is lossless below the bandgap~($E_g$).
		\textbf{c}, Lossless metal, satisfying the loss criteria~($W<G={\rm min}\{G_c,G_v\}$),
		have a lossless bandwith $ \hbar\Delta \omega$ from $W$ to $G$.
	}
\end{figure*}

Optical photons and electrons, the most important carriers for information and energy, have never been guided efficiently inside the same material due to the high absorption loss of metals.
Dielectrics transport photons as in optical fibers and metals conduct electrons as in copper wires.
At the photon energy much above that of the lattice phonons, the optical absorption is dominated by the electronic transitions~(direct or indirect) between the occupied and empty states. For all existing metals, at any optical frequency, there are always occupied and empty states available for the absorption process to take place~(Fig.~\ref{fig:banddos}a), causing the common misimpression that metals have to be lossy due to the so called free-carrier absorption.

Fortunately, the carriers in solids are never free electrons, since their density of sates~(DOS) can be highly modified by the lattice potentials.
It is indeed theoretically possible~\cite{medvedeva2005combining,khurgin2010search} for a metal to have zero single-electron transition rate in an energy bandwidth in which the photon absorption can be as low as that in insulators~(Fig.~\ref{fig:banddos}b).
This happens when there is a well-isolated metallic band~(of band width $W$ as in Fig.~\ref{fig:banddos}c) at the Fermi level with a lossless bandwidth of
\begin{equation}
	\hbar\Delta \omega={\rm min}\{G_c,G_v\}-W>0.
	\label{eq::LLcriteria}
\end{equation}
Eq.~\ref{eq::LLcriteria} is the lossless criteria in which the band separation, minimal values of $G_c, G_v$, have to be larger than $W$. $G_c, G_v$ are the energy differences from the Fermi level to the edges of conduction and valence bands.
Here, \emph{lossless} means the absence of single-electron absorption --- the same sense that insulators are considered lossless below the bandgap where the imaginary part of the dielectric constant~($\varepsilon_1+i \varepsilon_2$) vanishes~($\varepsilon_2=0$). Higher order processes, involving multiple electrons, multiple phonons and multiple photons, have much lower probabilities and are neglected.
This picture of lossless metals was proposed by Medvedeva and Freeman~\cite{medvedeva2005combining} in the context of ideal transparent conductors and by Khurgin and Sun~\cite{khurgin2010search} in the context of ideal plasmonic metamaterials. 
Although electrides~\cite{medvedeva2005combining} and 2D metals~\cite{gjerding2017band} have been suggested as possible directions, lossless metals remain elusive.

The scientific and technological impact of lossless metal is far-reaching.
For transparent electronics~\cite{medvedeva2005combining,zhang2015intrinsic,brunin2019transparent}, the inevitable trade-off between conductivity and transmissivity could be broken.
For plasmonics~\cite{khurgin2010search,khurgin2012reflecting,gjerding2017band} ($\varepsilon_1<0$), optical devices could be losslessly shrinked to deep subwavelength scales, enhancing light-matter interactions at unprecedented level~\cite{miller2016fundamental}.
For metamaterials~\cite{khurgin2010search,khurgin2012reflecting,gjerding2017band}, 
the numerous remarkable scientific demonstrations, such as cloaking and perfect lens, may deliver practical applications.
In reality, even if $\varepsilon_1$ is positive and $\Delta \omega$ is narrow, lossless metals still have exciting consequences, such as the waveguiding both electricity and light!
We notice that the three-band configuration of lossless metals is similar to, although having very different requirements, that of the intermediate-band materials~\cite{luque1997increasing}, a class of compounds expected to increase the solar-cell efficiency by absorbing a wider spectrum. The field of intermediate-band solar cells also lacks candidate materials~\cite{baquiao2019computational}.

In the rest of the paper, we first introduce the joint-density-of-states~(JDOS) picture~\cite{gjerding2017band} for understanding lossless metals. Then, a large-scale computational search is performed to find all potential candidates within the framework of the band theory. The candidates are ranked by the figures of merits~($W$ and $\Delta\omega$) and classified into three classes for their potential conduction paths in the real space. After compiling the previous experimental data on the candidate materials, we explain the deficiency of current theory in predicting lossless metals and discuss the next research agenda.

\section{JDOS model of absorption}
The dielectric constant~($\rm \varepsilon_2=0$) of a lossless metal cannot be described by the regular
Drude model, whose intra-band absorption is $\rm \varepsilon_2^{intra}(\omega)=\frac{\omega^2_p \Gamma}{\omega(\omega^2+\Gamma^2)}$, where $\rm\omega_p$ is the plasma frequency and $\Gamma$ is the constant electron scattering rate.
It is proposed in Ref.~\cite{gjerding2017band} that a frequency-dependant $\rm \Gamma(\omega) \propto JDOS(\omega)$, with JDOS proportionality, correctly gives $\rm \varepsilon_2=0$ inside the lossless bandwidth.
The scattering rate is roughly proportional to the JDOS of electrons, because the JDOS captures the phase space of the initial and final states.
Here, $\rm JDOS(\omega)= \int_{0}^{\omega}{\rm DOS}(\omega'-\omega){\rm DOS}(\omega')d\omega'$ relaxes the momentum conservation to account for the indirect transitions.

Our JDOS-Drude intra-band model is
\begin{align}
	\Gamma(\omega) = \frac{{\rm JDOS}(\omega)}{\omega\cdot{\rm DOS}^2(E_F)}\Gamma_{\rm dc},
\end{align}
where ${\rm DOS}(E_F)$ is the DOS at Fermi energy.
At zero frequency, $\Gamma(0)=	\Gamma_{\rm dc}=\epsilon_0\frac{  \omega_p^2}{4\pi }\rho_{\rm dc}~(\sim \rm 30THz~in~silver)$, in which $\rho_{\rm dc}$ is the measurable resistivity of direct current~(dc) and $\epsilon_0$ is the vacuum permittivity. Compared to that in Ref.~\cite{gjerding2017band}, there is no fitting parameter in our model.

We find that $\varepsilon_2 \propto \rm JDOS/\omega^2$ 
can be used as a qualitative estimation of material absorption~(as shown in Fig.~\ref{fig:banddos}) for both intra- and inter-band losses.
At low frequencies, the intra-band loss dominates and $\varepsilon^{\rm intra}_2(\omega\ll \Gamma_{\rm dc}) = \frac{\omega_p^2}{\Gamma_{\rm dc}{\rm DOS^2}(E_F)}\rm JDOS/\omega^2$.
At high frequencies, the inter-band loss dominates and the $\rm \varepsilon_2^{inter} \propto JDOS/\omega^2$ scaling is still valid. Since the transition rate, in Fermi's golden rule, is proportional to the vector potential squared $A^2=(E/\omega)^2$, where $E$ is the electric field~\cite{grosso2013solid}.
Since the band theory only works in the single-particle regime,
it is beneficial to formulate everything using DOS and JDOS that are valid for both interacting and non-interacting electron systems.
Equivalently, lossless means zero JDOS.

\section{High-throughput search}\label{sec:HTsearch}

\begin{figure*}[t]
	\centering
	\includegraphics{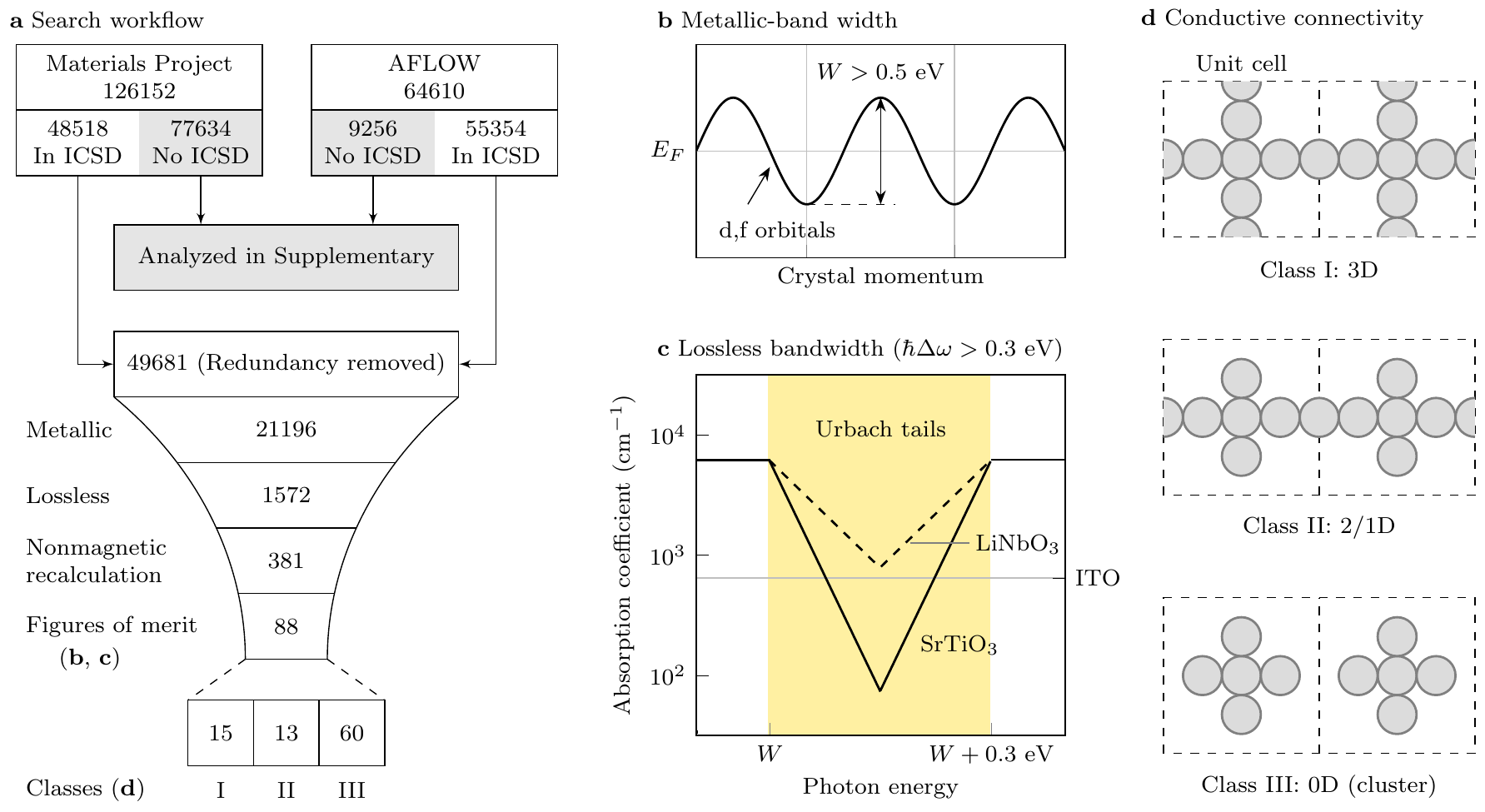}
	\caption{\label{fig:flow} \textbf{The search workflow and analysis of lossless-metal candidates}.
		\textbf{a}, Workflow diagram of the screening steps. \textbf{b}, First figure of merit: the metallic-band width containing d/f orbitals.
		\textbf{c}, Second figure of merit: the lossless bandwidth constrained by Urbach absorption tails.
		\textbf{d}, The 88 candidates are classified by their real-space conductive connectivity, whether the relevant atoms are close enough for current flow.}
\end{figure*}
We search for lossless metals in two computational online material databases of Materials Project (https://materialsproject.org) and AFLOW (http://aflow.org) with a total of 190762 material entries~(in September 2020), shown in Fig.\ref{fig:flow}a.
In the maintext, we analyze the materials that have experimental relevance --- those listed in the Inorganic Crystal Structure Database (ICSD: https://icsd.fiz-karlsruhe.de), a comprehensive collection of experimental crystal data.
The rest of the materials, not listed in ICSD, are analyzed in the Appendix~\ref{apdx:b}.
Among the 92153 ICSD IDs covered in the two databases, there are 49681 unique compounds after we merge the redundant entries of identical stoichiometries and space groups. The band-structure data are available for 42729 compounds and 21196 of them are metals --- having partially-filled metallic bands at the Fermi levels.

Each band structure in the databases represents a ground state, either magnetic or non-magnetic, predicted by the density functional theory~(DFT)~\cite{jain2011high,curtarolo2012aflow}.
For non-magnetic ground states, we can directly apply the lossless criteria of Eq.~\ref{eq::LLcriteria} and the energy limit of the search is up to 10~eV.
Among the 14463 nonmagnetic~(spin-degenerate) metals, we find 431 lossless-metal candidates.
For the rest metals with magnetic ground states~(spins splitting), we are actually more concerned about their properties at room temperature~(for optical applications) above most magnetic-transition temperatures.
In order to estimate their nonmagnetic properties from the available magnetic band structures while keeping enough potential candidates at this step, we relax the lossless criteria to be one of the spin bands satisfying the lossless criteria and obtain 1141 loss-metal candidates.
(791 of them satisfies the lossless criteria unconditionally considering both spins.)
This spin-criteria is justified by the fact that magnetism mostly split the two spins in the band structures.

Aiming at applications under room-temperature where most materials lose their magnetic ordering, we perform non-magnetic recalculations for all 1572 candidates~(431 non-magnetic plus 1141 magnetic candidates), with improved accuracy, to double-check whether they still satisfy the lossless criteria.
The high-throughput computations in the databases have compromised settings for speed, while we include the spin-orbit coupling, f-orbitals, as well as experimental lattice data from ICSD.
Our ab-initio recalculations are performed using the Vienna Ab initio Simulation Package~(VASP) with the generalized gradient approximation~(GGA) plus Coulomb repulsion~($U$), detailed in the Appendix~\ref{apdx:a}.

First, we drop the 306 candidates whose chemical formula is in fact different from their ICSD entries. They miss hydrogen atoms in the calculations due to the lack of the hydrogen positions in ICSD.
Second, we drop 242 candidates whose DFT calculations do not converge, mostly due to the inclusion of the f-orbitals from rare earth elements such as Terbium~(Tb) and Ytterbium~(Yb).
Third, in the remaining 1024 candidates with correct composition and converging band structures, 637 candidates no longer satisfy the lossless criteria.
Our recalculation is more restrictive for two major reasons.
On one hand, the magnetic candidates, from the previous screening step, are not necessarily lossless for their nonmagnetic band structures.
On the other hand, the database band structures are based on the computationally-relaxed structures~(with GGA DFT), which usually converge to larger lattice constants than the experimental values. An expanded lattice leads to flatter bands so the lossless criteria is easier to be met with the relaxed structures compared to the experimental structures.
After all, we have 381 lossless-metal candidates left after the nonmagnetic recalculations.

\section{Figures of merit}

The metallic-band width~($W$) and lossless bandwidth~($\hbar\Delta\omega$) are the two figures of merit for lossless metals.
Larger $W$ usually implies higher mobility and better conductors while larger $\hbar\Delta\omega$ means lower optical absorption.

We remove candidates whose metallic-band width~($W$) is too narrow to be metals in reality~(Fig.~\ref{fig:flow}b). It is well known that a narrow electronic band at Fermi level results in correlation effects~(low kinetic energy compared to Coulomb repulsion $U$) that DFT has a limited predictive power.
For example, $W<U$ is considered a condition for Mott insulators, where the $U$ values are mentioned in Appendix~\ref{apdx:a}.
This is especially true for d/f orbitals that are intrinsically narrow-banded and spatially localized.
Sadly, the lossless metal falls right into this category, because a narrow metallic band is required to satisfy the lossless criteria and 355 of the 381 candidates contain d/f orbitals~(from transition-metal elements) in their isolated metallic bands.
Consequently,
we set an empirical lower limit of 0.5~eV on the metallic-band width and remove 229 candidates whose metallic bands are made of d/f electrons and, at the same time, $W<0.5$~eV. We do not constrain s/p orbitals which are usually handled well by DFT.
It is worth-noting that, in the case when the metallic band~($W$) consists of several individually isolated sub-bands, only the bandwidth of the partially-filled isolated sub-bands are used as the figure of merit.

We also remove candidates whose lossless bandwidth~($\hbar\Delta\omega$) is too narrow to support low optical absorption in reality~(Fig.~\ref{fig:flow}c).
Although the electron DOS vanishes abruptly at the band edge in a perfectly periodic crystal, real crystals are imperfect and it is well known that the absorption edge drops exponentially with photon energy --- the Urbach tails~\cite{urbach1953the,mitra1975optical}.
The slope of the tail, reflecting crystal disorder, is quantified by the Urbach energy --- a temperature-dependent energy scale across which the absorption coefficient drops by 1/e. 
In Fig.~\ref{fig:flow}c, the room-temperature Urbach energies of two ternary compounds, 34~meV in $\rm SrTiO_3$~\cite{goldschmidt1987fundamental} and 78~meV in $\rm LiNbO_3$~\cite{bhatt2012urbach}, are used to estimate the minimum lossless bandwidth needed to achieve a low enough absorption loss. 
We do not use the smallest Urbach energies~($\sim$10meV) in Si and GaAs, because most other materials could not be grown with such high crystalline qualities. 
We find that $\hbar\Delta\omega>$0.3~eV is necessary to realize a lower absorption coefficient than that of the indium tin oxide~(ITO~\cite{konig2014elect}), the standard transparent conductor.
The 0.3~eV lower bound removes 113 candidates~(out of 381), including the $\rm LiTi_2O_4$ --- a novel transparent superconductor discussed in the later section.

\begin{figure*}[t]
	\centering
	\includegraphics{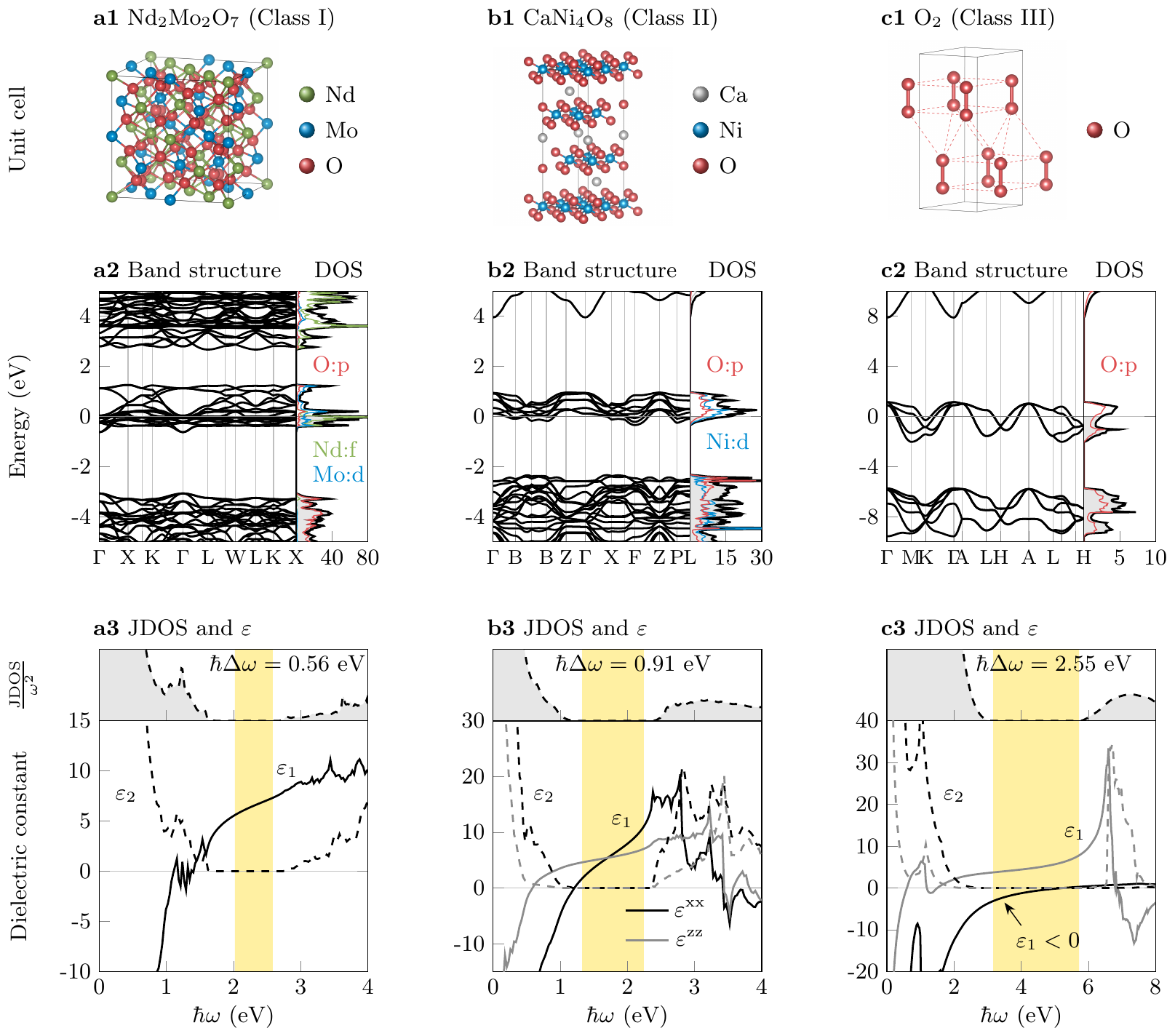}
	\caption{\label{fig:example}\textbf{Three examples of lossless-metal candidates}. \textbf{a1}-\textbf{a3}, $\rm Nd_2Mo_2O_7$ ($\rm{Fd\bar{3}m}$). \textbf{b1}-\textbf{b3}, $\rm CaNi_4O_8$ ($\rm{Fd\bar{3}m}$). \textbf{c1}-\textbf{c3}, $\rm O_2$ (P64/mmc). For each material, we plot the crystal structure, band structure, density of states, joint density of states, and dielectric constants. The atomic orbitals constitute the metallic bands are labeled in \textbf{a2-c2}.
	Data of the rest candidates are plotted in the Appendix.
	}
\end{figure*}

\begin{table*}[h!]
	\centering
	\small
	\renewcommand\arraystretch{1.2}
	\newcolumntype{L}[1]{>{\raggedright\arraybackslash}p{#1}}
	\newcolumntype{C}[1]{>{\centering\arraybackslash}p{#1}}
	\newcolumntype{R}[1]{>{\raggedleft\arraybackslash}p{#1}}
	\newcolumntype{Y}[1]{>{\centering\arraybackslash\columncolor[RGB]{255,240,162}}p{#1}}
		\caption{\label{tab:class} \textbf{Lossless-metal candidates}. Top candidates are given for each class, sorted by the lossless bandwidth~($\hbar\Delta\omega$). The experiment references of each material are provided in the Appendix~\ref{apdx:d}.}
	\begin{ruledtabular}
		\begin{tabular}{C{0.03\textwidth}L{0.15\textwidth}C{0.07\textwidth}C{0.07\textwidth}Y{0.07\textwidth}C{0.14\textwidth}C{0.15\textwidth}C{0.12\textwidth}C{0.10\textwidth}}
			&Formula	&Space group	&$W$ (eV)	&$\hbar\Delta\omega$ (eV)	&Conductive connectivity	&Magnetism~($T_C$/$T_N$) distortion	&Color	&Conduction	\\
			\hline
			\multirow{16}*{\begin{minipage}{0.025\textwidth}\rotatebox{90}{Class \romanNum{1}}\end{minipage}}&$\rm CoCO_3$	&167	&0.61	&1.95	&3D	&AFM~(18K)	&	&Insulator  \\
			&$\rm K_5CeCo_2{(NO_2)}_{12}$	&201	&0.85	&1.70	&3D	&	&	&	\\
			&$\rm Ba_2CoMoO_6$	&225	&1.12	&1.32	&3D	&AFM~(20K)	&Black	&	\\
			&$\rm Sr_2CoMoO_6$	&225	&1.30	&1.24	&3D	&AFM~(36K)	&Black	&Insulator	\\
			&$\rm NiMoO_4$	&12	&0.71	&0.91	&3D	&FM~(22K)	&Light green	&	\\
			&$\rm Ba_2GdMoO_6$	&225	&0.93	&0.88	&3D	&	&Black	&Insulator	\\
			&$\rm KCoF_3$	&221	&1.67	&0.80	&3D	&AFM~(114K)	&Rosy	&Insulator	\\
			&$\rm Nd_2Mo_2O_7$	&227	&2.02	&0.59	&3D	&FM~(90K)	&Black	&Metal	\\
			&$\rm CoTiO_3$	&148	&0.74	&0.44	&3D	&AFM~(38K)	&Green	&Insulator	\\
			&$\rm CaCr_2O_4$	&62	&1.69	&0.42	&3D	&AFM~(43K)	&	&Insulator	\\
			&$\rm CaCu_3Ti_4O_{12}$	&204	&0.89	&0.40	&3D	&AFM~(24K)	&	&Insulator	\\
			&$\rm CuSiO_3$	&148	&1.13	&0.37	&3D	&AFM~(110K)	&Black	&Insulator	\\
			&$\rm Li_2IrO_3$	&70	&2.75	&0.36	&3D	&FM~(38K)	&	&Insulator	\\
			&$\rm Bi_2Cu_5B_4O_{14}$	&1	&0.90	&0.35	&3D	&FM~(25K)	&Green	&Insulator	\\
			&$\rm NiF_2$	&58	&1.67	&0.30	&3D	&AFM~(73K)	&	&Insulator	\\\cdashline{2-9}
			&$\rm LiTi_2O_4$	&227	&2.37	&0.12	&3D	&Superconductor~(13.7K)	&Blue	&Metal	\\
			\hline
			\multirow{13}*{\begin{minipage}{0.025\textwidth}\rotatebox{90}{Class \romanNum{2}}\end{minipage}}&$\rm CaNi_4O_8$	&166	&1.33	&0.91	&2D	&	&	&	\\
			&$\rm Sc_2Cu_2O_5$	&33	&0.61	&0.80	&1D	&AFM~(16K)	&	&Insulator	\\
			&$\rm VCl_3$	&148	&1.19	&0.66	&2D	&FM	&Violet	&Insulator	\\
			&$\rm CuZrTiO_5$	&19	&0.94	&0.55	&1D	&AFM	&Green	&Insulator	\\
			&$\rm CuInOPO_4$	&62	&0.69	&0.51	&1D	&	&Green	&	\\
			&$\rm Li_2CuO_2$	&71	&1.13	&0.48	&1D	&AFM~(9K)	&Brown 	&Insulator	\\
			&$\rm BaCu_2Si_2O_7$	&62	&0.95	&0.42	&1D	&AFM~(9.2K)	&Dark blue	&Insulator	\\
			&$\rm Co_2B_2O_5$	&2	&1.55	&0.41	&1D	&AFM~(45K)	&Violet	&	\\
			&$\rm Ca_3Cu_5Si_9O_{26}$	&15	&0.96	&0.38	&2D	&	&Bluish-green	&Insulator	\\
			&$\rm CsNiBr_3$	&194	&1.14	&0.37	&1D	&AFM~(70K)	&Orange-brown	&	\\
			&$\rm CaCuGe_2O_6$	&14	&0.58	&0.36	&1D	&AFM, Jahn-Teller	&	&Insulator	\\
			&$\rm Cu(OH)F$	&14	&1.75	&0.36	&2D	&AFM, Jahn-Teller	&	&	\\
			&$\rm KSnS_2$	&166	&1.14	&0.36	&2D	&	&	&	\\
			\hline
			\multirow{18}*{\begin{minipage}{0.025\textwidth}\rotatebox{90}{Class \romanNum{3}}\end{minipage}}&$\rm NaO_2$	&205	&0.71	&3.57	&0D	&AFM~(193K)	&	&Insulator  \\
			&$\rm O_2$	&194	&3.17	&2.55	&0D	&High-pressure	&Dark red	&Insulator	\\
			&$\rm Rb_4O_6$	&220	&0.42	&2.44	&0D	&AFM	&Black	&Insulator	\\
			&$\rm LiO_2$	&58	&1.84	&2.28	&0D	&AFM~(7K)	&	&Insulator	\\
			&$\rm K_2BaCo{(NO_2)}_6$	&69	&1.14	&1.61	&0D	&Jahn-Teller	&	&Insulator	\\
			&$\rm Nb_2{(PO_4)}_3$	&167	&1.16	&1.52	&0D	&	&Black	&Insulator	\\
			&$\rm Rb_2NbCl_6$	&225	&0.71	&1.52	&0D	&Jahn-Teller	&	&	\\
			&$\rm RbSb$	&216	&0.96	&1.27	&0D	&	&	&Insulator	\\
			&$\rm K_2TaCl_6$	&225	&0.98	&1.03	&0D	&Jahn-Teller	&Black	&Insulator	\\
			&$\rm K_3Na{(RuO_4)}_2$	&15	&0.59	&0.97	&0D	&AFM~(70K)	&Black	&Insulator	\\
			&$\rm LiBa_2Cu_3O_6$	&69	&0.56	&0.96	&0D	&Jahn-Teller	&	&	\\
			&$\rm Sr_2CoWO_6$	&225	&1.65	&0.77	&0D	&AFM~(24K)	&Dark brown	&Insulator	\\
			&$\rm Ba_2MgReO_6$	&225	&1.50	&0.75	&0D	&AFM	&Black blue	&Insulator	\\
			&$\rm KRuO_4$	&88	&0.78	&0.74	&0D	&AFM~(150K)	&Black	&Insulator	\\
			&$\rm Cu{(HCOO)}_2$	&14	&0.97	&0.69	&0D	&AFM~(17K)	&Light blue	&Insulator	\\
			&$\rm SrCu_2{(BO_3)}_2$	&140	&0.99	&0.59	&0D	&AFM~(1.4K)	&Blue	&Insulator	\\
			&$\rm CuSe_2O_5$	&15	&0.84	&0.48	&0D	&AFM	&Green	&Insulator	\\
			&$\rm Ba_2CoWO_6$	&225	&1.93	&0.44	&0D	&AFM~(17K)	&Brown	&Insulator	\\
		\end{tabular}
	\end{ruledtabular}
	\begin{flushleft}{
			$T_C$/$T_N$: Curie/Neel temperature of magnetic transition. 
			FM: ferromagnetism.
			AFM: antiferromagnetism.
	}\end{flushleft}
\end{table*}

\section{Real-space analysis}

88 candidates are left with reasonably wide metallic-band width~($W$) and lossless bandwidth~($\hbar\Delta\omega$).
Theoretically, a large $\hbar\Delta\omega$ ensures optical transparency, but a wide $W$ cannot always ensure electrical conduction.
Because conduction depends more on the real-space wave-functions than the reciprocal-space band dispersions.
For example, localized unpaired electrons cannot flow in the lattice, but they appear as ``metals''~(partially-filled bands across Fermi level) in the band theory~(assuming global Bloch modes across the crystal).
So we study whether the conductive atoms --- the atoms contributing to the electron DOS at the Fermi level~($\pm$50~meV)--- are closely-spaced enough to connect a current path in the crystal.
The connectivity of these conductive atoms is determined by the spatial overlap of their atomic radii~\cite{slater1964atomic}.
Illustrated in Fig.~\ref{fig:flow}d, the \emph{conductive connectivity} can be three dimensional~(3D), low-dimensional~(2D, 1D), or in isolated clusters~(0D), according to which we classify the 88 candidates into three classes with 15, 13, and 60 materials, respectively.
The candidates of class I, II and most candidates of class III are tabulated in Table~\ref{tab:class}.

The materials in Class I have 3D conductive connectivity~(for potential current paths) in vertex‐sharing or edge-sharing polyhedrons. 
The narrow metallic band originates from the localized d/f orbitals. The isolation of the metallic bands is partly due to the large and uniform splitting of the d/f bands in the crystal fields of high symmetries~(high space groups).
A typical material example is pyrochlore molybdate~($\rm Nd_2Mo_2O_7$) shown in Fig.~\ref{fig:example}a, that is experimentally verified to be a metal~(resistance decreases with the decrease of temperature) and is a ferromagnet below the Curie temperature of $\sim90$K~\cite{kezsmarki2004charge, kezsmarki2005magneto}.

The materials in Class~\romanNum{2} have low-dimensional conductive connectivity, in which the atoms connect more closely in one or two dimensions. The weaker coupling in certain directions facilitates bands of less dispersions and narrower bandwidths~\cite{gjerding2017band}.
A typical example is calcium nickelate $\rm CaNi_4O_8$ in Fig.~\ref{fig:example}b.
Nickel dioxide ($\rm NiO_2$) is a layered insulator having an isolated narrow conduction band~(see Appendix~\ref{apdx:c}).
Calcium intercalation, in $\rm NiO_2$, supplies itinerant electrons and raises the Fermi level into the middle of this narrow conduction band, thus satisfying the lossless-metal criteria. 
The same mechanism applies to $\rm KSnS_2$~(detailed in Appendix~\ref{apdx:c}), in which Potassium intercalates into the tin disulfide ($\rm SnS_2$) --- a common layered material.
Neither $\rm CaNi_4O_8$~\cite{bityutskij1984electron} nor $\rm KSnS_2$~\cite{bronold1991alkali} have been synthesized in single-crystal forms.

The materials in Class~\romanNum{3} consist of distanced atomic clusters~(forming narrow metallic bands), which are easier, than the other two classes, to satisfy the lossless criteria in DFT calculations. However, the weak couplings between the clusters impede electric conduction.
A typical example is the solid oxygen $\rm O_2$~(P64/mmc) in Fig.~\ref{fig:example}c, showing a huge lossless bandwidth of $\hbar\Delta\omega=$2.55~eV with an isolated metallic-band width of $W=$3.17~eV.
This molecular crystal consists of dense arrangement of diatomic molecules under high pressure.
In experiments,
this phase forms at $\sim$17GPa and is not conducting.
The oxygen metallizes under a much higher pressure around 100GPa~\cite{akahama1995new,shimizu1998oxygen} (see Appendix~\ref{apdx:c} for the $\rm O_2$ phase diagram).

\section{Existing experiments}
We go through the existing experimental literature on the candidates in Table~\ref{tab:class} and note their key feedback information (the references are listed in Appendix~\ref{apdx:c}).
The cold truth is that, experimentally, most candidates are false metals and real insulators~\cite{malyi2020false}, except $\rm Nd_2Mo_2O_7$ (and the $\rm LiTi_2O_4$ discussed in the next section).
The false positive prediction of metals is a common problem of DFT for complex materials involving transition-metal elements, d/f electrons, narrow bands, magnetism, or correlation effects.
In reality, these materials usually find insulating ground states of lower energies, than the predicted metallic states, by structure distortions, magnetic orderings, or electron-electron interactions.
As can be seen in Table~\ref{tab:class}, magnetism and cooperative Jahn-Teller distortion are observed for most candidates.

Magnetic orders are difficult to predict.
Even the para-magnetic states above the temperatures of magnetic phase transitions, containing disordered local magnetic moments, remains a challenge for DFT~\cite{malyi2020false}.
So far, our recalculations are nonmagnetic, assuming zero local magnetic moments.
The above facts cast doubts on whether more candidates would turn out to be true metals or whether the electronic state of $\rm Nd_2Mo_2O_7$~(Fig.~\ref{fig:example}a) is predicted accurately enough to satisfy the lossless criteria in experiments.

Optical loss is a property not having much data in the literature. In most reports, only the sample color is mentioned. But we do require the broad-band dielectric constants $\varepsilon(\omega)$, using tools such as the ellipsometry or transmission/reflection, to verify the predictions.
Note that the optical absorption~($\varepsilon_2$) is sensitive to the sample quality, so low loss is generally harder to verify experimentally than electric conduction.

\section{The case of $\rm \mathbf{LiTi_2O_4}$}

One positive prediction of our search is lithium titanate $\rm{LiTi_2O_4}$~($\rm{Fd\bar{3}m}$), a well studied metal having a superconducting ground state~\cite{johnston1973high}, rather than the magnetic orders or Jahn-Teller distortions observed for the false-metal candidates or the magnetic metal $\rm{Nd_2Mo_2O_7}$ in Table~\ref{tab:class}.
Thin-film $\rm{LiTi_2O_4}$ is transparent~\cite{kumatani2012growth} with a reasonably low optical loss~\cite{zhao2016investigation}. Prepared by the pulsed laser deposition, the film deteriorates in air and its quality is usually limited by oxygen vacancies. Bulk single crystals of $\rm{LiTi_2O_4}$  are difficult to grow~\cite{chen2003crystal}.

Although the above experimental feedback of $\rm{LiTi_2O_4}$ agrees with our predictions in Appendix~\ref{apdx:c}, its theoretical lossless bandwidth is too narrow~($\hbar\Delta\omega=0.12$~eV) to support an optical absorption much lower than that of ITO.
Nevertheless, 
it is encouraging to find that realistic metal satisfying the lossless criteria can exist.
A metal of a larger lossless bandwidth awaits identification.

\section{Future directions}

Although the feedback from existing experiments indicate the drawbacks of the current high-throughput approach in finding lossless metals, the candidates in Table~\ref{tab:class} still serve as reasonable starting points.
Efforts should be made to study their failure modes, their actual electronic structures and their optical properties.
These efforts involve ab-initio calculations beyond simple DFT~\cite{malyi2020false} such as the dynamical mean-field theory~(DMFT), as well as the experimental attempts in growing and characterizing the high-quality single-crystal samples that most of the candidates lack.

The candidates also indicate promising directions for future searches.
Obviously, majority of the entries in Table~\ref{tab:class} are oxides~\cite{tsuda2000electronic}.
In Class I, transition-metal compounds with high spatial symmetries are promising material systems, for examples the double perovskites (such as $\rm Ba_2CoMoO_6$), pyrochlore~($\rm Nd_2Mo_2O_7$), and spinel~($\rm LiTi_2O_4$).
In Class II, layered insulators of a narrow band near Fermi level~(such as the $\rm SnS_2$ and $\rm NiO_2$ presented in Appendix~\ref{apdx:c}) worth the attention, since their Fermi levels could be tuned by intercalation.

Inspired by the solid oxygen in Class III, we propose a novel high-pressure route to lossless metals.
Conceptually, there are two ways to obtain narrow-band metals. One way is expanding the lattice constant of a metal~\cite{khurgin2010search} to narrow its bandwidth while maintaining its conduction.
However, there is no experimental technique to do that.
The other way is shrinking the lattice constant of a false-metal candidate in order to widen its metallic bandwidth into a metal, through an insulator-metal transition, while maintaining a finite lossless bandwidth.
The standard high-pressure technique using diamond-anvil cell is well suited for this purpose, since both the transmission/reflection and the resistance can be monitored when the pressure is applied to the sample~\cite{jayaraman1983diamond}.

\section{Conclusion}
We perform the first high-throughput screening for the elusive lossless metals.
Starting from 44660 distinctive inorganic materials in ICSD, we obtain 88 high-quality candidates, while most of them are found to be insulating in experiments.
Lossless metals are difficult to predict using the current condensed-matter theory due to the complexity of the candidate material systems.
Nevertheless, our results shine light on a few hopeful directions including oxide conductors, low-dimensional metals, and compressing the false-metal candidates.
Finally, we emphasize that our current search is far from complete, because the data quantity and accuracy in the databases are still under development. 
The search could also be extended to organic materials~\cite{xie2020electrically}.

\section*{acknowledgment}
We acknowledge the SC2 group at our institute led by Kui Jin for investigating $\rm LiTi_2O_4$. 
This work is supported 
by the the Chinese Academy of Sciences through
the Youth Innovation Promotion Association (2021008),
the Project for Young Scientists in Basic Research (YSBR-021), the Strategic Priority Research Program (XDB33000000), the Informatization Plan (CAS-WX2021SF-0102),
the Interdisciplinary Innovation Team,
and the International Partnership Program with the Croucher Foundation  (112111KYSB20200024), 
by the National Key R\&D Program of China (2017YFA0303800, 2017YFA0303800, 2018YFA0305700), by Natural Science Foundation of China (12025409, 11721404, 11974415, 12025409, 11721404, 11974415, 11925408, 11921004, 12188101), 
and by Beijing Natural Science Foundation (Z200008).

	
	\clearpage
	\onecolumngrid
	\appendix
	\setcounter{figure}{0}
\setcounter{table}{0}
\captionsetup[figure]{labelfont={bf},name={Fig.~S\!},labelsep=period,font=small}
\captionsetup[table]{labelfont={bf},name={Table.~S\!},labelsep=period,font=small}

\section{Settings for first-principle calculations}\label{apdx:a}
The calculations in the workflows are performed by VASP~\cite{kresse1996efficient} with standard generalized gradient approximation (GGA) of the Perdew–Burke–Ernzerhof (PBE)-type exchange-correlation potential~\cite{perdew1996generalized}.
The pseudopotential files that we used are from PAW~(projector augmented wave) datasets v.54 in VASP and are selected to be consistent with what are used in both databases~(Materials Project and AFLOW). The DFT + U method is based on the simplest rotationally invariant formulation while the values of U are consistent with those used in both databases~\cite{dudarev1998electron,curtarolo2012aflow}.
The cut-off energy of the plane wave basis set is set to be the ENMAX value in the pseudopotential file plus 25\%.
A $\Gamma$--centered Monkhorst-Pack grid with 40 k-points per $\text{\AA}^{-1}$ is used for the self-consistent calculations.
A doubly dense grid is used for the density-of-states calculations with an energy interval of 5~meV.
The k-points files along high symmetry lines for band structure calculations are consistent with the choices in Materials Project.
The spin-orbital coupling is included.
The optical dielectric constants are computed by WIEN2k~\cite{Blaha2001wien,ambrosch2006linear} using our JDOS model in the maintext.

\newpage
\section{Search workflow without ICSD ids}\label{apdx:b}
\begin{figure}[th!]
	\centering
	\includegraphics{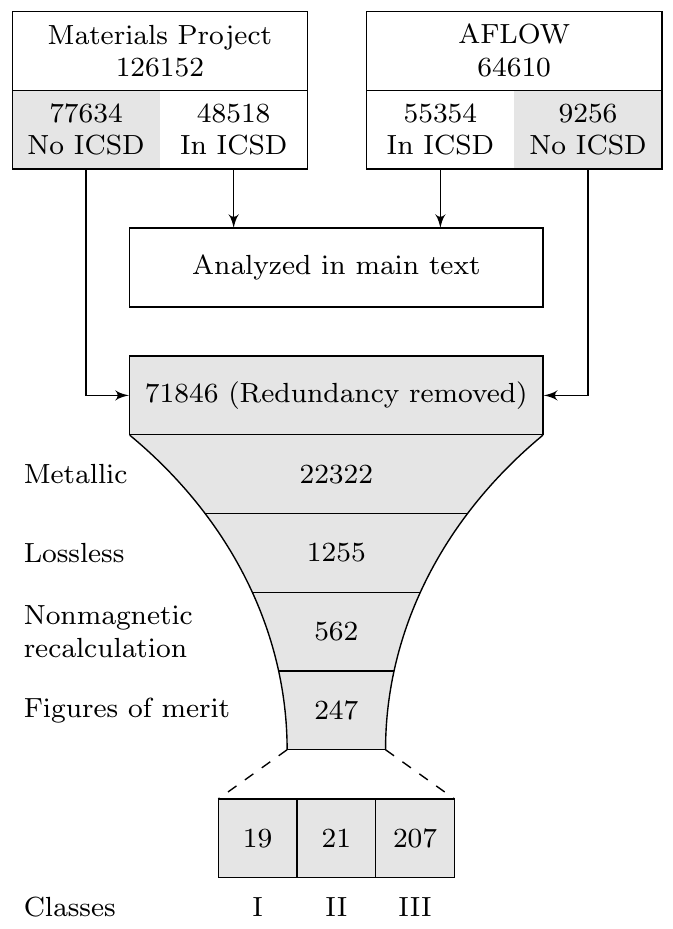}
	\captionsetup{justification=raggedright}
	\caption{\textbf{Workflow of the search for lossless metal among materials in the computational database of Material Project and AFlow but not in the experimental database of ICSD.}}
	\label{fig:noicsdflow}
\end{figure}

In the main text, we presented the workflow~(Fig.~2a) and candidates~(Table~I and Table~s\ref{tab:references}) from the high-throughput search among materials in ICSD, those have been reported in experiments.
Here we present the workflow~(Fig.~s\ref{fig:noicsdflow}) and candidates~(Table~s\ref{tab:noicsd}) from the high-throughput search among materials NOT in ICSD, those have NOT been reported in experiments.

\begin{table}[h]
	\centering
	\small
	\renewcommand\arraystretch{1.1}
	\newcolumntype{L}[1]{>{\raggedright\arraybackslash}p{#1}}
	\newcolumntype{C}[1]{>{\centering\arraybackslash}p{#1}}
	\newcolumntype{R}[1]{>{\raggedleft\arraybackslash}p{#1}}
	\newcolumntype{Y}[1]{>{\centering\arraybackslash\columncolor[RGB]{255,240,162}}p{#1}}
	\caption{\label{tab:noicsd} \textbf{Lossless-metal candidates without ICSD IDs}. Representative candidates are given for each class, ranked by the lossless bandwidth. The full list are tabulated in the Supplementary Excel file. These materials may not exist.}
	\begin{ruledtabular}
		\begin{tabular}{L{0.11\textwidth}|C{0.06\textwidth}|C{0.06\textwidth}|Y{0.06\textwidth}||L{0.11\textwidth}|C{0.06\textwidth}|C{0.06\textwidth}|Y{0.06\textwidth}||L{0.11\textwidth}|C{0.06\textwidth}|C{0.06\textwidth}|Y{0.06\textwidth}}
			\multicolumn{4}{c||}{Class~\romanNum{1}}&\multicolumn{4}{c||}{Class~\romanNum{2}}&\multicolumn{4}{c}{Class~\romanNum{3}}\\
			\hline
			Formula	&Space group	&$W$ (eV)	&$\hbar\Delta\omega$ (eV)	&Formula	&Space group	&$W$ (eV)	&$\hbar\Delta\omega$ (eV)	&Formula	&Space group	&$W$ (eV)	&$\hbar\Delta\omega$ (eV)\\
			\hline
			$\rm YbF_3$	&225	&4.76	&2.81		&$\rm ZrCoO_3$	&148	&0.61	&0.89		&$\rm F$	&194	&4.21	&7.90	\\
			$\rm Ba_2HoMoO_6$	&225	&1.09	&1.00		&$\rm Li_2Co_3TeO_8$	&166	&1.67	&0.83		&$\rm BaF_3$	&139	&2.20	&5.62	\\
			$\rm KLuO_3$	&221	&3.23	&0.79		&$\rm CrF_5$	&71	&0.78	&0.79		&$\rm RbNaO_3$	&221	&2.42	&4.74	\\
			$\rm MnNF_3$	&62	&1.45	&0.73		&$\rm Li_2CuF_4$	&65	&0.99	&0.77		&$\rm Cs_2NaLiF_6$	&225	&4.50	&3.38	\\
			$\rm CsTbO_3$	&221	&2.48	&0.60		&$\rm NaMnO_2$	&194	&1.08	&0.64		&$\rm CaSbF_6$	&148	&0.58	&3.20	\\
			$\rm Cu_2OF_2$	&141	&1.32	&0.56		&$\rm LiTiO_2$	&194	&1.59	&0.58		&$\rm LiSn{(PO_3)}_4$	&60	&0.56	&2.39	\\
			$\rm V_2Si_2O_7$	&227	&0.81	&0.47		&$\rm TlCoF_3$	&221	&1.55	&0.46		&$\rm RbS$	&123	&2.04	&2.25	\\
			$\rm LiAg_2F_4$	&227	&0.77	&0.40		&$\rm CuCl_2$	&166	&0.68	&0.45		&$\rm SbP_2O_7$	&14	&0.28	&1.68	\\
			$\rm Na_2Ni_5O_{10}$	&2	&1.54	&0.38		&$\rm NaNb_2O_4$	&57	&1.90	&0.38		&$\rm Li_8BiO_6$	&148	&0.65	&1.06	\\
			$\rm SrC_2$	&166	&5.04	&0.37		&$\rm Ni{(OH)}_2$	&12	&0.99	&0.34		&$\rm Ba_2ScReO_6$	&225	&1.84	&0.44	\\
		\end{tabular}
	\end{ruledtabular}
\end{table}

\clearpage
\section{More examples}\label{apdx:c}
\begin{figure}[h]
	\centering
	\includegraphics{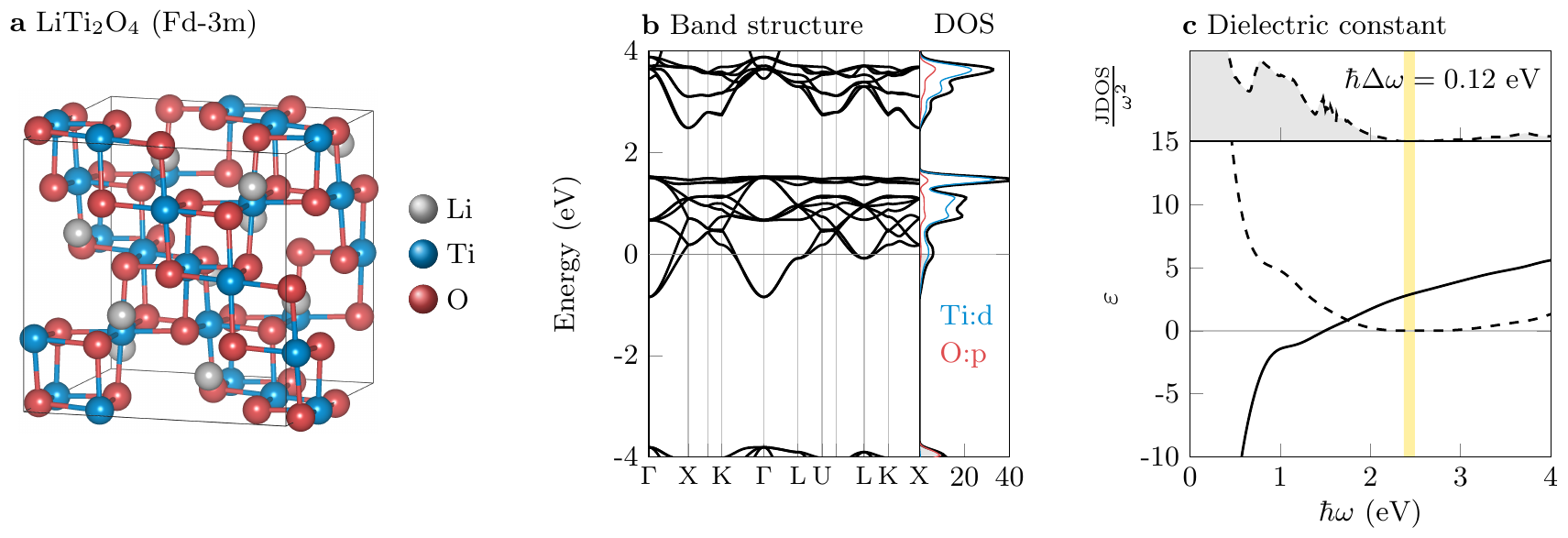}
	\caption{\label{fig:lto}\textbf{Class-I lossless metal candidate $\rm \mathbf{Li Ti_2O_4}$~(space group 227)}. The narrow lossless bandwidth of 0.12eV could be enlarged by pulling down the Fermi level.}
\end{figure}

\begin{figure}[h]
	\centering
	\includegraphics{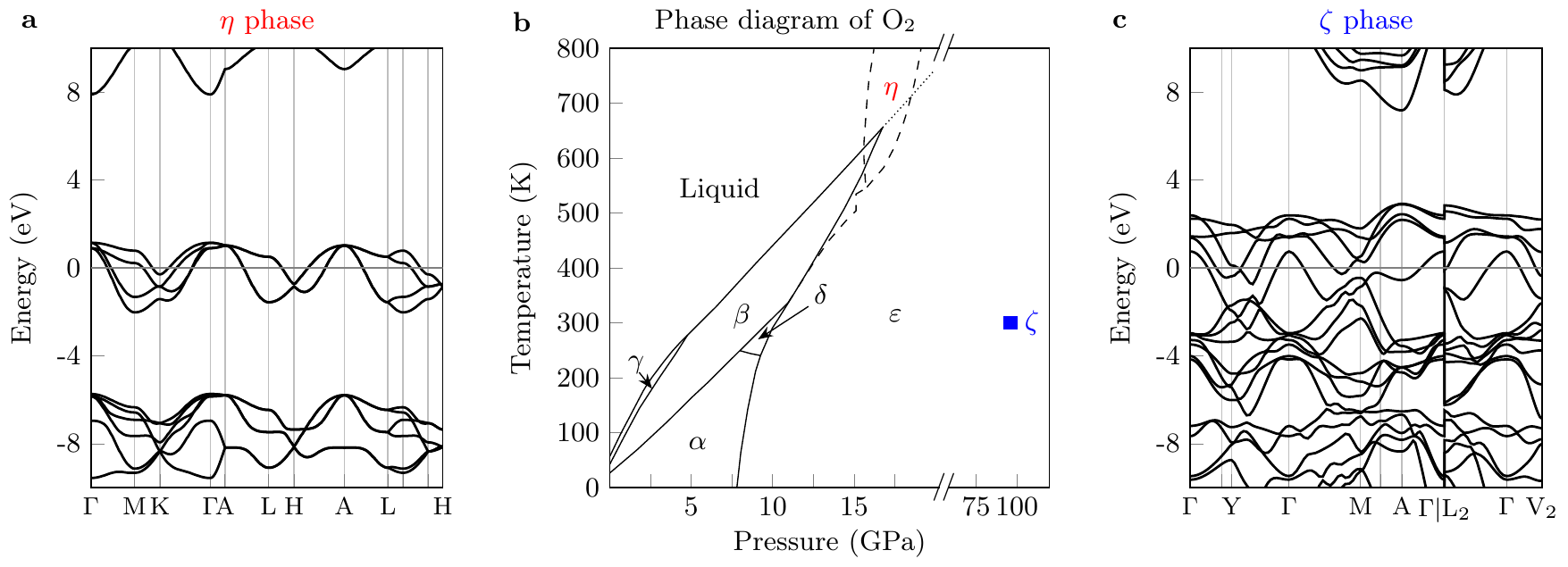}
	\caption{\label{fig:o2phase}\textbf{Class-III lossless metal candidate $\rm \mathbf{O_2}$~($\eta$-phase, space group 194)}. 
	\textbf{a}, $\eta$-phase band structure.
	\textbf{b}, The phase diagram of high-pressed $\rm O_2$~\cite{lundegaard2009structure}.
	\textbf{c}, Structure of metallic $\zeta$-$\rm O_2$ is still unknown. The crystal structure we use here is isomorphic to $\varepsilon$-$\rm O_2$~(space group 12) with the lattice parameters: $a=3.332$~\AA, $b=4.426$~\AA, $c=6.866$~\AA, $\beta=116.4^{\circ}$ \cite{akahama1995new}.}
\end{figure}

\clearpage
\begin{figure}[h]
	\centering
	\includegraphics[width=\textwidth]{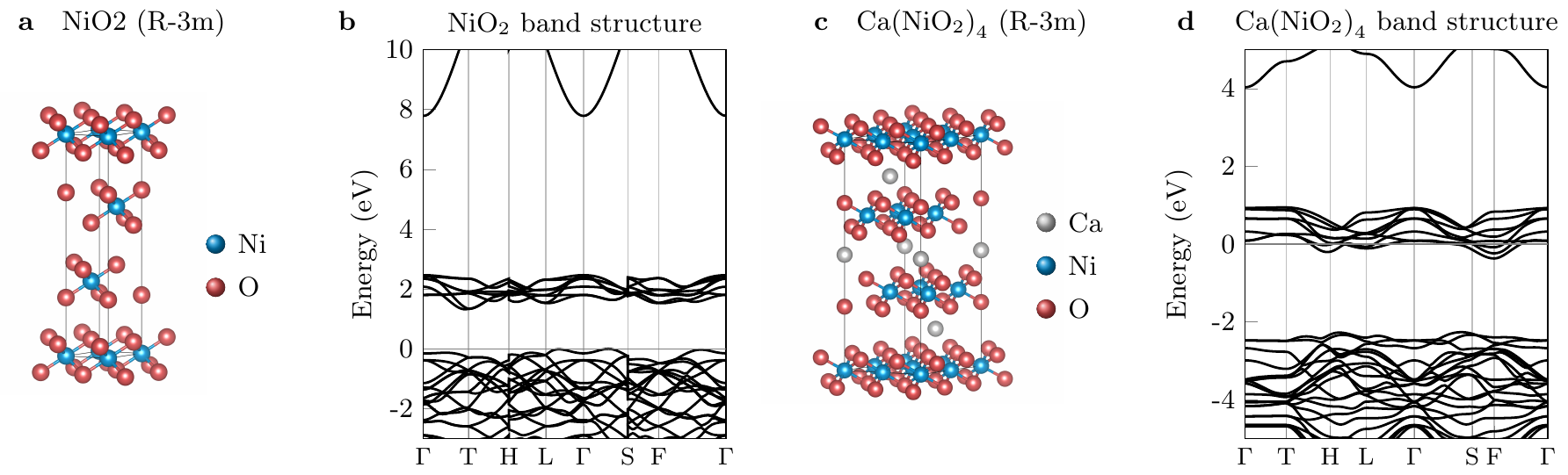}
	\caption{\label{fig:nio2}\textbf{Class-II lossless metal candidate $\rm \mathbf{Ca(NiO_2)_4}$~(space group 166), compared with $\rm \mathbf{NiO_2}$~(space group 166)}.}
\end{figure}

\begin{figure}[h]
	\centering
	\includegraphics[width=\textwidth]{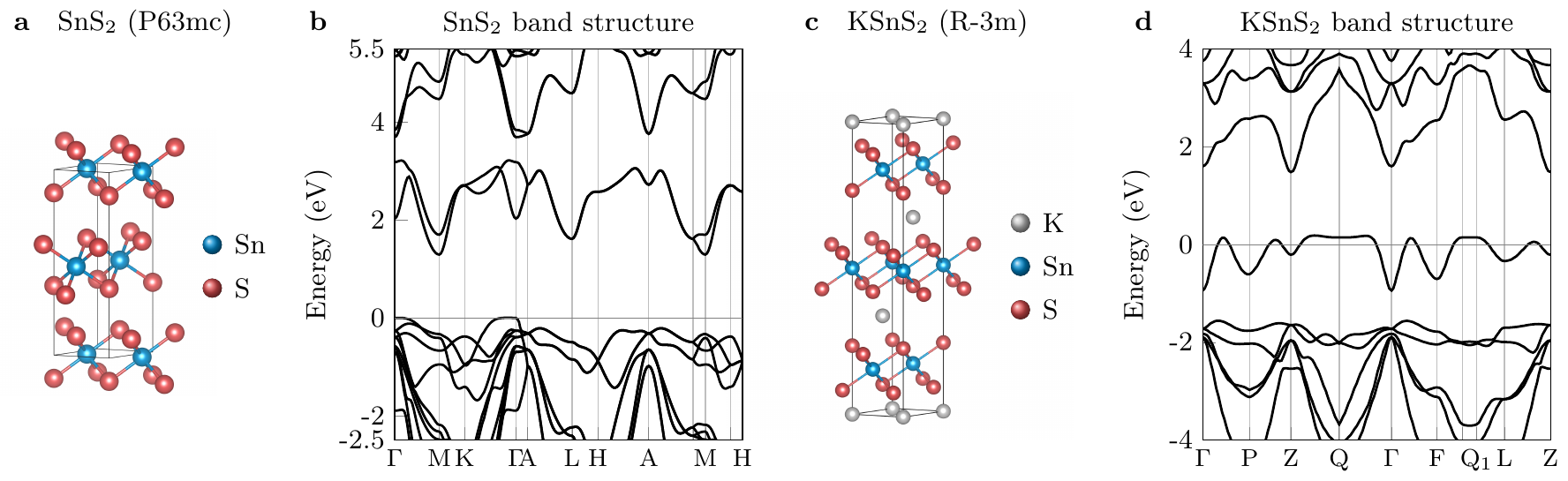} 
	\caption{\label{fig:sns}\textbf{Class-II lossless metal candidate $\rm \mathbf{KSnS_2}$~(space group 166), compared with $\rm \mathbf{SnS_2}$~(space group 186)}.}
\end{figure}

\vfill
\section{Experimental literature of ICSD candidates}\label{apdx:d}
\begin{table}[b]
	\centering
	\small
	\renewcommand\arraystretch{1.2}
	\newcolumntype{L}[1]{>{\raggedright\arraybackslash}p{#1}}
	\newcolumntype{C}[1]{>{\centering\arraybackslash}p{#1}}
	\newcolumntype{R}[1]{>{\raggedleft\arraybackslash}p{#1}}
	\newcolumntype{Y}[1]{>{\centering\arraybackslash\columncolor[RGB]{255,240,162}}p{#1}}
	\begin{ruledtabular}
		\begin{tabular}{C{0.02\textwidth}L{0.13\textwidth}C{0.05\textwidth}C{0.05\textwidth}Y{0.05\textwidth}C{0.12\textwidth}C{0.15\textwidth}C{0.12\textwidth}C{0.12\textwidth}C{0.09\textwidth}}
			&Formula	&Space group	&$W$ (eV)	&$\hbar\Delta\omega$ (eV)	&Conductive connectivity	&Magnetism~($T_C$/$T_N$) distortion	&Color	&Conduction	&Reference	\\
			\hline
			\multirow{16}*{\begin{minipage}{0.025\textwidth}\rotatebox{90}{Class \romanNum{1}}\end{minipage}}&$\rm CoCO_3$	&167	&0.61	&1.95	&3D	&AFM~(18K)	&	&Insulator  &\cite{pertlik1986structures,srivastava1970effect}\\
			&$\rm K_5CeCo_2{(NO_2)}_{12}$	&201	&0.85	&1.70	&3D	&	&	&	&\cite{ferrari1951metall}\\
			&$\rm Ba_2CoMoO_6$	&225	&1.12	&1.32	&3D	&AFM~(20K)	&Black	&	&\cite{kang2001electronic, martinez2002preparation}\\
			&$\rm Sr_2CoMoO_6$	&225	&1.30	&1.24	&3D	&AFM~(36K)	&Black	&Insulator	&\cite{gagulin2003synthesis, ivanov2005magnetoelectric}\\
			&$\rm NiMoO_4$	&12	&0.71	&0.91	&3D	&FM~(22K)	&Light green	&	&\cite{ehrenberg1995crystal}\\
			&$\rm Ba_2GdMoO_6$	&225	&0.93	&0.88	&3D	&	&Black	&Insulator	&\cite{mclaughlin2006magnetic, wallace2013variable}\\
			&$\rm KCoF_3$	&221	&1.67	&0.80	&3D	&AFM~(114K)	&Rosy	&Insulator	&\cite{markovin1976observation, ratuszna1979temperature}\\
			&$\rm Nd_2Mo_2O_7$	&227	&2.02	&0.59	&3D	&FM~(90K)	&Black	&Metal	&\cite{moritomo2001chemical,kezsmarki2004charge,kezsmarki2005magneto}\\
			&$\rm CoTiO_3$	&148	&0.74	&0.44	&3D	&AFM~(38K)	&Green	&Insulator	&\cite{newnham1964crystal, roset1986mossbauer, lin2006synthesis}\\
			&$\rm CaCr_2O_4$	&62	&1.69	&0.42	&3D	&AFM~(43K)	&	&Insulator	&\cite{toth2011120}\\
			&$\rm CaCu_3Ti_4O_{12}$	&204	&0.89	&0.40	&3D	&AFM~(24K)	&	&Insulator	&\cite{adams2006influence, subramanian2000high, kim2002neutron}\\
			&$\rm CuSiO_3$	&148	&1.13	&0.37	&3D	&AFM~(110K)	&Black	&Insulator	&\cite{wintenberger1993magnetic}\\
			&$\rm Li_2IrO_3$	&70	&2.75	&0.36	&3D	&FM~(38K)	&	&Insulator	&\cite{takayama2015hyperhoneycomb}\\
			&$\rm Bi_2Cu_5B_4O_{14}$	&1	&0.90	&0.35	&3D	&FM~(25K)	&Green	&Insulator	&\cite{pan2008synthesis, arjun2016structural}\\
			&$\rm NiF_2$	&58	&1.67	&0.30	&3D	&AFM~(73K)	&	&Insulator	&\cite{austin1969high,cooke1965magnetic}\\
			&$\rm LiTi_2O_4$	&227	&2.37	&0.12	&3D	&Superconductor~(13.7K)	&Blue	&Metal	&\cite{johnston1973high,kumatani2012growth}	\\
			\hline
			\multirow{13}*{\begin{minipage}{0.025\textwidth}\rotatebox{90}{Class \romanNum{2}}\end{minipage}}&$\rm CaNi_4O_8$	&166	&1.33	&0.91	&2D	&	&	&	&\cite{bityutskij1984electron}\\
			&$\rm Sc_2Cu_2O_5$	&33	&0.61	&0.80	&1D	&AFM~(16K)	&	&Insulator	&\cite{sannigrahi2019microscopic}\\
			&$\rm VCl_3$	&148	&1.19	&0.66	&2D	&FM	&Violet	&Insulator	&\cite{emeis1975far, he2016unusual}\\
			&$\rm CuZrTiO_5$	&19	&0.94	&0.55	&1D	&AFM	&Green	&Insulator	&\cite{troitzsch2010synthesis, alyahyaei2011magnetic}\\
			&$\rm CuInOPO_4$	&62	&0.69	&0.51	&1D	&	&Green	&	&\cite{schwunck1999kupfer}\\
			&$\rm Li_2CuO_2$	&71	&1.13	&0.48	&1D	&AFM~(9K)	&Brown 	&Insulator	&\cite{sapina1990crystal, hauck1989phase, ebisu1998extremely}\\
			&$\rm BaCu_2Si_2O_7$	&62	&0.95	&0.42	&1D	&AFM~(9.2K)	&Dark blue	&Insulator	&\cite{sologubenko2003universal, chen2014hydrothermal}\\
			&$\rm Co_2B_2O_5$	&2	&1.55	&0.41	&1D	&AFM~(45K)	&Violet	&	&\cite{berger1950crystal,kazak2021spin}\\
			&$\rm Ca_3Cu_5Si_9O_{26}$	&15	&0.96	&0.38	&2D	&	&Bluish-green	&Insulator	&\cite{zoller1992liebauite}\\
			&$\rm CsNiBr_3$	&194	&1.14	&0.37	&1D	&AFM~(70K)	&Orange-brown	&	&\cite{witteveen1974magnetic,raw2012syntheses}\\
			&$\rm CaCuGe_2O_6$	&14	&0.58	&0.36	&1D	&AFM, Jahn-Teller	&	&Insulator	&\cite{sasago1995discovery, koo2005importance, redhammer2005structure}\\
			&$\rm Cu(OH)F$	&14	&1.75	&0.36	&2D	&AFM, Jahn-Teller	&	&	&\cite{giester2003crystal, danilovich2016vehement}\\
			&$\rm KSnS_2$	&166	&1.14	&0.36	&2D	&	&	&	&\cite{bronold1991alkali}\\
			\hline
			\multirow{18}*{\begin{minipage}{0.025\textwidth}\rotatebox{90}{Class \romanNum{3}}\end{minipage}}&$\rm NaO_2$	&205	&0.71	&3.57	&0D	&AFM~(193K)	&	&Insulator  &\cite{carter1953polymorphism,sparks1966magnetic}\\
			&$\rm O_2$	&194	&3.17	&2.55	&0D	&High-pressure	&Dark red	&Insulator	&\cite{lundegaard2009structure, desgreniers1990optical, shimizu1998superconductivity}\\
			&$\rm Rb_4O_6$	&220	&0.42	&2.44	&0D	&AFM	&Black	&Insulator	&\cite{winterlik2009challenge,winterlik2009exotic}\\
			&$\rm LiO_2$	&58	&1.84	&2.28	&0D	&AFM~(7K)	&	&Insulator	&\cite{andrews1969infrared, smith1966antiferromagnetism}\\
			&$\rm K_2BaCo{(NO_2)}_6$	&69	&1.14	&1.61	&0D	&Jahn-Teller	&	&Insulator	&\cite{bertrand1966structure,bertrand1971structure,morioka1981far}\\
			&$\rm Nb_2{(PO_4)}_3$	&167	&1.16	&1.52	&0D	&	&Black	&Insulator	&\cite{sugantha1994synthesis}\\
			&$\rm Rb_2NbCl_6$	&225	&0.71	&1.52	&0D	&Jahn-Teller	&	&	&\cite{henke2007significance}\\
			&$\rm RbSb$	&216	&0.96	&1.27	&0D	&	&	&Insulator	&\cite{collaboration1998nasb}\\
			&$\rm K_2TaCl_6$	&225	&0.98	&1.03	&0D	&Jahn-Teller	&Black	&Insulator	&\cite{jongen2004dipotassium,ishikawa2019ordering}\\
			&$\rm K_3Na{(RuO_4)}_2$	&15	&0.59	&0.97	&0D	&AFM~(70K)	&Black	&Insulator	&\cite{mogare2006k3na}\\
			&$\rm LiBa_2Cu_3O_6$	&69	&0.56	&0.96	&0D	&Jahn-Teller	&	&	&\cite{burdett1995charge}\\
			&$\rm Sr_2CoWO_6$	&225	&1.65	&0.77	&0D	&AFM~(24K)	&Dark brown	&Insulator	&\cite{viola2003structure}\\
			&$\rm Ba_2MgReO_6$	&225	&1.50	&0.75	&0D	&AFM	&Black blue	&Insulator	&\cite{longo1961magnetic,sleight1972magnetic}\\
			&$\rm KRuO_4$	&88	&0.78	&0.74	&0D	&AFM~(150K)	&Black	&Insulator	&\cite{marjerrison2016structure}\\
			&$\rm Cu{(HCOO)}_2$	&14	&0.97	&0.69	&0D	&AFM~(17K)	&Light blue	&Insulator	&\cite{gunter1980crystal,castner1993critical}\\
			&$\rm SrCu_2{(BO_3)}_2$	&140	&0.99	&0.59	&0D	&AFM~(1.4K)	&Blue	&Insulator	&\cite{smith1991synthesis,liu2006plane}\\
			&$\rm CuSe_2O_5$	&15	&0.84	&0.48	&0D	&AFM	&Green	&Insulator	&\cite{becker2006reinvestigation}\\
			&$\rm Ba_2CoWO_6$	&225	&1.93	&0.44	&0D	&AFM~(17K)	&Brown	&Insulator	&\cite{khattak1975crystal}\\
		\end{tabular}
	\end{ruledtabular}
	\begin{flushleft}{$T_C$/$T_N$: Curie/Neel temperature of magnetic transition. 
			FM: ferromagnetism.
			AFM: antiferromagnetism.}\end{flushleft}
	\caption{\textbf{Lossless-metal candidates in ICSD with experimental references.} The band structures and dielectric constants of the candidates are appended at the end of this Supplementary Material.}
				\label{tab:references}
\end{table}

\clearpage

\newcolumntype{L}[1]{>{\raggedright\arraybackslash}p{#1}}
\newcolumntype{C}[1]{>{\centering\arraybackslash}p{#1}}
\newcolumntype{R}[1]{>{\raggedleft\arraybackslash}p{#1}}
\newcommand\ygrow[2][0]{\addstackgap[.5\dimexpr#2\relax]{\vphantom{#1}}}
\begin{table}
	\centering
	\renewcommand\arraystretch{2.6}
	\begin{tabular}{|C{0.13\columnwidth}|C{0.15\columnwidth}|C{0.67\columnwidth}|}
		\hline
		Class&\romanNum{1}&\multirow{4}[10]*{\begin{minipage}{0.65\columnwidth}\includegraphics{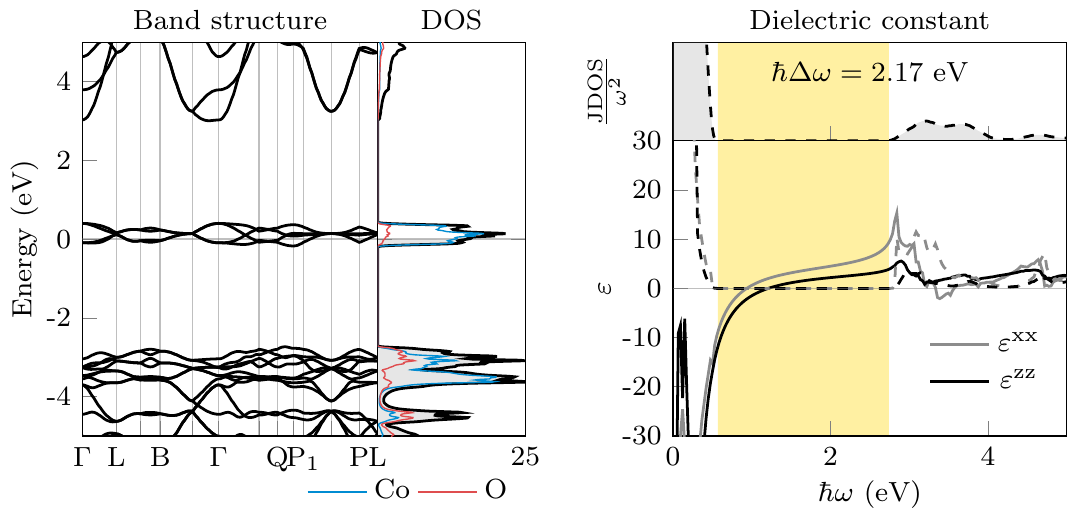}\end{minipage}}\\
		\cline{1-2}Formula&$\rm CoCO_3$&\\
		\cline{1-2}Space group&167~(R-3c)&\\
		\cline{1-2}\emph{W}&0.57~eV&\\
		\cline{1-2}$\hbar\Delta\omega$&2.17~eV&\\\hline
	\end{tabular}
\end{table}

\begin{table}
	\centering
	\renewcommand\arraystretch{2.6}
	\begin{tabular}{|C{0.13\columnwidth}|C{0.15\columnwidth}|C{0.67\columnwidth}|}
		\hline
		Class&\romanNum{1}&\multirow{4}[10]*{\begin{minipage}{0.65\columnwidth}\includegraphics{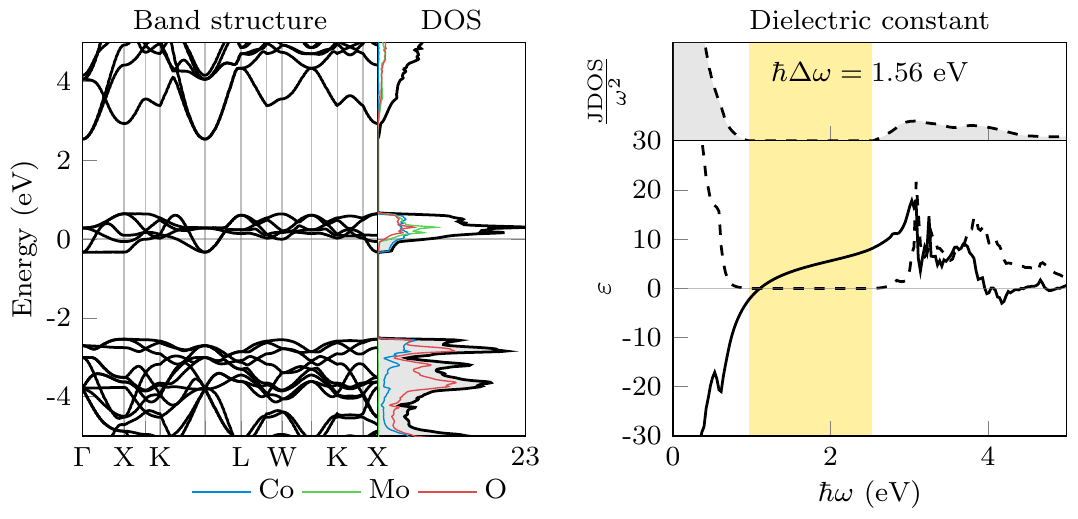}\end{minipage}}\\
		\cline{1-2}Formula&$\rm Ba_2CoMoO_6$&\\
		\cline{1-2}Space group&225~(Fm-3m)&\\
		\cline{1-2}\emph{W}&0.97~eV&\\
		\cline{1-2}$\hbar\Delta\omega$&1.56~eV&\\\hline
	\end{tabular}
\end{table}

\begin{table}
	\centering
	\renewcommand\arraystretch{2.6}
	\begin{tabular}{|C{0.13\columnwidth}|C{0.15\columnwidth}|C{0.67\columnwidth}|}
		\hline
		Class&\romanNum{1}&\multirow{4}[10]*{\begin{minipage}{0.65\columnwidth}\includegraphics{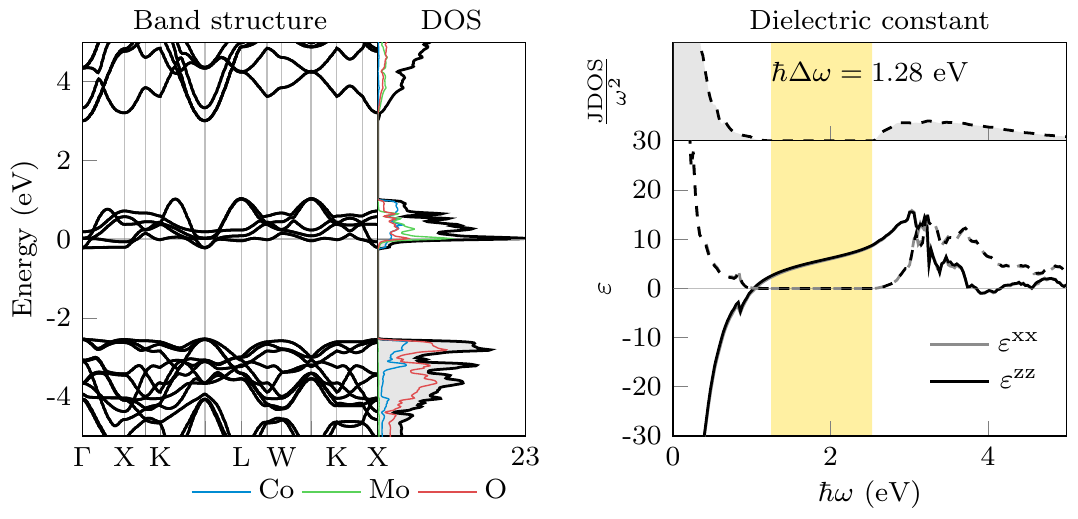}\end{minipage}}\\
		\cline{1-2}Formula&$\rm Sr_2CoMoO_6$&\\
		\cline{1-2}Space group&225~(Fm-3m)&\\
		\cline{1-2}\emph{W}&1.25~eV&\\
		\cline{1-2}$\hbar\Delta\omega$&1.28~eV&\\\hline
	\end{tabular}
\end{table}

\begin{table}
	\centering
	\renewcommand\arraystretch{2.6}
	\begin{tabular}{|C{0.13\columnwidth}|C{0.15\columnwidth}|C{0.67\columnwidth}|}
		\hline
		Class&\romanNum{1}&\multirow{4}[10]*{\begin{minipage}{0.65\columnwidth}\includegraphics{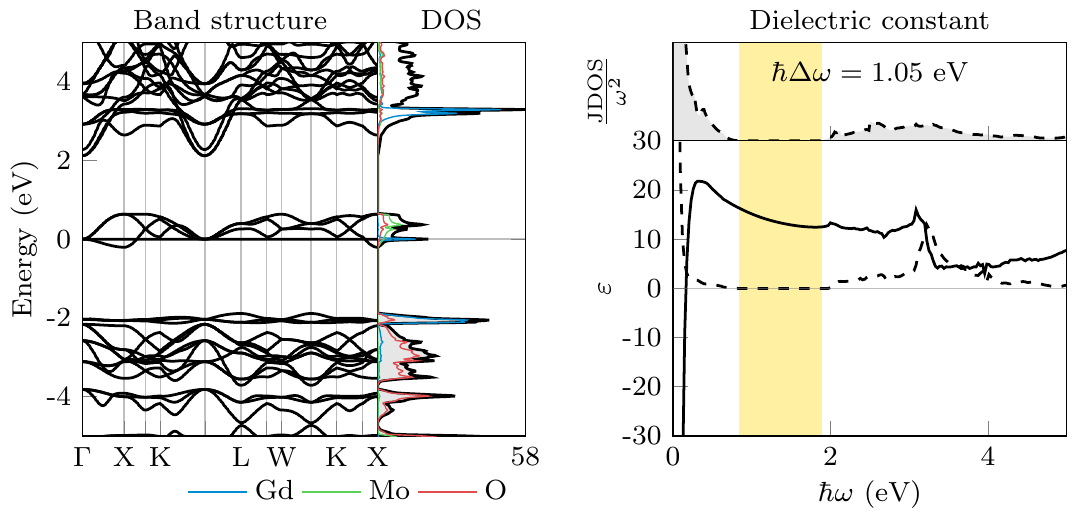}\end{minipage}}\\
		\cline{1-2}Formula&$\rm Ba_2GdMoO_6$&\\
		\cline{1-2}Space group&225~(Fm-3m)&\\
		\cline{1-2}\emph{W}&0.84~eV&\\
		\cline{1-2}$\hbar\Delta\omega$&1.05~eV&\\\hline
	\end{tabular}
\end{table}

\clearpage
\begin{table}
	\centering
	\renewcommand\arraystretch{2.6}
	\begin{tabular}{|C{0.13\columnwidth}|C{0.15\columnwidth}|C{0.67\columnwidth}|}
		\hline
		Class&\romanNum{1}&\multirow{4}[10]*{\begin{minipage}{0.65\columnwidth}\includegraphics{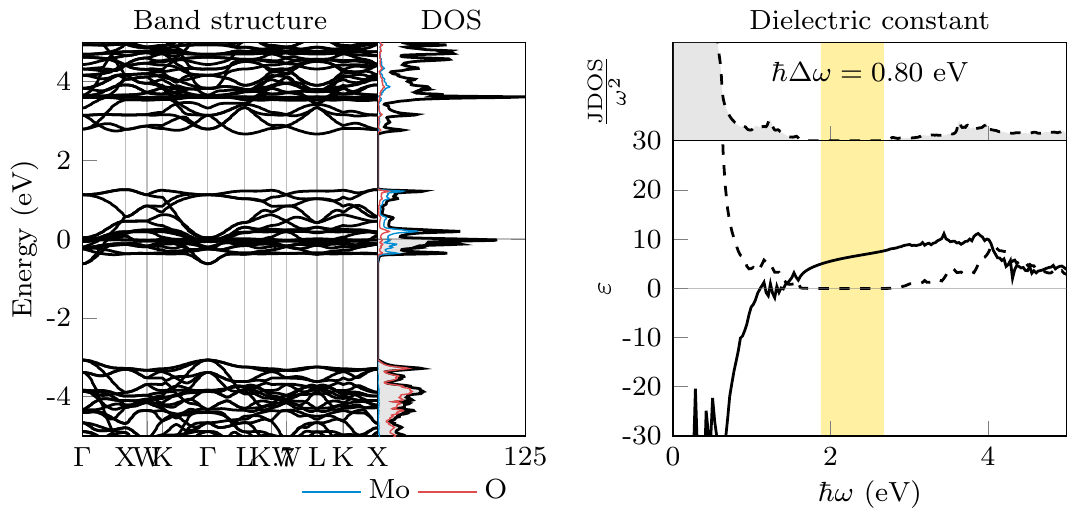}\end{minipage}}\\
		\cline{1-2}Formula&$\rm Nd_2Mo_2O_7$&\\
		\cline{1-2}Space group&227~(Fd-3m)&\\
		\cline{1-2}\emph{W}&1.88~eV&\\
		\cline{1-2}$\hbar\Delta\omega$&0.80~eV&\\\hline
	\end{tabular}
\end{table}

\begin{table}
	\centering
	\renewcommand\arraystretch{2.6}
	\begin{tabular}{|C{0.13\columnwidth}|C{0.15\columnwidth}|C{0.67\columnwidth}|}
		\hline
		Class&\romanNum{1}&\multirow{4}[10]*{\begin{minipage}{0.65\columnwidth}\includegraphics{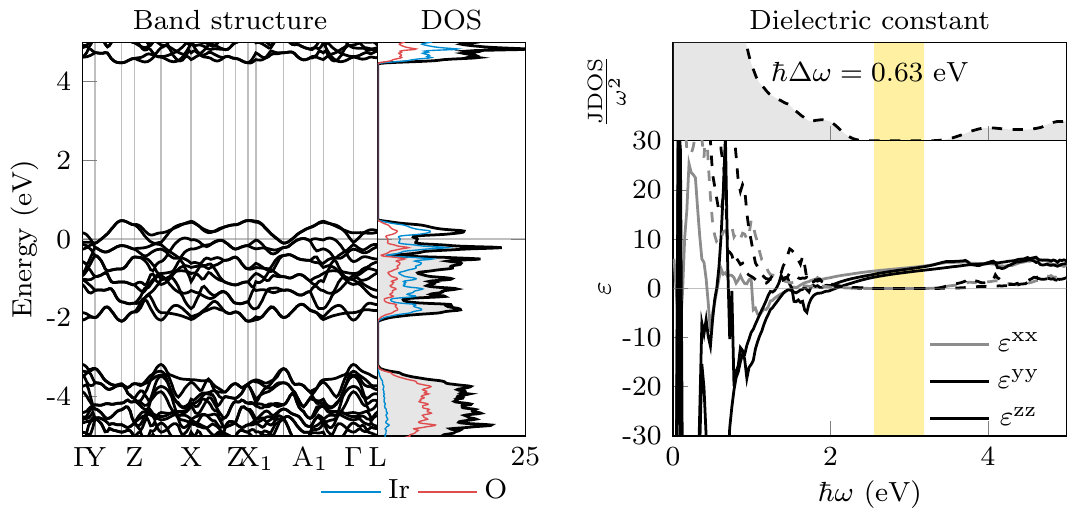}\end{minipage}}\\
		\cline{1-2}Formula&$\rm Li_2IrO_3$&\\
		\cline{1-2}Space group&70~(Fddd)&\\
		\cline{1-2}\emph{W}&2.56~eV&\\
		\cline{1-2}$\hbar\Delta\omega$&0.63~eV&\\\hline
	\end{tabular}
\end{table}

\begin{table}
	\centering
	\renewcommand\arraystretch{2.6}
	\begin{tabular}{|C{0.13\columnwidth}|C{0.15\columnwidth}|C{0.67\columnwidth}|}
		\hline
		Class&\romanNum{1}&\multirow{4}[10]*{\begin{minipage}{0.65\columnwidth}\includegraphics{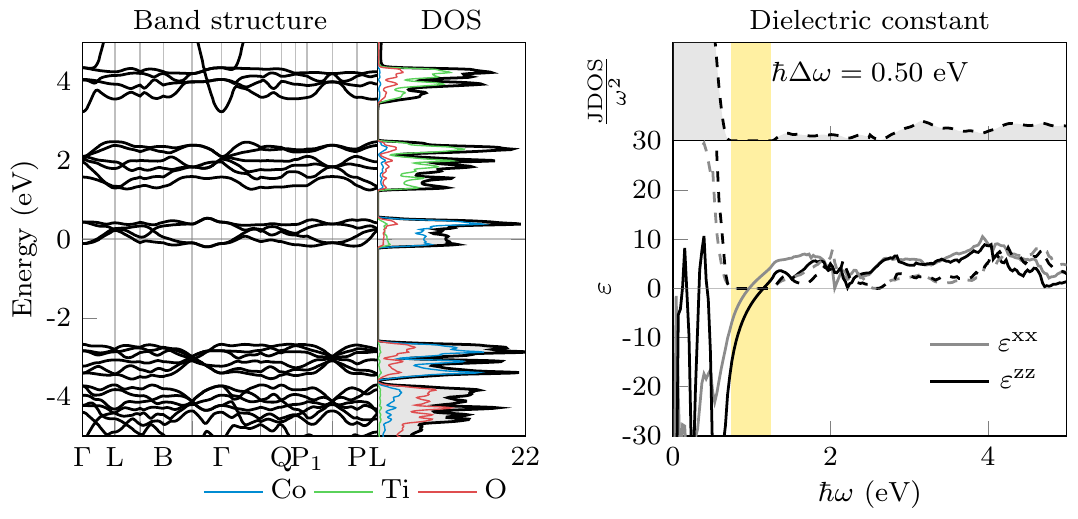}\end{minipage}}\\
		\cline{1-2}Formula&$\rm CoTiO_3$&\\
		\cline{1-2}Space group&148~(R-3)&\\
		\cline{1-2}\emph{W}&0.74~eV&\\
		\cline{1-2}$\hbar\Delta\omega$&0.50~eV&\\\hline
	\end{tabular}
\end{table}

\begin{table}
	\centering
	\renewcommand\arraystretch{2.6}
	\begin{tabular}{|C{0.13\columnwidth}|C{0.15\columnwidth}|C{0.67\columnwidth}|}
		\hline
		Class&\romanNum{1}&\multirow{4}[10]*{\begin{minipage}{0.65\columnwidth}\includegraphics{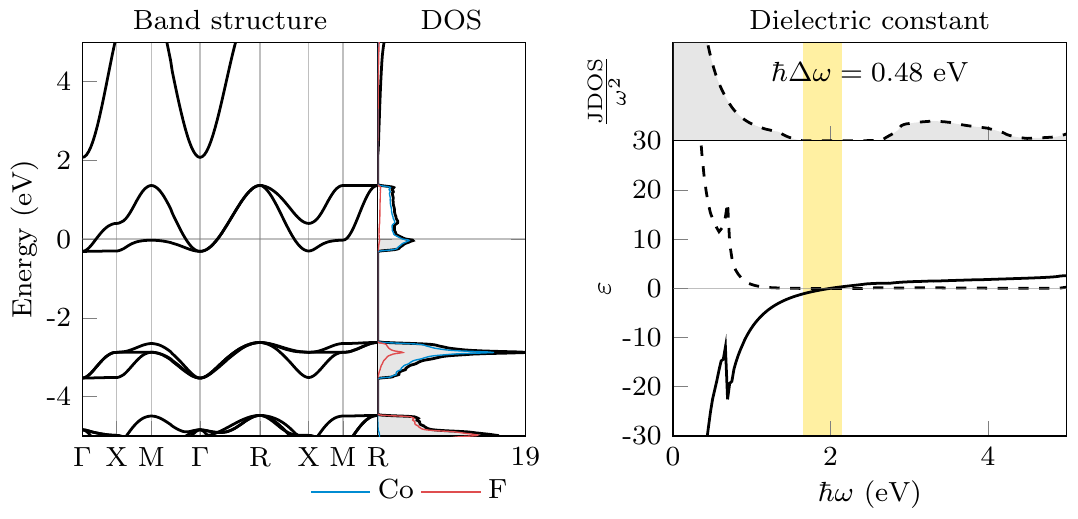}\end{minipage}}\\
		\cline{1-2}Formula&$\rm KCoF_3$&\\
		\cline{1-2}Space group&221~(Pm-3m)&\\
		\cline{1-2}\emph{W}&1.66~eV&\\
		\cline{1-2}$\hbar\Delta\omega$&0.48~eV&\\\hline
	\end{tabular}
\end{table}

\clearpage
\begin{table}
	\centering
	\renewcommand\arraystretch{2.6}
	\begin{tabular}{|C{0.13\columnwidth}|C{0.15\columnwidth}|C{0.67\columnwidth}|}
		\hline
		Class&\romanNum{1}&\multirow{4}[10]*{\begin{minipage}{0.65\columnwidth}\includegraphics{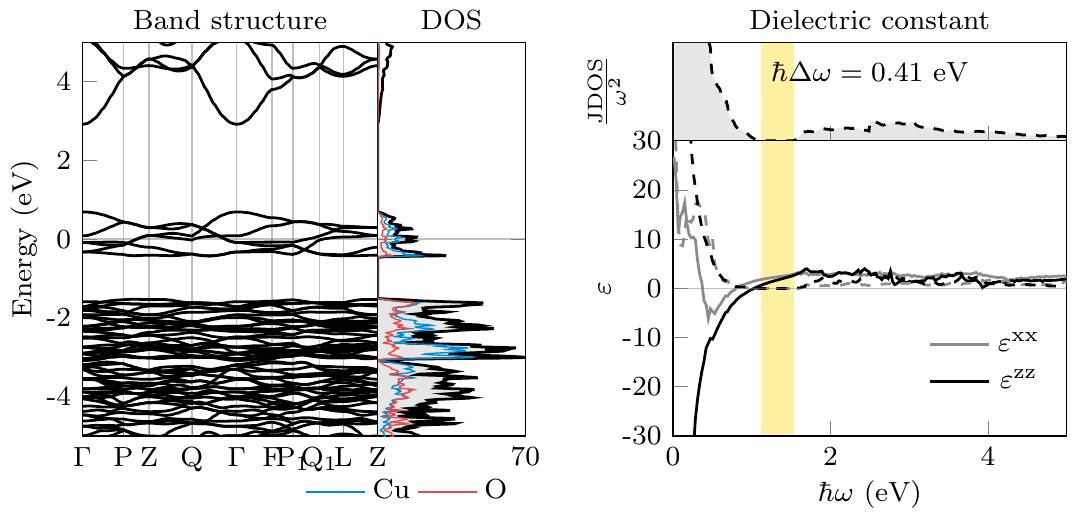}\end{minipage}}\\
		\cline{1-2}Formula&$\rm CuSiO_3$&\\
		\cline{1-2}Space group&148~(R-3)&\\
		\cline{1-2}\emph{W}&1.12~eV&\\
		\cline{1-2}$\hbar\Delta\omega$&0.41~eV&\\\hline
	\end{tabular}
\end{table}

\begin{table}
	\centering
	\renewcommand\arraystretch{2.6}
	\begin{tabular}{|C{0.13\columnwidth}|C{0.15\columnwidth}|C{0.67\columnwidth}|}
		\hline
		Class&\romanNum{1}&\multirow{4}[10]*{\begin{minipage}{0.65\columnwidth}\includegraphics{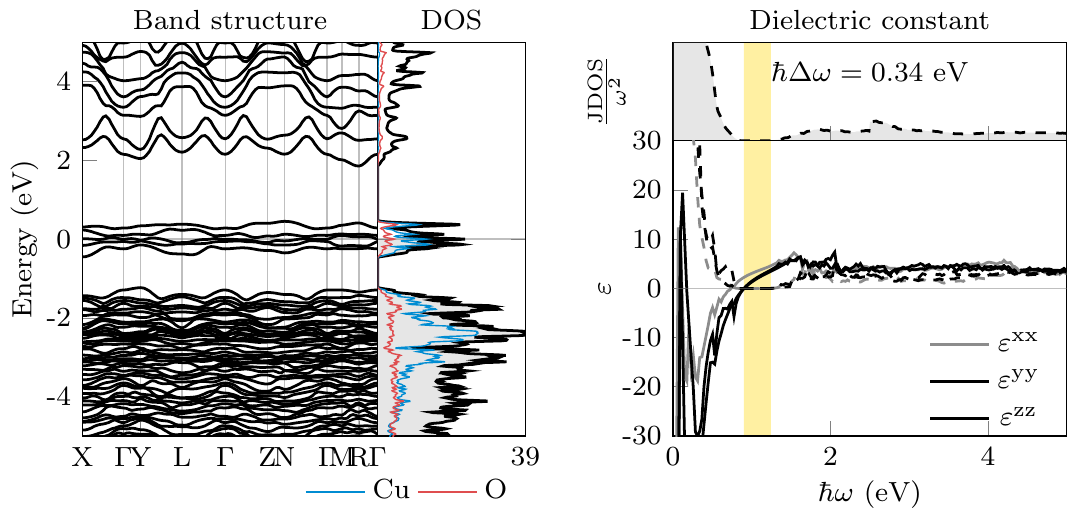}\end{minipage}}\\
		\cline{1-2}Formula&$\rm Bi_2Cu_5B_4O_{14}$&\\
		\cline{1-2}Space group&1~(P1)&\\
		\cline{1-2}\emph{W}&0.90~eV&\\
		\cline{1-2}$\hbar\Delta\omega$&0.34~eV&\\\hline
	\end{tabular}
\end{table}

\begin{table}
	\centering
	\renewcommand\arraystretch{2.6}
	\begin{tabular}{|C{0.13\columnwidth}|C{0.15\columnwidth}|C{0.67\columnwidth}|}
		\hline
		Class&\romanNum{1}&\multirow{4}[10]*{\begin{minipage}{0.65\columnwidth}\includegraphics{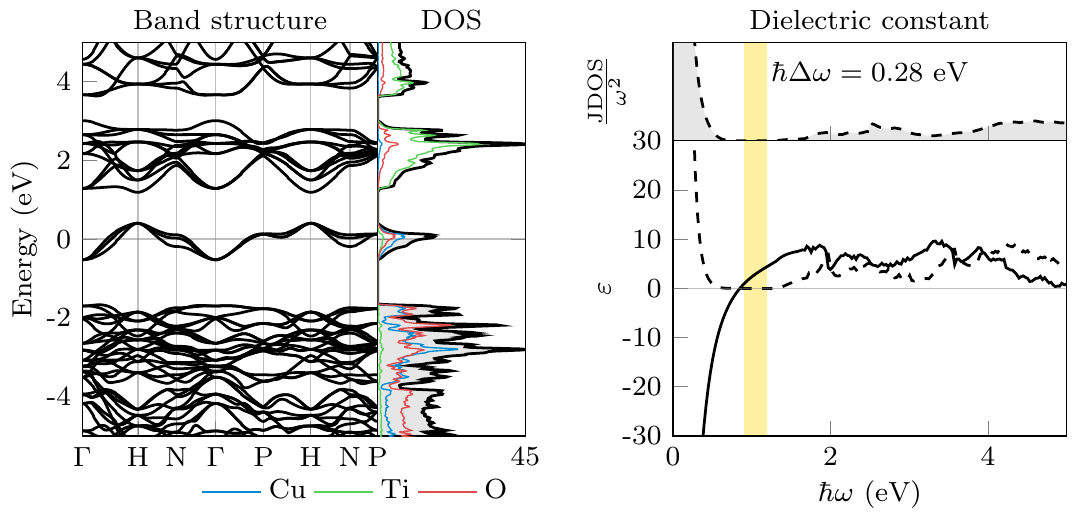}\end{minipage}}\\
		\cline{1-2}Formula&$\rm CaCu_3Ti_4O_{12}$&\\
		\cline{1-2}Space group&204~(Im-3)&\\
		\cline{1-2}\emph{W}&0.91~eV&\\
		\cline{1-2}$\hbar\Delta\omega$&0.28~eV&\\\hline
	\end{tabular}
\end{table}

\begin{table}
	\centering
	\renewcommand\arraystretch{2.6}
	\begin{tabular}{|C{0.13\columnwidth}|C{0.15\columnwidth}|C{0.67\columnwidth}|}
		\hline
		Class&\romanNum{2}&\multirow{4}[10]*{\begin{minipage}{0.65\columnwidth}\includegraphics{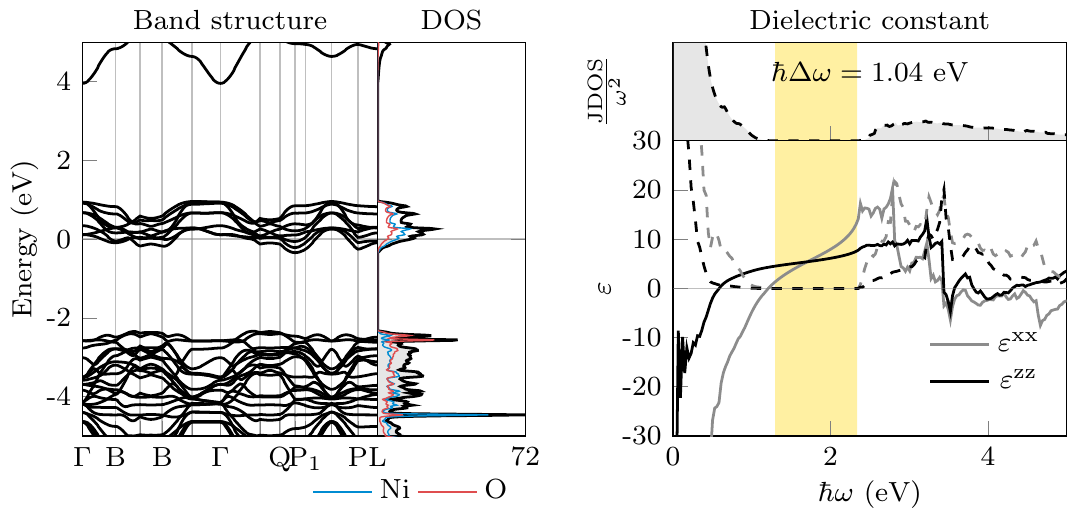}\end{minipage}}\\
		\cline{1-2}Formula&$\rm CaNi_4O_8$&\\
		\cline{1-2}Space group&166~(R-3m)&\\
		\cline{1-2}\emph{W}&1.30~eV&\\
		\cline{1-2}$\hbar\Delta\omega$&1.04~eV&\\\hline
	\end{tabular}
\end{table}

\clearpage
\begin{table}
	\centering
	\renewcommand\arraystretch{2.6}
	\begin{tabular}{|C{0.13\columnwidth}|C{0.15\columnwidth}|C{0.67\columnwidth}|}
		\hline
		Class&\romanNum{2}&\multirow{4}[10]*{\begin{minipage}{0.65\columnwidth}\includegraphics{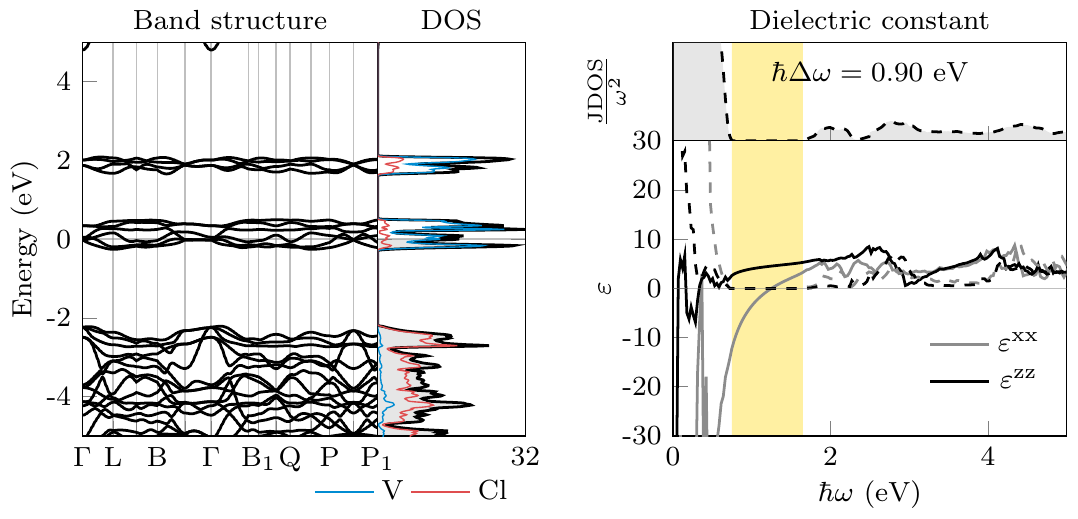}\end{minipage}}\\
		\cline{1-2}Formula&$\rm VCl_3$&\\
		\cline{1-2}Space group&148~(R-3)&\\
		\cline{1-2}\emph{W}&0.75~eV&\\
		\cline{1-2}$\hbar\Delta\omega$&0.90~eV&\\\hline
	\end{tabular}
\end{table}

\begin{table}
	\centering
	\renewcommand\arraystretch{2.6}
	\begin{tabular}{|C{0.13\columnwidth}|C{0.15\columnwidth}|C{0.67\columnwidth}|}
		\hline
		Class&\romanNum{2}&\multirow{4}[10]*{\begin{minipage}{0.65\columnwidth}\includegraphics{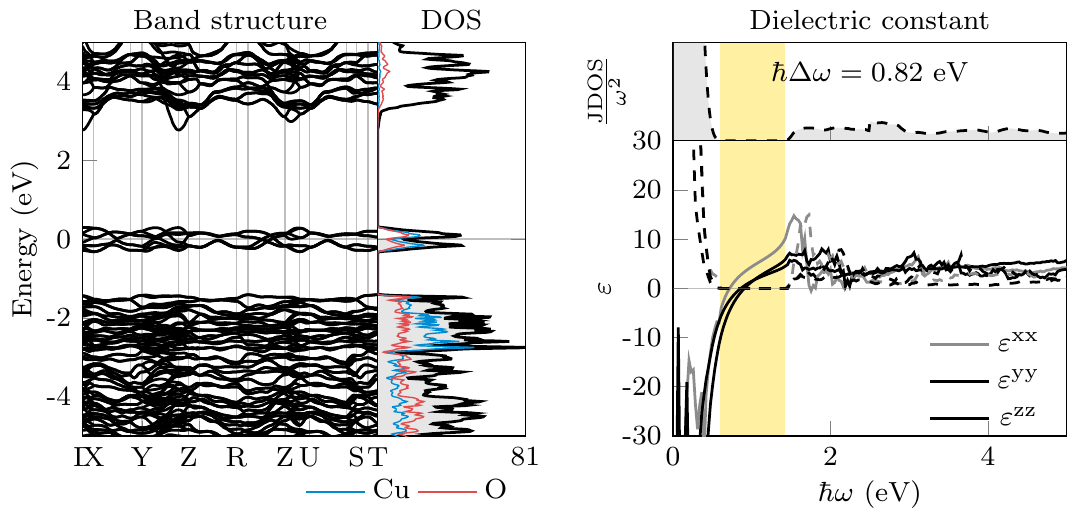}\end{minipage}}\\
		\cline{1-2}Formula&$\rm Sc_2Cu_2O_5$&\\
		\cline{1-2}Space group&33~(Pna21)&\\
		\cline{1-2}\emph{W}&0.60~eV&\\
		\cline{1-2}$\hbar\Delta\omega$&0.82~eV&\\\hline
	\end{tabular}
\end{table}

\begin{table}
	\centering
	\renewcommand\arraystretch{2.6}
	\begin{tabular}{|C{0.13\columnwidth}|C{0.15\columnwidth}|C{0.67\columnwidth}|}
		\hline
		Class&\romanNum{2}&\multirow{4}[10]*{\begin{minipage}{0.65\columnwidth}\includegraphics{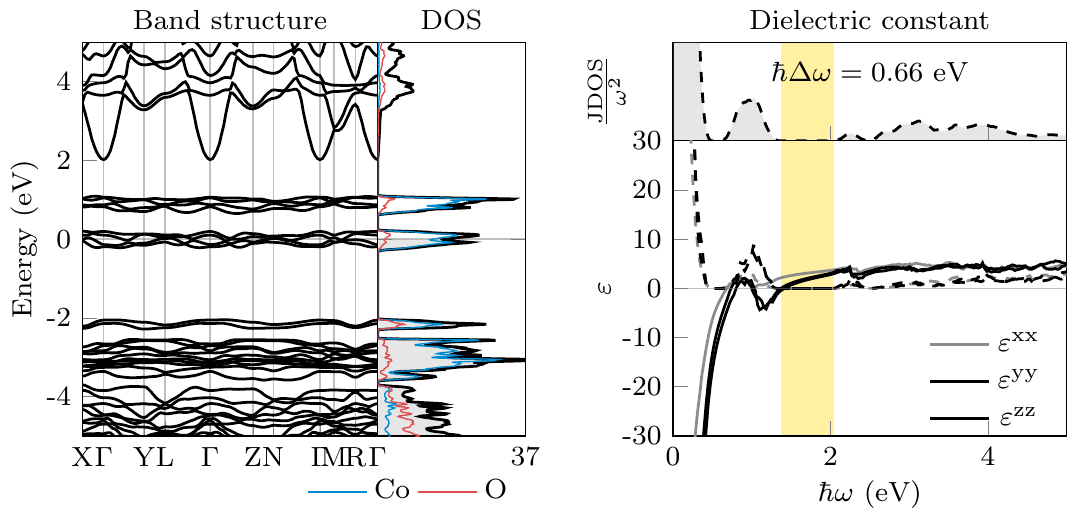}\end{minipage}}\\
		\cline{1-2}Formula&$\rm Co_2B_2O_5$&\\
		\cline{1-2}Space group&2~(P-1)&\\
		\cline{1-2}\emph{W}&1.38~eV&\\
		\cline{1-2}$\hbar\Delta\omega$&0.66~eV&\\\hline
	\end{tabular}
\end{table}

\begin{table}
	\centering
	\renewcommand\arraystretch{2.6}
	\begin{tabular}{|C{0.13\columnwidth}|C{0.15\columnwidth}|C{0.67\columnwidth}|}
		\hline
		Class&\romanNum{2}&\multirow{4}[10]*{\begin{minipage}{0.65\columnwidth}\includegraphics{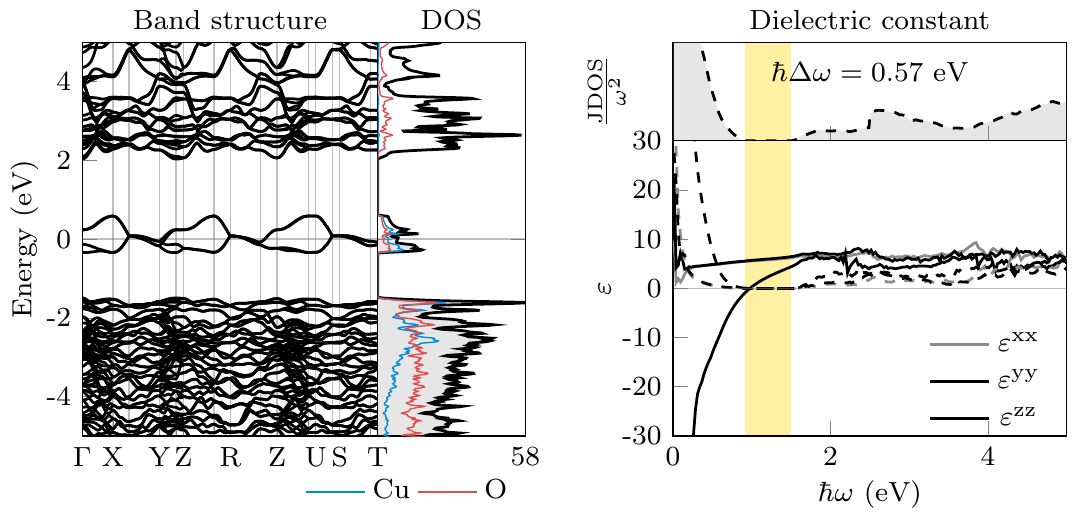}\end{minipage}}\\
		\cline{1-2}Formula&$\rm CuZrTiO_5$&\\
		\cline{1-2}Space group&19~(P212121)&\\
		\cline{1-2}\emph{W}&0.92~eV&\\
		\cline{1-2}$\hbar\Delta\omega$&0.57~eV&\\\hline
	\end{tabular}
\end{table}

\clearpage
\begin{table}
	\centering
	\renewcommand\arraystretch{2.6}
	\begin{tabular}{|C{0.13\columnwidth}|C{0.15\columnwidth}|C{0.67\columnwidth}|}
		\hline
		Class&\romanNum{2}&\multirow{4}[10]*{\begin{minipage}{0.65\columnwidth}\includegraphics{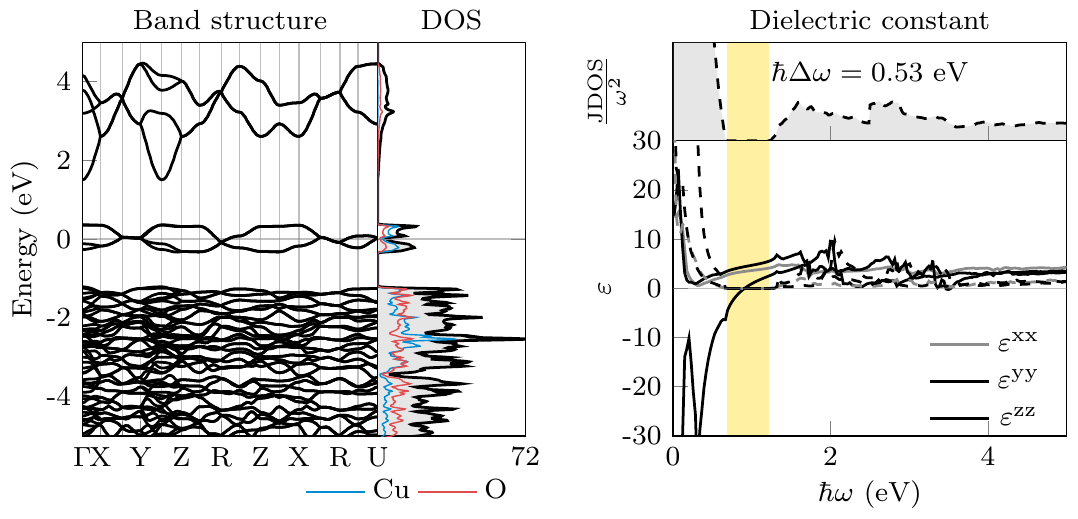}\end{minipage}}\\
		\cline{1-2}Formula&$\rm CuInOPO_4$&\\
		\cline{1-2}Space group&62~(Pnma)&\\
		\cline{1-2}\emph{W}&0.69~eV&\\
		\cline{1-2}$\hbar\Delta\omega$&0.53~eV&\\\hline
	\end{tabular}
\end{table}

\begin{table}
	\centering
	\renewcommand\arraystretch{2.6}
	\begin{tabular}{|C{0.13\columnwidth}|C{0.15\columnwidth}|C{0.67\columnwidth}|}
		\hline
		Class&\romanNum{2}&\multirow{4}[10]*{\begin{minipage}{0.65\columnwidth}\includegraphics{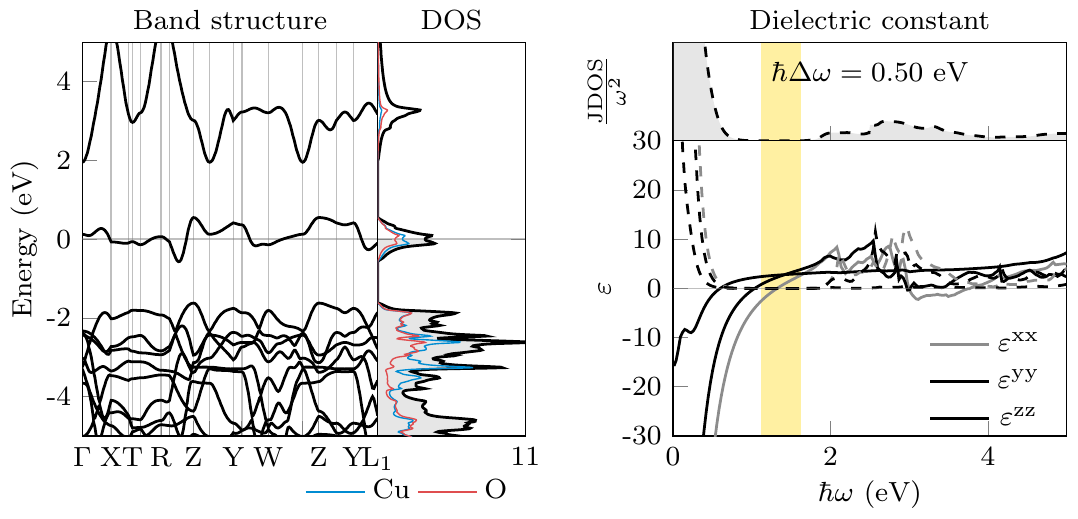}\end{minipage}}\\
		\cline{1-2}Formula&$\rm Li_2CuO_2$&\\
		\cline{1-2}Space group&71~(Immm)&\\
		\cline{1-2}\emph{W}&1.12~eV&\\
		\cline{1-2}$\hbar\Delta\omega$&0.50~eV&\\\hline
	\end{tabular}
\end{table}

\begin{table}
	\centering
	\renewcommand\arraystretch{2.6}
	\begin{tabular}{|C{0.13\columnwidth}|C{0.15\columnwidth}|C{0.67\columnwidth}|}
		\hline
		Class&\romanNum{2}&\multirow{4}[10]*{\begin{minipage}{0.65\columnwidth}\includegraphics{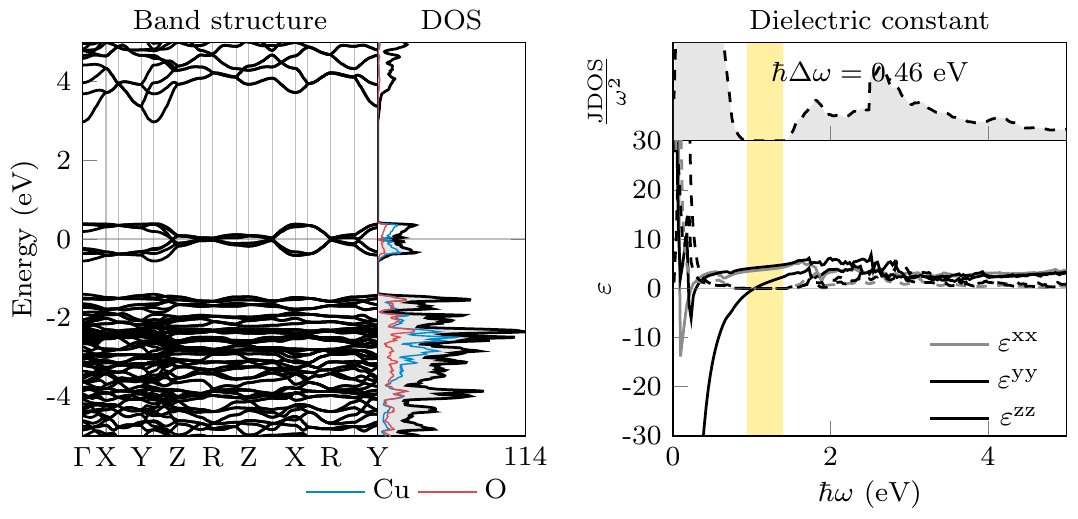}\end{minipage}}\\
		\cline{1-2}Formula&$\rm BaCu_2Si_2O_7$&\\
		\cline{1-2}Space group&62~(Pnma)&\\
		\cline{1-2}\emph{W}&0.94~eV&\\
		\cline{1-2}$\hbar\Delta\omega$&0.46~eV&\\\hline
	\end{tabular}
\end{table}

\begin{table}
	\centering
	\renewcommand\arraystretch{2.6}
	\begin{tabular}{|C{0.13\columnwidth}|C{0.15\columnwidth}|C{0.67\columnwidth}|}
		\hline
		Class&\romanNum{2}&\multirow{4}[10]*{\begin{minipage}{0.65\columnwidth}\includegraphics{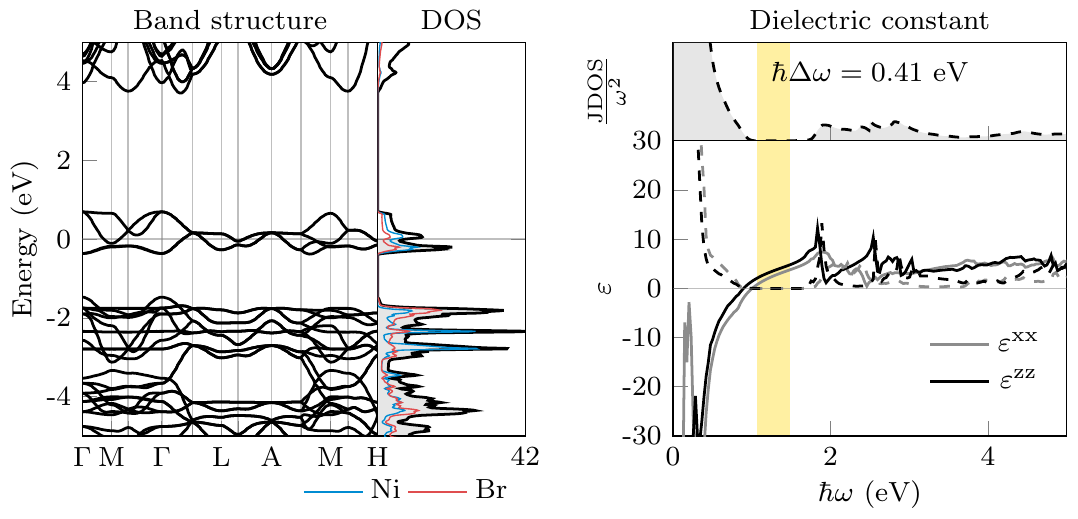}\end{minipage}}\\
		\cline{1-2}Formula&$\rm CsNiBr_3$&\\
		\cline{1-2}Space group&194~(P63/mmc)&\\
		\cline{1-2}\emph{W}&1.07~eV&\\
		\cline{1-2}$\hbar\Delta\omega$&0.41~eV&\\\hline
	\end{tabular}
\end{table}

\clearpage
\begin{table}
	\centering
	\renewcommand\arraystretch{2.6}
	\begin{tabular}{|C{0.13\columnwidth}|C{0.15\columnwidth}|C{0.67\columnwidth}|}
		\hline
		Class&\romanNum{2}&\multirow{4}[10]*{\begin{minipage}{0.65\columnwidth}\includegraphics{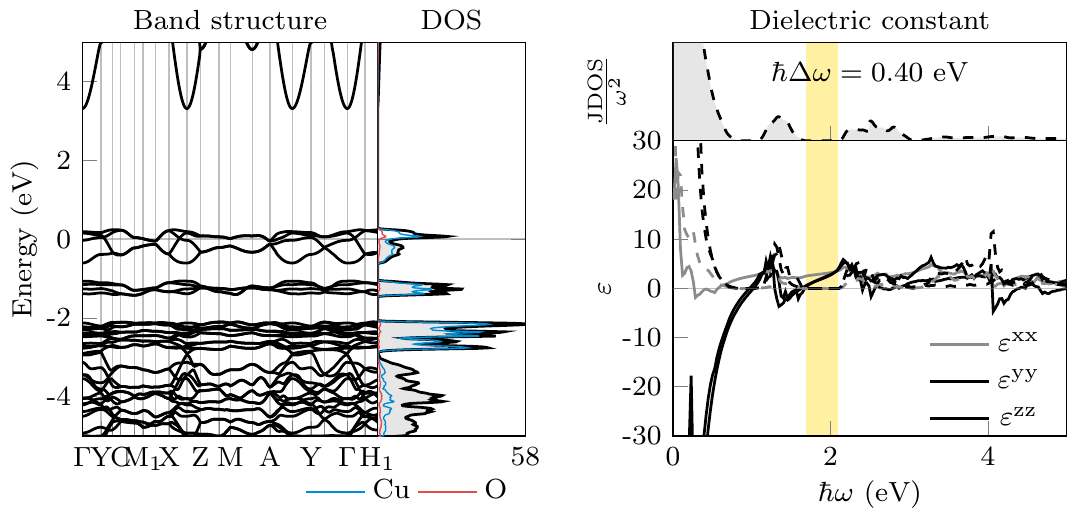}\end{minipage}}\\
		\cline{1-2}Formula&$\rm Cu(OH)F$&\\
		\cline{1-2}Space group&14~(P21/c)&\\
		\cline{1-2}\emph{W}&1.69~eV&\\
		\cline{1-2}$\hbar\Delta\omega$&0.40~eV&\\\hline
	\end{tabular}
\end{table}

\begin{table}
	\centering
	\renewcommand\arraystretch{2.6}
	\begin{tabular}{|C{0.13\columnwidth}|C{0.15\columnwidth}|C{0.67\columnwidth}|}
		\hline
		Class&\romanNum{2}&\multirow{4}[10]*{\begin{minipage}{0.65\columnwidth}\includegraphics{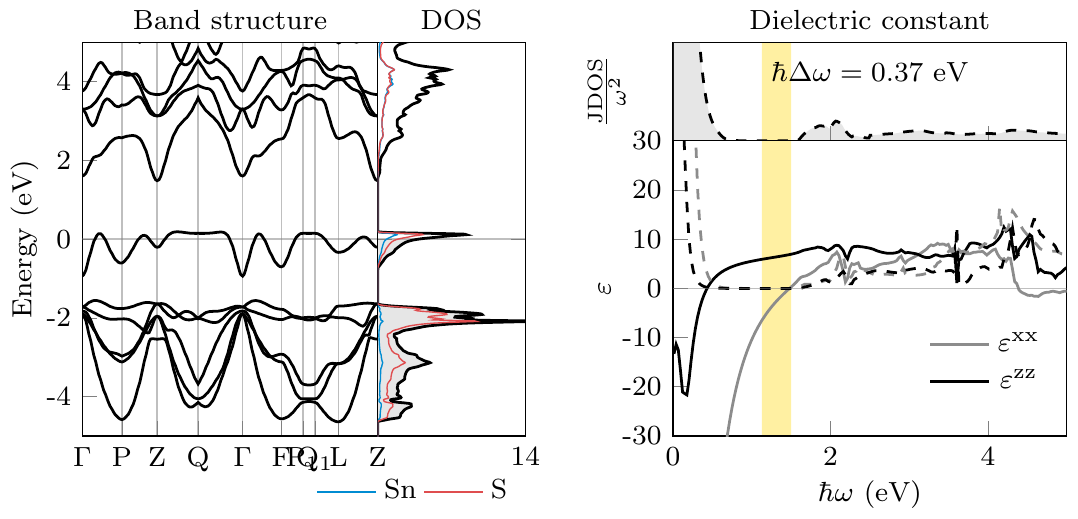}\end{minipage}}\\
		\cline{1-2}Formula&$\rm KSnS_2$&\\
		\cline{1-2}Space group&166~(R-3m)&\\
		\cline{1-2}\emph{W}&1.13~eV&\\
		\cline{1-2}$\hbar\Delta\omega$&0.37~eV&\\\hline
	\end{tabular}
\end{table}

\begin{table}
	\centering
	\renewcommand\arraystretch{2.6}
	\begin{tabular}{|C{0.13\columnwidth}|C{0.15\columnwidth}|C{0.67\columnwidth}|}
		\hline
		Class&\romanNum{2}&\multirow{4}[10]*{\begin{minipage}{0.65\columnwidth}\includegraphics{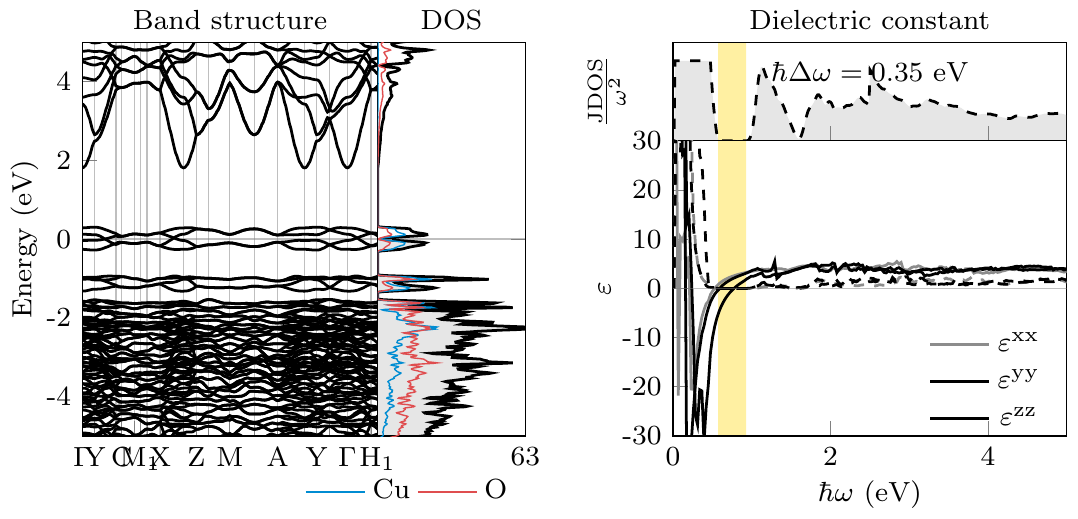}\end{minipage}}\\
		\cline{1-2}Formula&$\rm CaCuGe_2O_6$&\\
		\cline{1-2}Space group&14~(P21/c)&\\
		\cline{1-2}\emph{W}&0.58~eV&\\
		\cline{1-2}$\hbar\Delta\omega$&0.35~eV&\\\hline
	\end{tabular}
\end{table}

\begin{table}
	\centering
	\renewcommand\arraystretch{2.6}
	\begin{tabular}{|C{0.13\columnwidth}|C{0.15\columnwidth}|C{0.67\columnwidth}|}
		\hline
		Class&\romanNum{3}&\multirow{4}[10]*{\begin{minipage}{0.65\columnwidth}\includegraphics{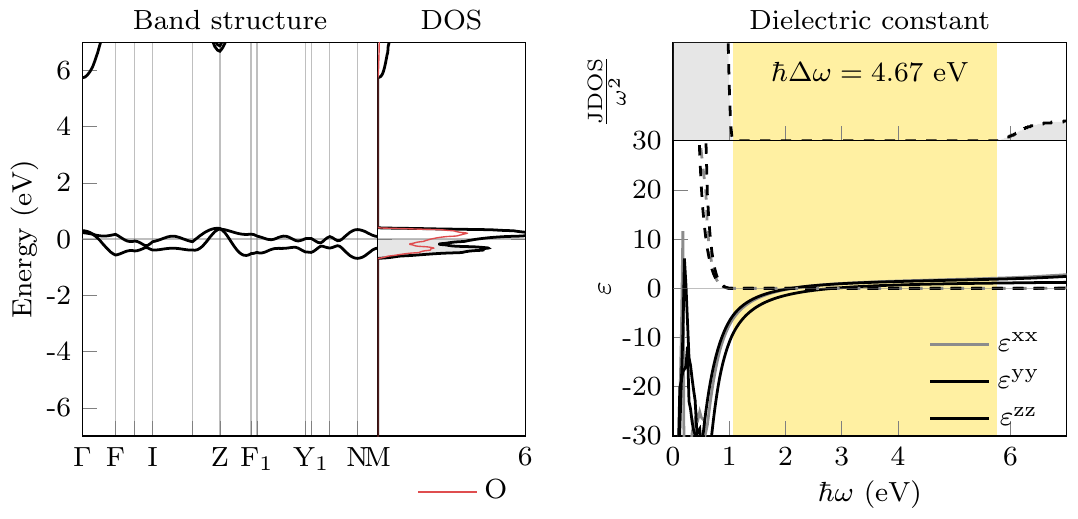}\end{minipage}}\\
		\cline{1-2}Formula&$\rm O_2$&\\
		\cline{1-2}Space group&12~(C2/m)&\\
		\cline{1-2}\emph{W}&1.08~eV&\\
		\cline{1-2}$\hbar\Delta\omega$&4.67~eV&\\\hline
	\end{tabular}
\end{table}

\clearpage
\begin{table}
	\centering
	\renewcommand\arraystretch{2.6}
	\begin{tabular}{|C{0.13\columnwidth}|C{0.15\columnwidth}|C{0.67\columnwidth}|}
		\hline
		Class&\romanNum{3}&\multirow{4}[10]*{\begin{minipage}{0.65\columnwidth}\includegraphics{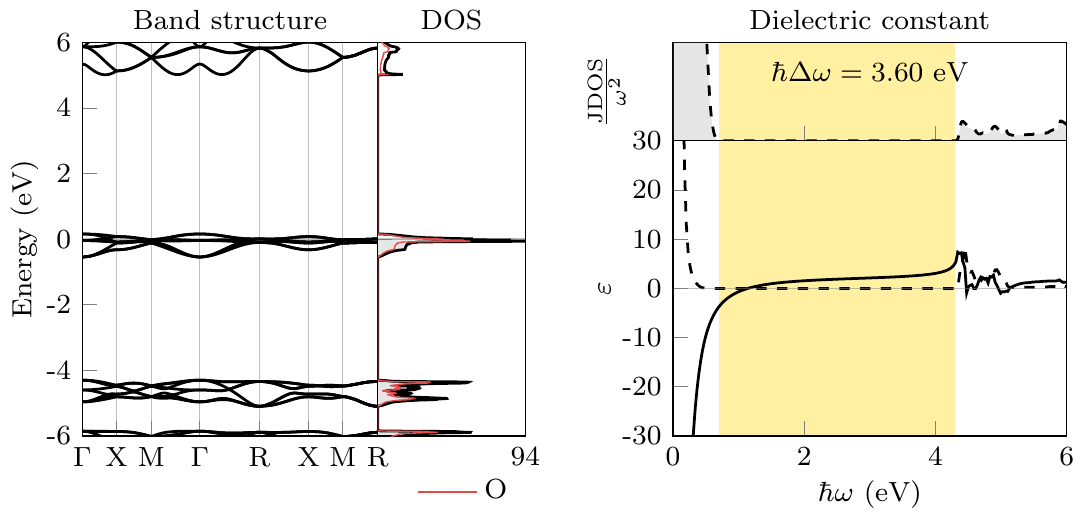}\end{minipage}}\\
		\cline{1-2}Formula&$\rm NaO_2$&\\
		\cline{1-2}Space group&205~(Pa-3)&\\
		\cline{1-2}\emph{W}&0.70~eV&\\
		\cline{1-2}$\hbar\Delta\omega$&3.60~eV&\\\hline
	\end{tabular}
\end{table}

\begin{table}
	\centering
	\renewcommand\arraystretch{2.6}
	\begin{tabular}{|C{0.13\columnwidth}|C{0.15\columnwidth}|C{0.67\columnwidth}|}
		\hline
		Class&\romanNum{3}&\multirow{4}[10]*{\begin{minipage}{0.65\columnwidth}\includegraphics{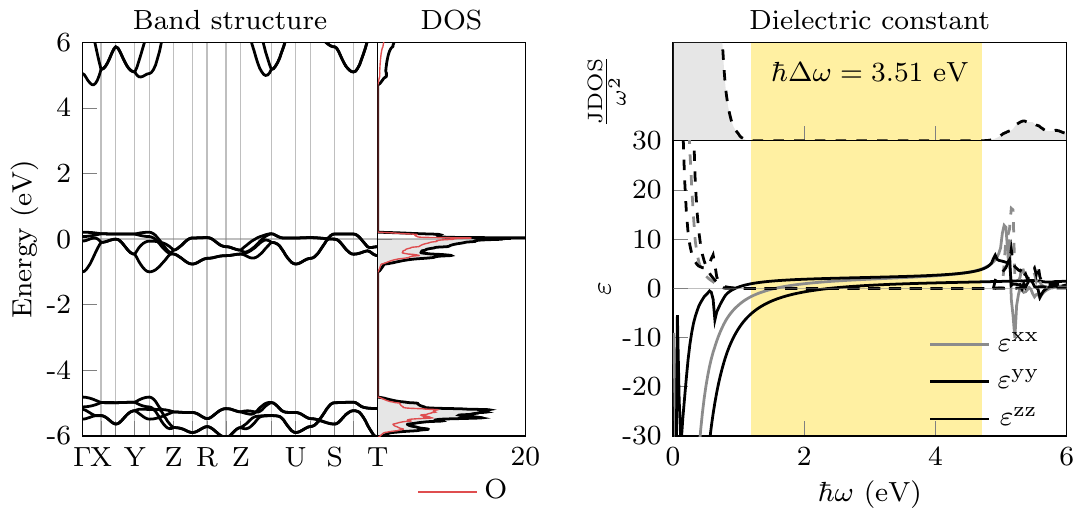}\end{minipage}}\\
		\cline{1-2}Formula&$\rm NaO_2$&\\
		\cline{1-2}Space group&58~(Pnnm)&\\
		\cline{1-2}\emph{W}&1.20~eV&\\
		\cline{1-2}$\hbar\Delta\omega$&3.51~eV&\\\hline
	\end{tabular}
\end{table}

\begin{table}
	\centering
	\renewcommand\arraystretch{2.6}
	\begin{tabular}{|C{0.13\columnwidth}|C{0.15\columnwidth}|C{0.67\columnwidth}|}
		\hline
		Class&\romanNum{3}&\multirow{4}[10]*{\begin{minipage}{0.65\columnwidth}\includegraphics{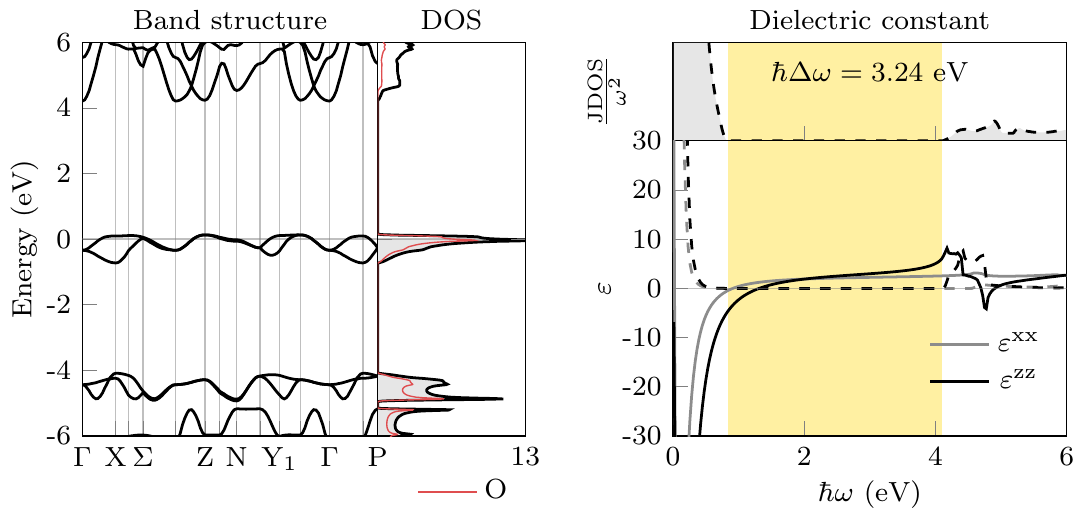}\end{minipage}}\\
		\cline{1-2}Formula&$\rm CsO_2$&\\
		\cline{1-2}Space group&139~(I4/mmm)&\\
		\cline{1-2}\emph{W}&0.85~eV&\\
		\cline{1-2}$\hbar\Delta\omega$&3.24~eV&\\\hline
	\end{tabular}
\end{table}

\begin{table}
	\centering
	\renewcommand\arraystretch{2.6}
	\begin{tabular}{|C{0.13\columnwidth}|C{0.15\columnwidth}|C{0.67\columnwidth}|}
		\hline
		Class&\romanNum{3}&\multirow{4}[10]*{\begin{minipage}{0.65\columnwidth}\includegraphics{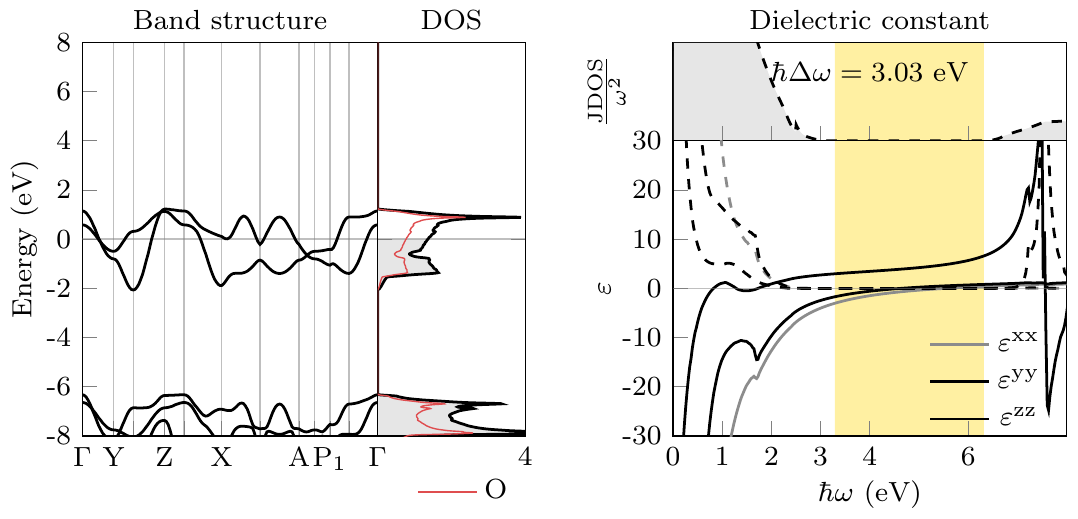}\end{minipage}}\\
		\cline{1-2}Formula&$\rm O_2$&\\
		\cline{1-2}Space group&69~(Fmmm)&\\
		\cline{1-2}\emph{W}&3.29~eV&\\
		\cline{1-2}$\hbar\Delta\omega$&3.03~eV&\\\hline
	\end{tabular}
\end{table}

\clearpage
\begin{table}
	\centering
	\renewcommand\arraystretch{2.6}
	\begin{tabular}{|C{0.13\columnwidth}|C{0.15\columnwidth}|C{0.67\columnwidth}|}
		\hline
		Class&\romanNum{3}&\multirow{4}[10]*{\begin{minipage}{0.65\columnwidth}\includegraphics{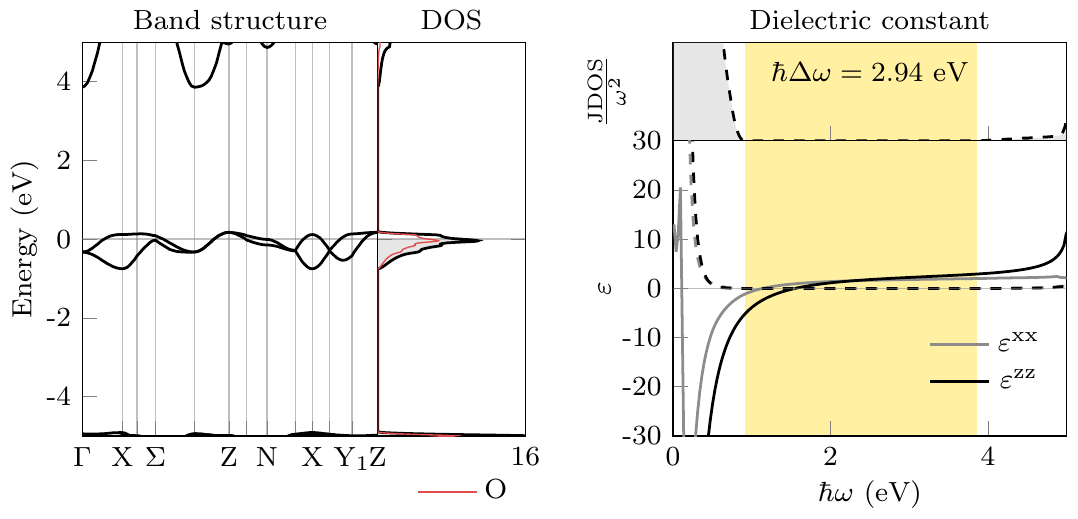}\end{minipage}}\\
		\cline{1-2}Formula&$\rm RbO_2$&\\
		\cline{1-2}Space group&139~(I4/mmm)&\\
		\cline{1-2}\emph{W}&0.92~eV&\\
		\cline{1-2}$\hbar\Delta\omega$&2.94~eV&\\\hline
	\end{tabular}
\end{table}

\begin{table}
	\centering
	\renewcommand\arraystretch{2.6}
	\begin{tabular}{|C{0.13\columnwidth}|C{0.15\columnwidth}|C{0.67\columnwidth}|}
		\hline
		Class&\romanNum{3}&\multirow{4}[10]*{\begin{minipage}{0.65\columnwidth}\includegraphics{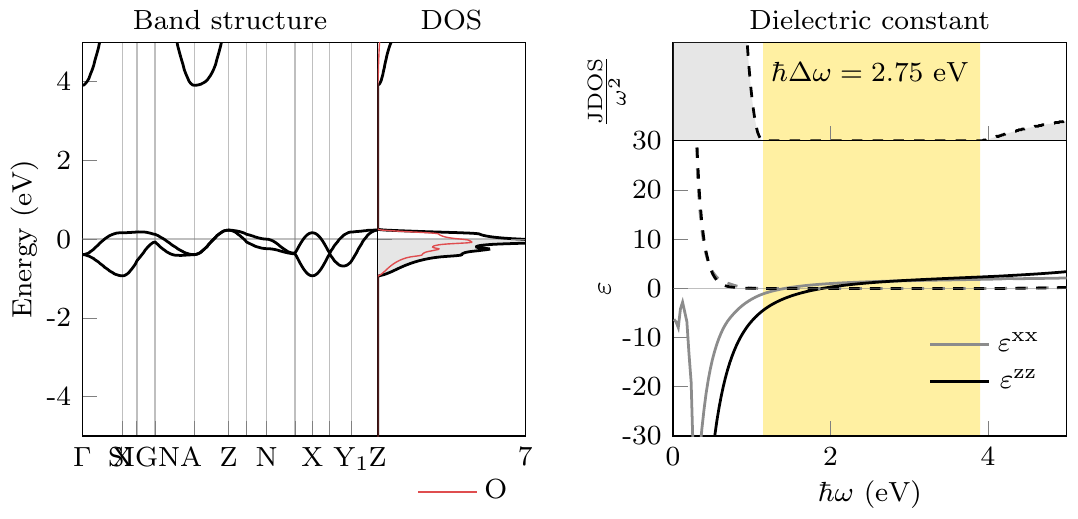}\end{minipage}}\\
		\cline{1-2}Formula&$\rm KO_2$&\\
		\cline{1-2}Space group&139~(I4/mmm)&\\
		\cline{1-2}\emph{W}&1.15~eV&\\
		\cline{1-2}$\hbar\Delta\omega$&2.75~eV&\\\hline
	\end{tabular}
\end{table}

\begin{table}
	\centering
	\renewcommand\arraystretch{2.6}
	\begin{tabular}{|C{0.13\columnwidth}|C{0.15\columnwidth}|C{0.67\columnwidth}|}
		\hline
		Class&\romanNum{3}&\multirow{4}[10]*{\begin{minipage}{0.65\columnwidth}\includegraphics{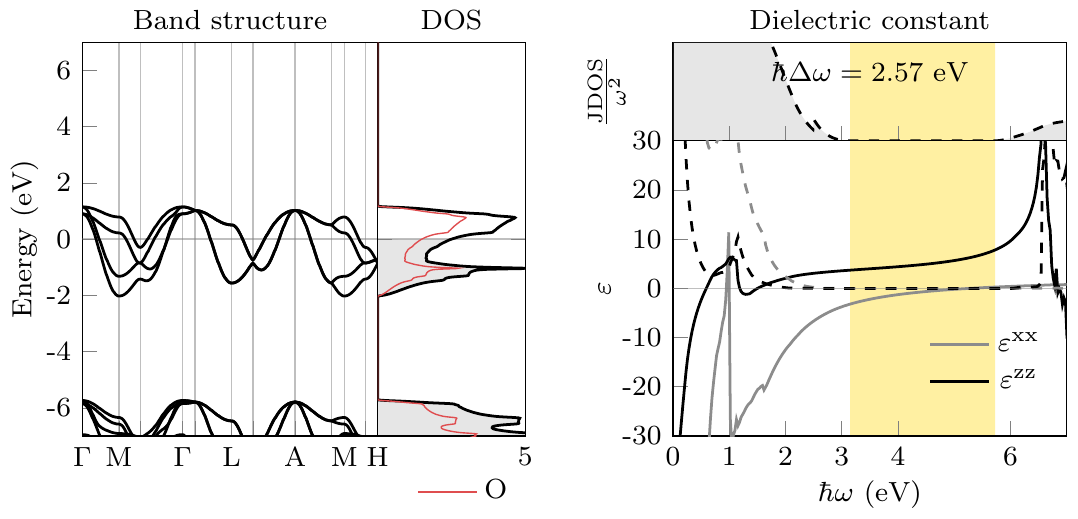}\end{minipage}}\\
		\cline{1-2}Formula&$\rm O_2$&\\
		\cline{1-2}Space group&194~(P63/mmc)&\\
		\cline{1-2}\emph{W}&3.16~eV&\\
		\cline{1-2}$\hbar\Delta\omega$&2.57~eV&\\\hline
	\end{tabular}
\end{table}

\begin{table}
	\centering
	\renewcommand\arraystretch{2.6}
	\begin{tabular}{|C{0.13\columnwidth}|C{0.15\columnwidth}|C{0.67\columnwidth}|}
		\hline
		Class&\romanNum{3}&\multirow{4}[10]*{\begin{minipage}{0.65\columnwidth}\includegraphics{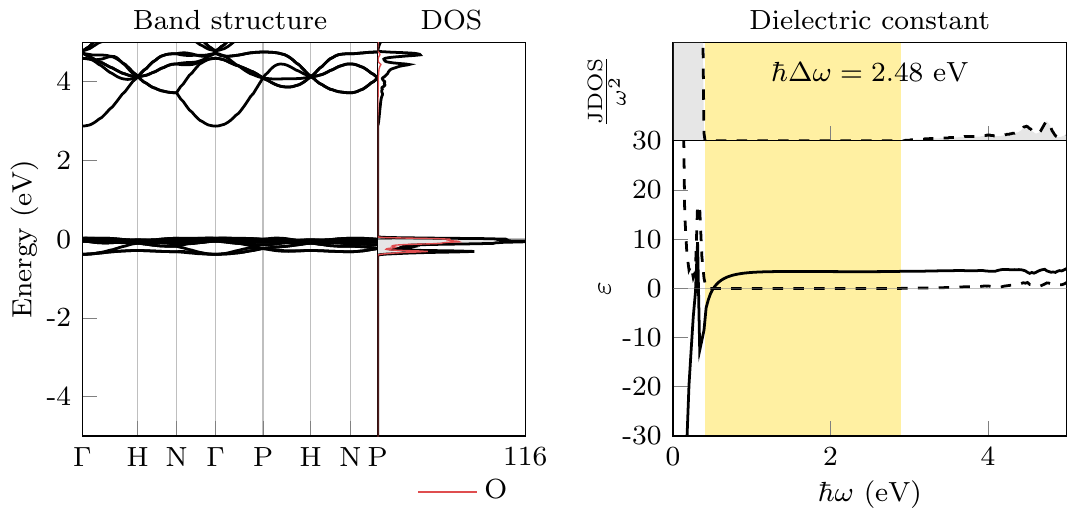}\end{minipage}}\\
		\cline{1-2}Formula&$\rm Rb_4O_6$&\\
		\cline{1-2}Space group&220~(I-43d)&\\
		\cline{1-2}\emph{W}&0.41~eV&\\
		\cline{1-2}$\hbar\Delta\omega$&2.48~eV&\\\hline
	\end{tabular}
\end{table}

\clearpage
\begin{table}
	\centering
	\renewcommand\arraystretch{2.6}
	\begin{tabular}{|C{0.13\columnwidth}|C{0.15\columnwidth}|C{0.67\columnwidth}|}
		\hline
		Class&\romanNum{3}&\multirow{4}[10]*{\begin{minipage}{0.65\columnwidth}\includegraphics{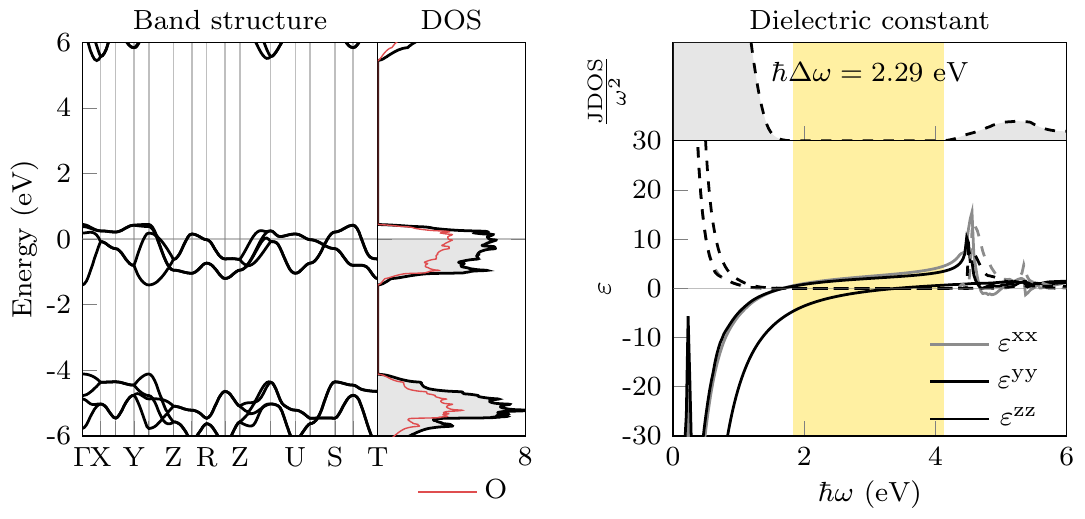}\end{minipage}}\\
		\cline{1-2}Formula&$\rm LiO_2$&\\
		\cline{1-2}Space group&58~(Pnnm)&\\
		\cline{1-2}\emph{W}&1.83~eV&\\
		\cline{1-2}$\hbar\Delta\omega$&2.29~eV&\\\hline
	\end{tabular}
\end{table}

\begin{table}
	\centering
	\renewcommand\arraystretch{2.6}
	\begin{tabular}{|C{0.13\columnwidth}|C{0.15\columnwidth}|C{0.67\columnwidth}|}
		\hline
		Class&\romanNum{3}&\multirow{4}[10]*{\begin{minipage}{0.65\columnwidth}\includegraphics{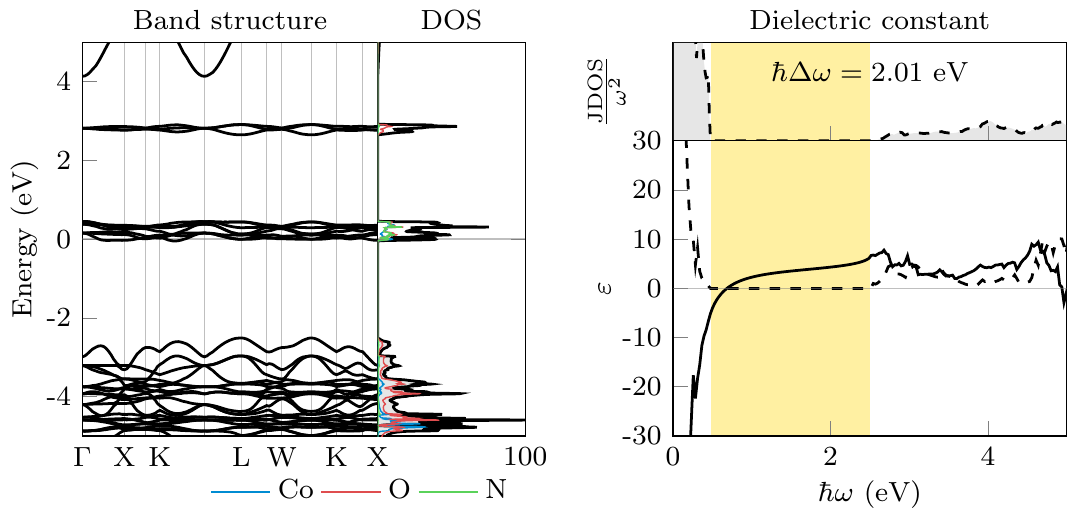}\end{minipage}}\\
		\cline{1-2}Formula&$\rm Rb_2CoPb{(NO_2)}_6$&\\
		\cline{1-2}Space group&202~(Fm-3)&\\
		\cline{1-2}\emph{W}&0.49~eV&\\
		\cline{1-2}$\hbar\Delta\omega$&2.01~eV&\\\hline
	\end{tabular}
\end{table}

\begin{table}
	\centering
	\renewcommand\arraystretch{2.6}
	\begin{tabular}{|C{0.13\columnwidth}|C{0.15\columnwidth}|C{0.67\columnwidth}|}
		\hline
		Class&\romanNum{3}&\multirow{4}[10]*{\begin{minipage}{0.65\columnwidth}\includegraphics{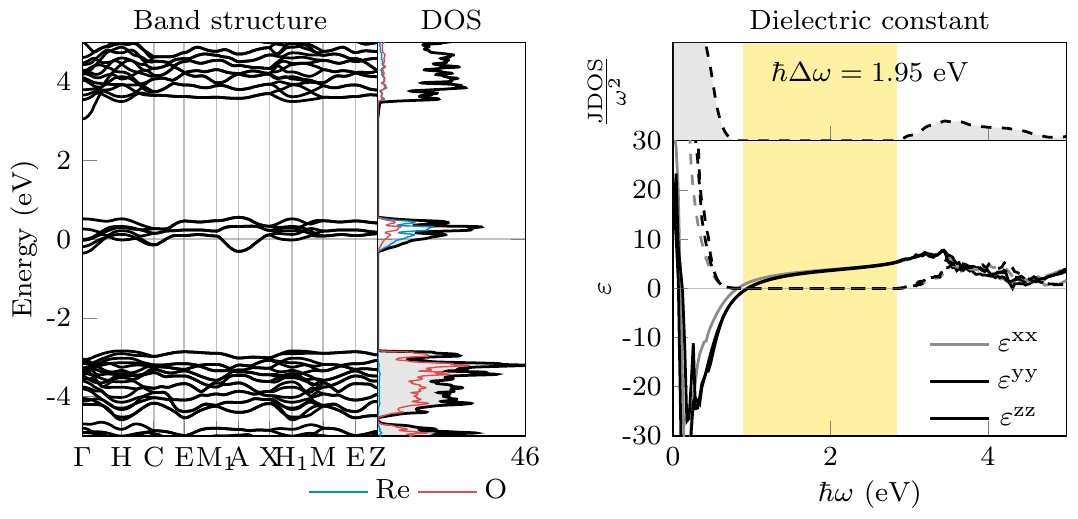}\end{minipage}}\\
		\cline{1-2}Formula&$\rm Ca_3ReO_6$&\\
		\cline{1-2}Space group&14~(P21/c)&\\
		\cline{1-2}\emph{W}&0.89~eV&\\
		\cline{1-2}$\hbar\Delta\omega$&1.95~eV&\\\hline
	\end{tabular}
\end{table}

\begin{table}
	\centering
	\renewcommand\arraystretch{2.6}
	\begin{tabular}{|C{0.13\columnwidth}|C{0.15\columnwidth}|C{0.67\columnwidth}|}
		\hline
		Class&\romanNum{3}&\multirow{4}[10]*{\begin{minipage}{0.65\columnwidth}\includegraphics{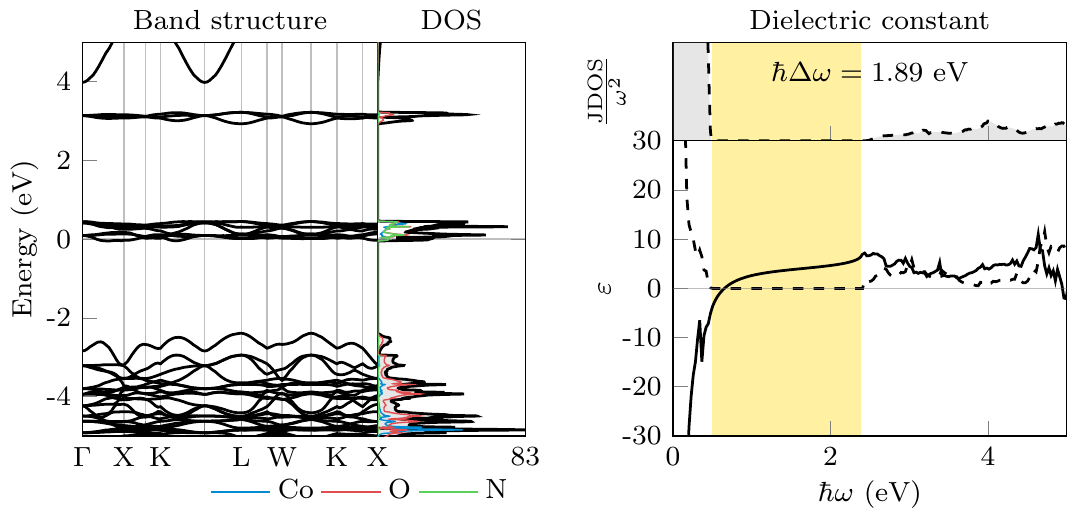}\end{minipage}}\\
		\cline{1-2}Formula&$\rm K_2CoPb{(NO_2)}_6$&\\
		\cline{1-2}Space group&202~(Fm-3)&\\
		\cline{1-2}\emph{W}&0.50~eV&\\
		\cline{1-2}$\hbar\Delta\omega$&1.89~eV&\\\hline
	\end{tabular}
\end{table}

\clearpage
\begin{table}
	\centering
	\renewcommand\arraystretch{2.6}
	\begin{tabular}{|C{0.13\columnwidth}|C{0.15\columnwidth}|C{0.67\columnwidth}|}
		\hline
		Class&\romanNum{3}&\multirow{4}[10]*{\begin{minipage}{0.65\columnwidth}\includegraphics{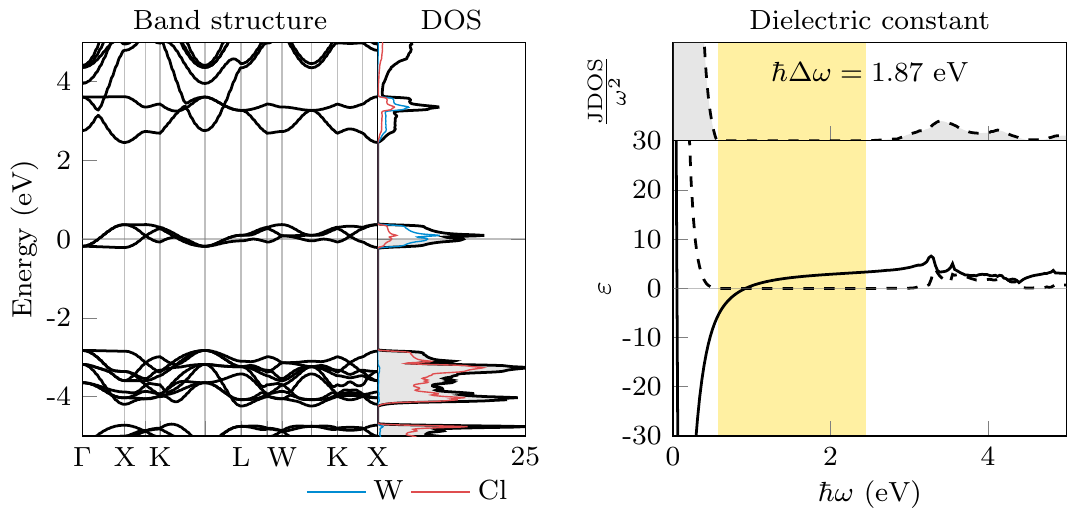}\end{minipage}}\\
		\cline{1-2}Formula&$\rm K_2WCl_6$&\\
		\cline{1-2}Space group&225~(Fm-3m)&\\
		\cline{1-2}\emph{W}&0.58~eV&\\
		\cline{1-2}$\hbar\Delta\omega$&1.87~eV&\\\hline
	\end{tabular}
\end{table}

\begin{table}
	\centering
	\renewcommand\arraystretch{2.6}
	\begin{tabular}{|C{0.13\columnwidth}|C{0.15\columnwidth}|C{0.67\columnwidth}|}
		\hline
		Class&\romanNum{3}&\multirow{4}[10]*{\begin{minipage}{0.65\columnwidth}\includegraphics{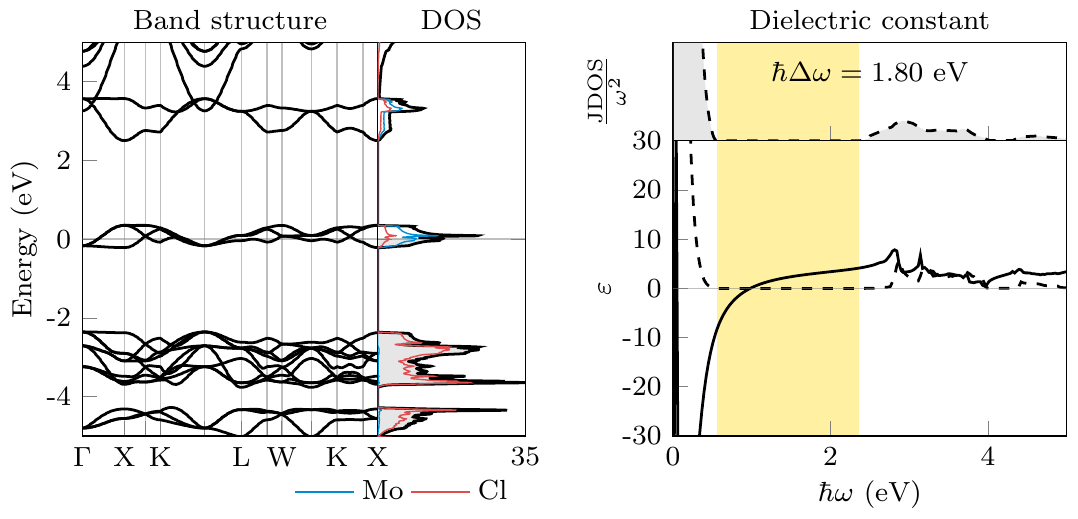}\end{minipage}}\\
		\cline{1-2}Formula&$\rm K2MoCl_6$&\\
		\cline{1-2}Space group&225~(Fm-3m)&\\
		\cline{1-2}\emph{W}&0.56~eV&\\
		\cline{1-2}$\hbar\Delta\omega$&1.80~eV&\\\hline
	\end{tabular}
\end{table}

\begin{table}
	\centering
	\renewcommand\arraystretch{2.6}
	\begin{tabular}{|C{0.13\columnwidth}|C{0.15\columnwidth}|C{0.67\columnwidth}|}
		\hline
		Class&\romanNum{3}&\multirow{4}[10]*{\begin{minipage}{0.65\columnwidth}\includegraphics{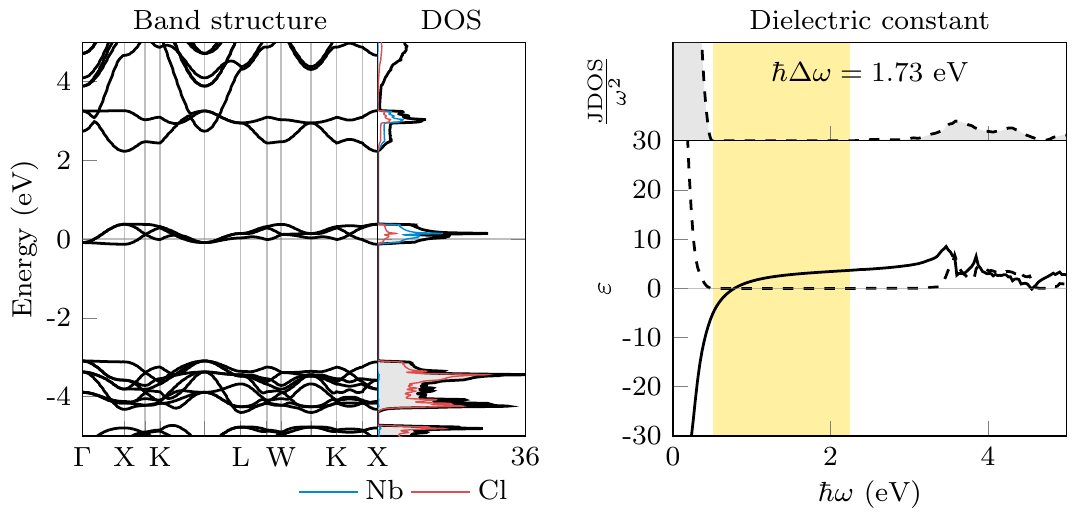}\end{minipage}}\\
		\cline{1-2}Formula&$\rm Rb_2NbCl_6$&\\
		\cline{1-2}Space group&225~(Fm-3m)&\\
		\cline{1-2}\emph{W}&0.51~eV&\\
		\cline{1-2}$\hbar\Delta\omega$&1.73~eV&\\\hline
	\end{tabular}
\end{table}

\begin{table}
	\centering
	\renewcommand\arraystretch{2.6}
	\begin{tabular}{|C{0.13\columnwidth}|C{0.15\columnwidth}|C{0.67\columnwidth}|}
		\hline
		Class&\romanNum{3}&\multirow{4}[10]*{\begin{minipage}{0.65\columnwidth}\includegraphics{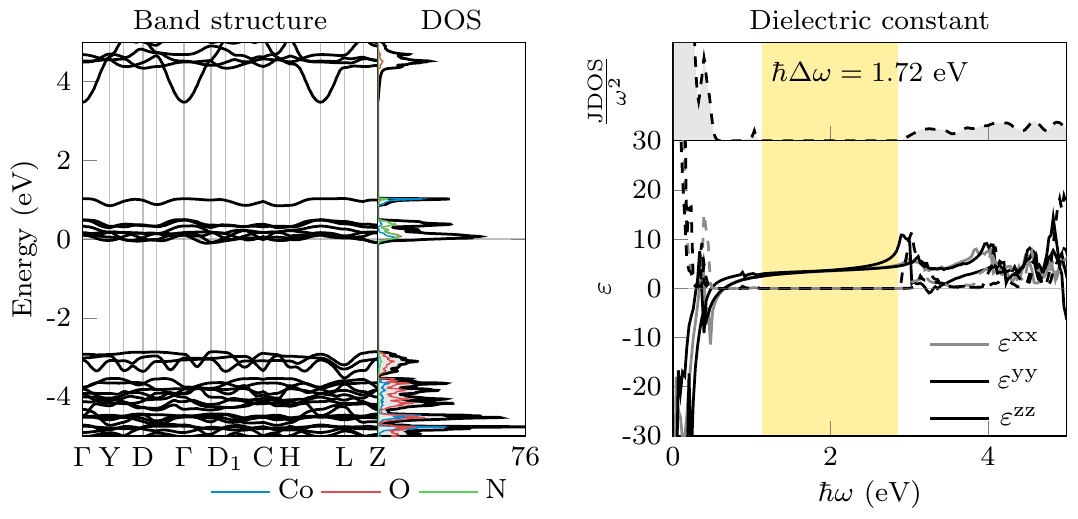}\end{minipage}}\\
		\cline{1-2}Formula&$\rm K_2BaCo{(NO_2)}_6$&\\
		\cline{1-2}Space group&69~(Fmmm)&\\
		\cline{1-2}\emph{W}&1.13~eV&\\
		\cline{1-2}$\hbar\Delta\omega$&1.72~eV&\\\hline
	\end{tabular}
\end{table}

\clearpage
\begin{table}
	\centering
	\renewcommand\arraystretch{2.6}
	\begin{tabular}{|C{0.13\columnwidth}|C{0.15\columnwidth}|C{0.67\columnwidth}|}
		\hline
		Class&\romanNum{3}&\multirow{4}[10]*{\begin{minipage}{0.65\columnwidth}\includegraphics{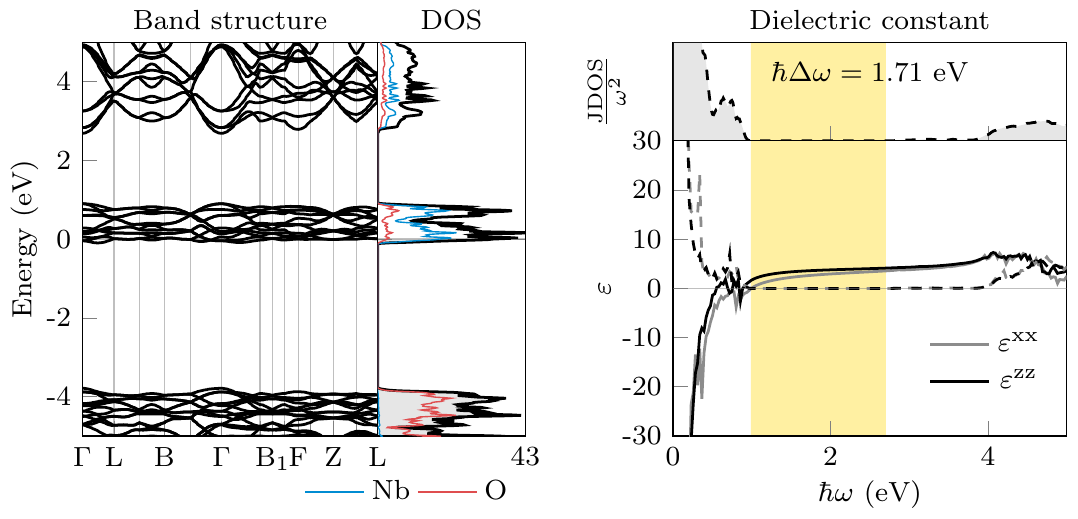}\end{minipage}}\\
		\cline{1-2}Formula&$\rm Nb_2{(PO_4)}_3$&\\
		\cline{1-2}Space group&167~(R-3c)&\\
		\cline{1-2}\emph{W}&0.99~eV&\\
		\cline{1-2}$\hbar\Delta\omega$&1.71~eV&\\\hline
	\end{tabular}
\end{table}

\begin{table}
	\centering
	\renewcommand\arraystretch{2.6}
	\begin{tabular}{|C{0.13\columnwidth}|C{0.15\columnwidth}|C{0.67\columnwidth}|}
		\hline
		Class&\romanNum{3}&\multirow{4}[10]*{\begin{minipage}{0.65\columnwidth}\includegraphics{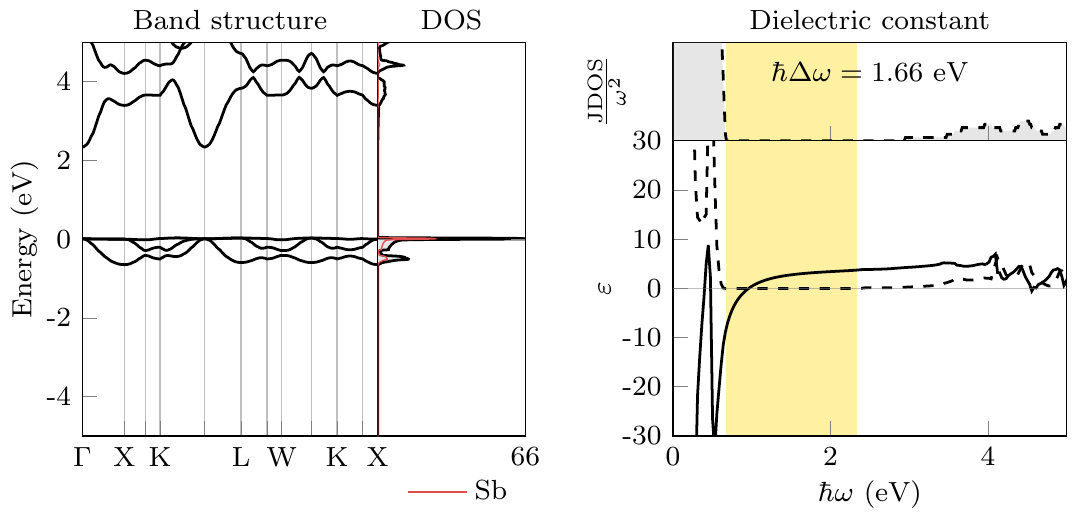}\end{minipage}}\\
		\cline{1-2}Formula&$\rm RbSb$&\\
		\cline{1-2}Space group&216~(F-43m)&\\
		\cline{1-2}\emph{W}&0.68~eV&\\
		\cline{1-2}$\hbar\Delta\omega$&1.66~eV&\\\hline
	\end{tabular}
\end{table}

\begin{table}
	\centering
	\renewcommand\arraystretch{2.6}
	\begin{tabular}{|C{0.13\columnwidth}|C{0.15\columnwidth}|C{0.67\columnwidth}|}
		\hline
		Class&\romanNum{3}&\multirow{4}[10]*{\begin{minipage}{0.65\columnwidth}\includegraphics{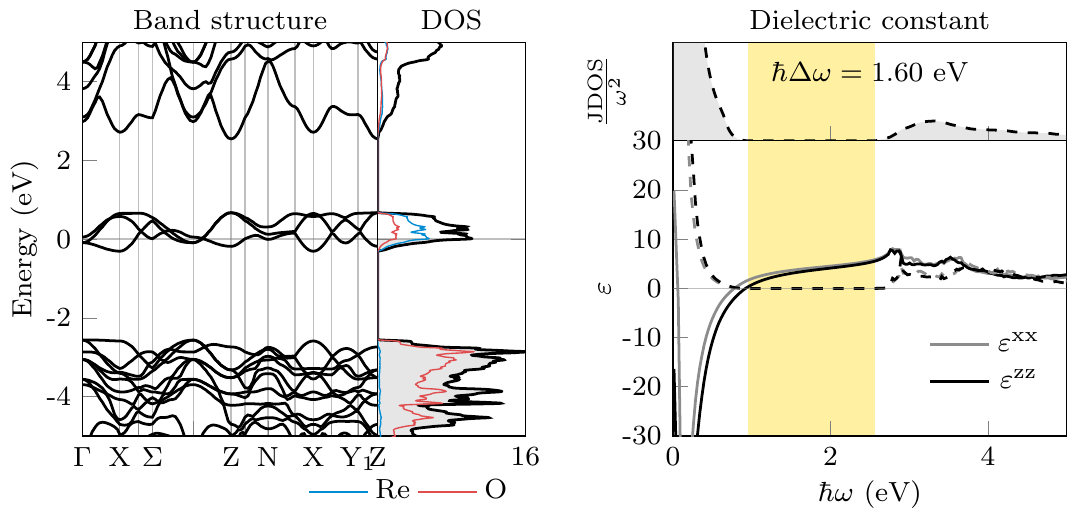}\end{minipage}}\\
		\cline{1-2}Formula&$\rm Sr_2ZnReO_6$&\\
		\cline{1-2}Space group&87~(I4/m)&\\
		\cline{1-2}\emph{W}&0.96~eV&\\
		\cline{1-2}$\hbar\Delta\omega$&1.60~eV&\\\hline
	\end{tabular}
\end{table}

\begin{table}
	\centering
	\renewcommand\arraystretch{2.6}
	\begin{tabular}{|C{0.13\columnwidth}|C{0.15\columnwidth}|C{0.67\columnwidth}|}
		\hline
		Class&\romanNum{3}&\multirow{4}[10]*{\begin{minipage}{0.65\columnwidth}\includegraphics{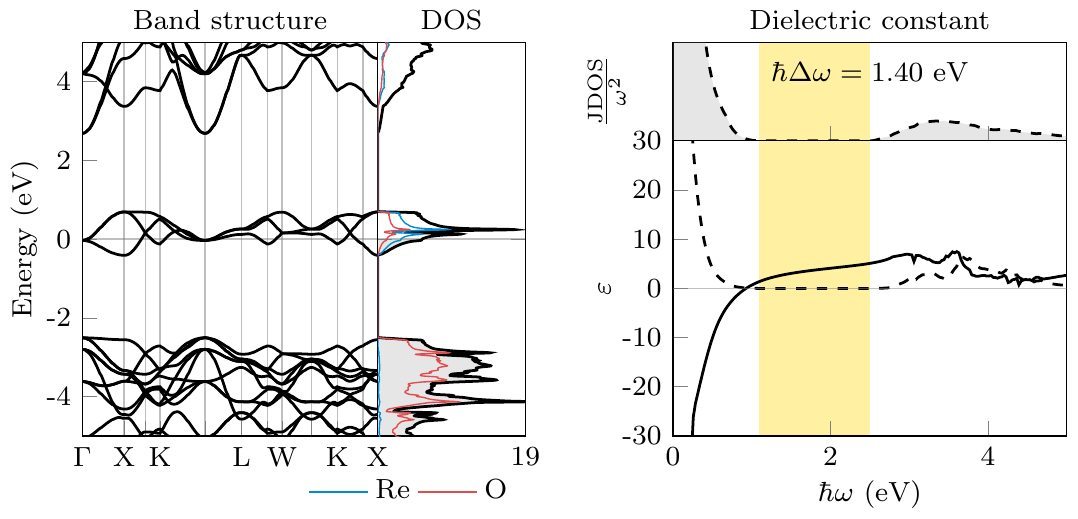}\end{minipage}}\\
		\cline{1-2}Formula&$\rm Ba_2MgReO_6$&\\
		\cline{1-2}Space group&225~(Fm-3m)&\\
		\cline{1-2}\emph{W}&1.10~eV&\\
		\cline{1-2}$\hbar\Delta\omega$&1.40~eV&\\\hline
	\end{tabular}
\end{table}

\clearpage
\begin{table}
	\centering
	\renewcommand\arraystretch{2.6}
	\begin{tabular}{|C{0.13\columnwidth}|C{0.15\columnwidth}|C{0.67\columnwidth}|}
		\hline
		Class&\romanNum{3}&\multirow{4}[10]*{\begin{minipage}{0.65\columnwidth}\includegraphics{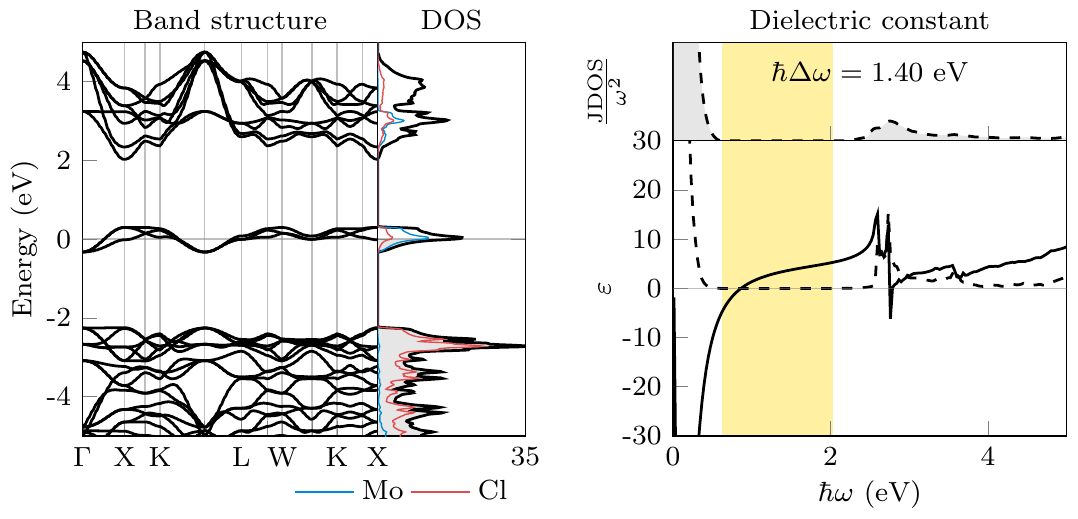}\end{minipage}}\\
		\cline{1-2}Formula&$\rm Tl_2MoCl_6$&\\
		\cline{1-2}Space group&225~(Fm-3m)&\\
		\cline{1-2}\emph{W}&0.63~eV&\\
		\cline{1-2}$\hbar\Delta\omega$&1.40~eV&\\\hline
	\end{tabular}
\end{table}

\begin{table}
	\centering
	\renewcommand\arraystretch{2.6}
	\begin{tabular}{|C{0.13\columnwidth}|C{0.15\columnwidth}|C{0.67\columnwidth}|}
		\hline
		Class&\romanNum{3}&\multirow{4}[10]*{\begin{minipage}{0.65\columnwidth}\includegraphics{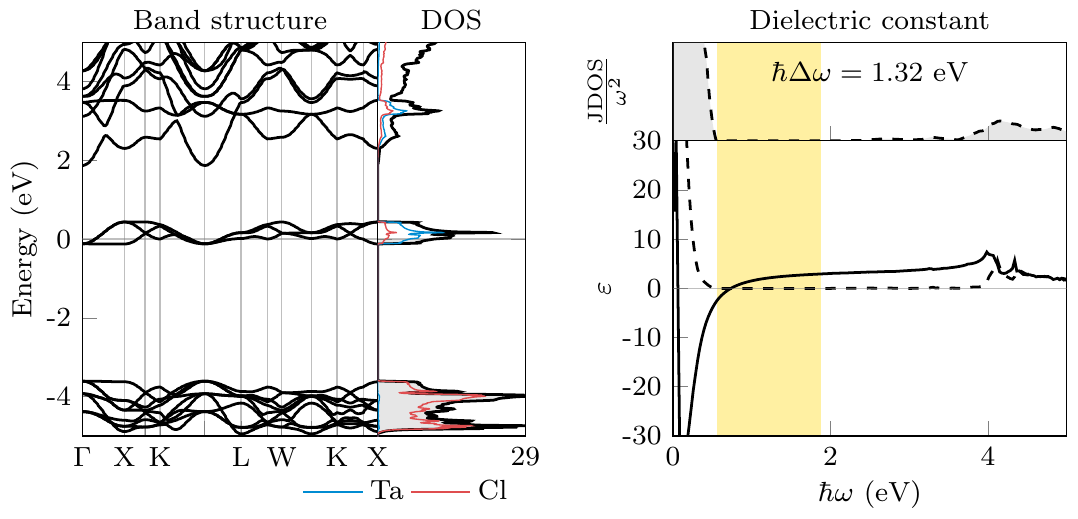}\end{minipage}}\\
		\cline{1-2}Formula&$\rm K_2TaCl_6$&\\
		\cline{1-2}Space group&225~(Fm-3m)&\\
		\cline{1-2}\emph{W}&0.56~eV&\\
		\cline{1-2}$\hbar\Delta\omega$&1.32~eV&\\\hline
	\end{tabular}
\end{table}

\begin{table}
	\centering
	\renewcommand\arraystretch{2.6}
	\begin{tabular}{|C{0.13\columnwidth}|C{0.15\columnwidth}|C{0.67\columnwidth}|}
		\hline
		Class&\romanNum{3}&\multirow{4}[10]*{\begin{minipage}{0.65\columnwidth}\includegraphics{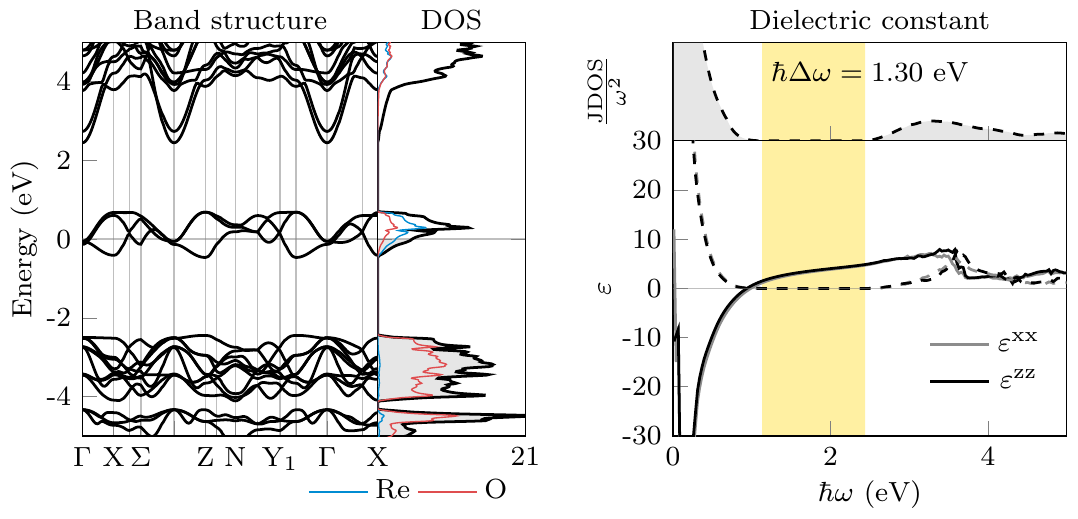}\end{minipage}}\\
		\cline{1-2}Formula&$\rm Ba_2CaReO_6$&\\
		\cline{1-2}Space group&87~(I4/m)&\\
		\cline{1-2}\emph{W}&1.14~eV&\\
		\cline{1-2}$\hbar\Delta\omega$&1.30~eV&\\\hline
	\end{tabular}
\end{table}

\begin{table}
	\centering
	\renewcommand\arraystretch{2.6}
	\begin{tabular}{|C{0.13\columnwidth}|C{0.15\columnwidth}|C{0.67\columnwidth}|}
		\hline
		Class&\romanNum{3}&\multirow{4}[10]*{\begin{minipage}{0.65\columnwidth}\includegraphics{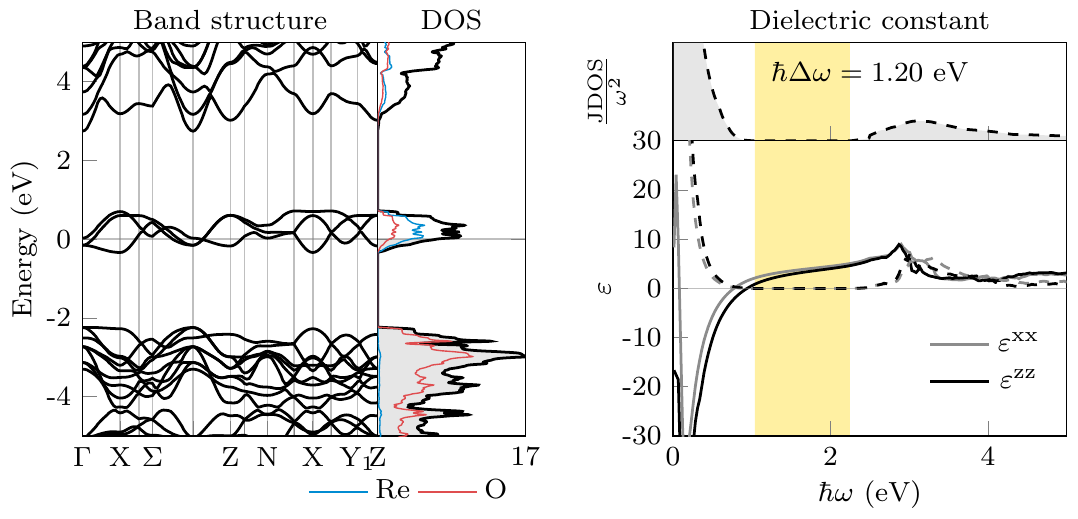}\end{minipage}}\\
		\cline{1-2}Formula&$\rm Sr_2MgReO_6$&\\
		\cline{1-2}Space group&87~(I4/m)&\\
		\cline{1-2}\emph{W}&1.04~eV&\\
		\cline{1-2}$\hbar\Delta\omega$&1.20~eV&\\\hline
	\end{tabular}
\end{table}

\clearpage
\begin{table}
	\centering
	\renewcommand\arraystretch{2.6}
	\begin{tabular}{|C{0.13\columnwidth}|C{0.15\columnwidth}|C{0.67\columnwidth}|}
		\hline
		Class&\romanNum{3}&\multirow{4}[10]*{\begin{minipage}{0.65\columnwidth}\includegraphics{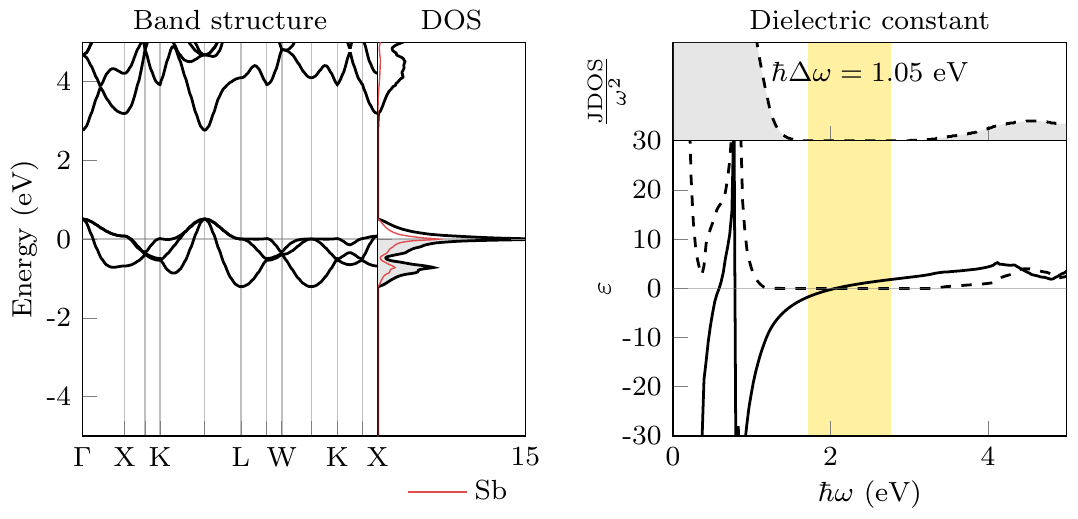}\end{minipage}}\\
		\cline{1-2}Formula&$\rm RbSb$&\\
		\cline{1-2}Space group&225~(Fm-3m)&\\
		\cline{1-2}\emph{W}&1.72~eV&\\
		\cline{1-2}$\hbar\Delta\omega$&1.05~eV&\\\hline
	\end{tabular}
\end{table}

\begin{table}
	\centering
	\renewcommand\arraystretch{2.6}
	\begin{tabular}{|C{0.13\columnwidth}|C{0.15\columnwidth}|C{0.67\columnwidth}|}
		\hline
		Class&\romanNum{3}&\multirow{4}[10]*{\begin{minipage}{0.65\columnwidth}\includegraphics{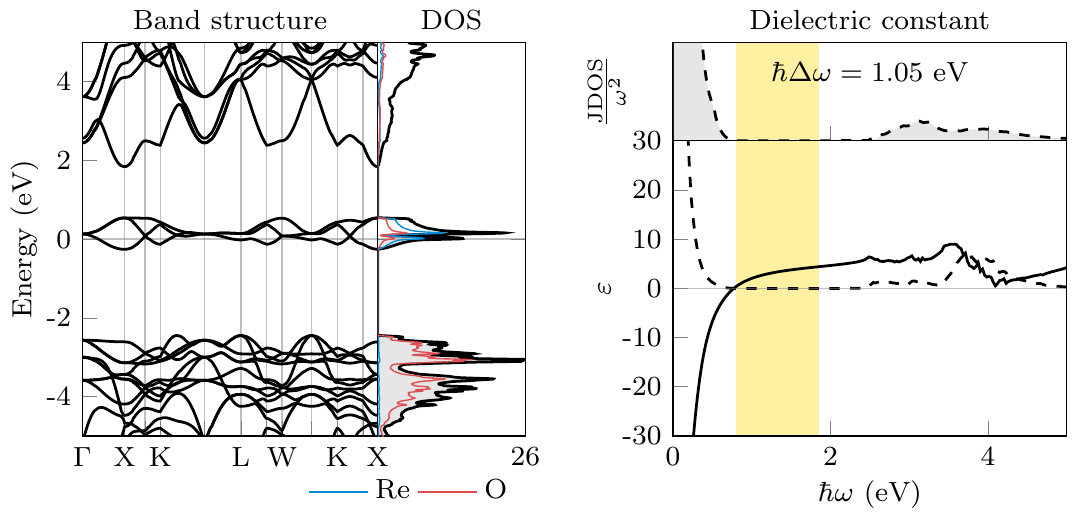}\end{minipage}}\\
		\cline{1-2}Formula&$\rm Ba_2CdReO_6$&\\
		\cline{1-2}Space group&225~(Fm-3m)&\\
		\cline{1-2}\emph{W}&0.80~eV&\\
		\cline{1-2}$\hbar\Delta\omega$&1.05~eV&\\\hline
	\end{tabular}
\end{table}

\begin{table}
	\centering
	\renewcommand\arraystretch{2.6}
	\begin{tabular}{|C{0.13\columnwidth}|C{0.15\columnwidth}|C{0.67\columnwidth}|}
		\hline
		Class&\romanNum{3}&\multirow{4}[10]*{\begin{minipage}{0.65\columnwidth}\includegraphics{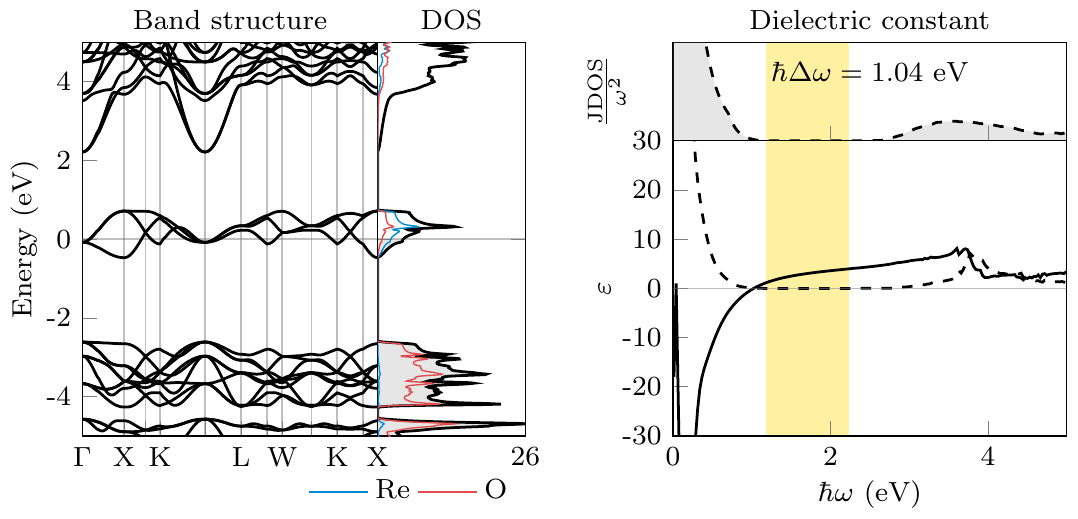}\end{minipage}}\\
		\cline{1-2}Formula&$\rm Ba_2CaReO_6$&\\
		\cline{1-2}Space group&225~(Fm-3m)&\\
		\cline{1-2}\emph{W}&1.19~eV&\\
		\cline{1-2}$\hbar\Delta\omega$&1.04~eV&\\\hline
	\end{tabular}
\end{table}

\begin{table}
	\centering
	\renewcommand\arraystretch{2.6}
	\begin{tabular}{|C{0.13\columnwidth}|C{0.15\columnwidth}|C{0.67\columnwidth}|}
		\hline
		Class&\romanNum{3}&\multirow{4}[10]*{\begin{minipage}{0.65\columnwidth}\includegraphics{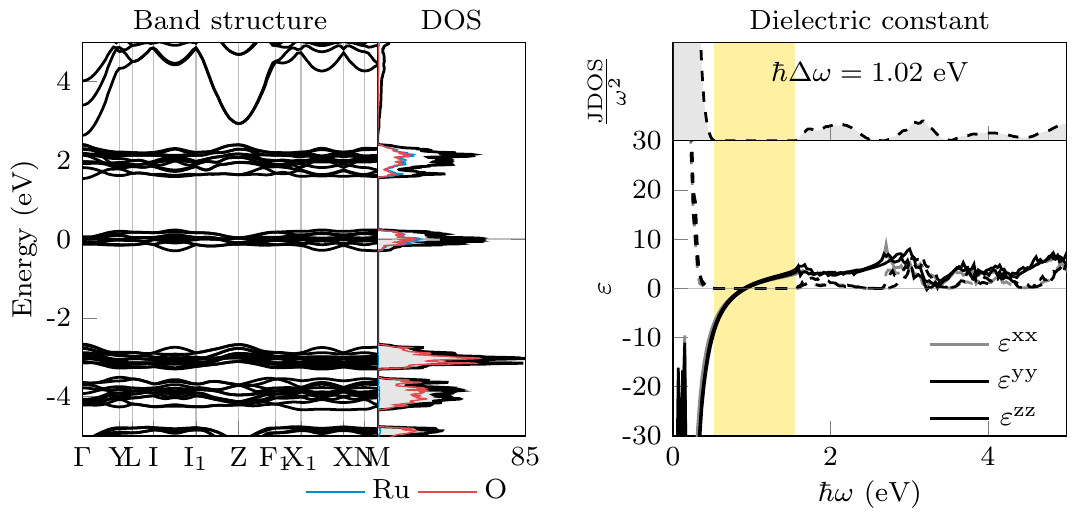}\end{minipage}}\\
		\cline{1-2}Formula&$\rm K_3Na{(RuO_4)}_2$&\\
		\cline{1-2}Space group&15~(C2/c)&\\
		\cline{1-2}\emph{W}&0.53~eV&\\
		\cline{1-2}$\hbar\Delta\omega$&1.02~eV&\\\hline
	\end{tabular}
\end{table}

\clearpage
\begin{table}
	\centering
	\renewcommand\arraystretch{2.6}
	\begin{tabular}{|C{0.13\columnwidth}|C{0.15\columnwidth}|C{0.67\columnwidth}|}
		\hline
		Class&\romanNum{3}&\multirow{4}[10]*{\begin{minipage}{0.65\columnwidth}\includegraphics{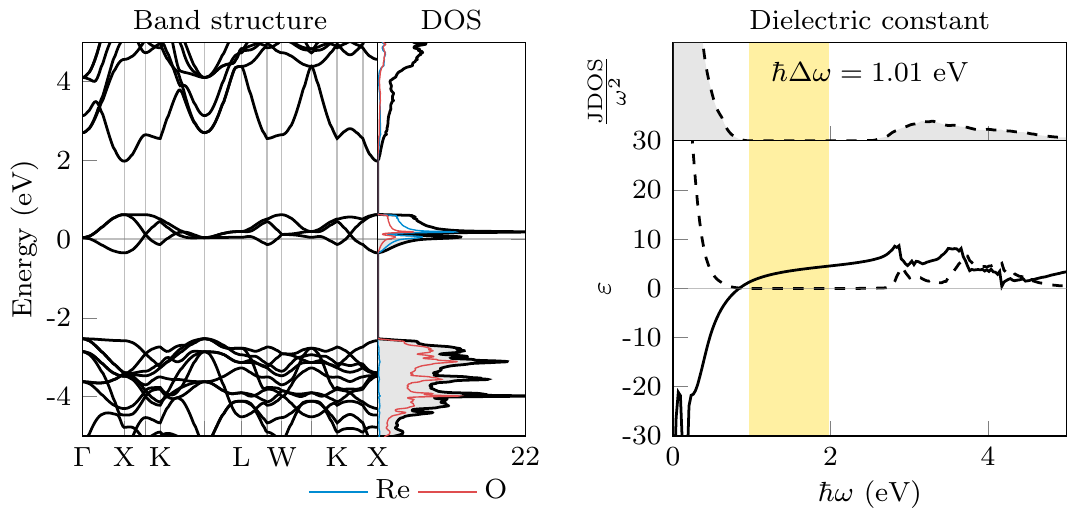}\end{minipage}}\\
		\cline{1-2}Formula&$\rm Ba_2ZnReO_6$&\\
		\cline{1-2}Space group&225~(Fm-3m)&\\
		\cline{1-2}\emph{W}&0.97~eV&\\
		\cline{1-2}$\hbar\Delta\omega$&1.01~eV&\\\hline
	\end{tabular}
\end{table}

\begin{table}
	\centering
	\renewcommand\arraystretch{2.6}
	\begin{tabular}{|C{0.13\columnwidth}|C{0.15\columnwidth}|C{0.67\columnwidth}|}
		\hline
		Class&\romanNum{3}&\multirow{4}[10]*{\begin{minipage}{0.65\columnwidth}\includegraphics{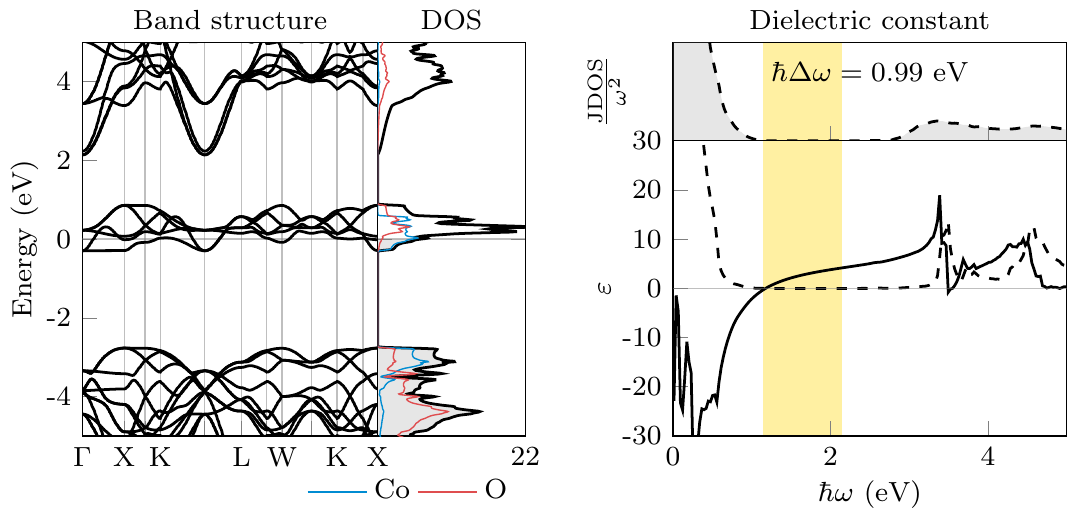}\end{minipage}}\\
		\cline{1-2}Formula&$\rm Sr_2CoWO_6$&\\
		\cline{1-2}Space group&225~(Fm-3m)&\\
		\cline{1-2}\emph{W}&1.15~eV&\\
		\cline{1-2}$\hbar\Delta\omega$&0.99~eV&\\\hline
	\end{tabular}
\end{table}

\begin{table}
	\centering
	\renewcommand\arraystretch{2.6}
	\begin{tabular}{|C{0.13\columnwidth}|C{0.15\columnwidth}|C{0.67\columnwidth}|}
		\hline
		Class&\romanNum{3}&\multirow{4}[10]*{\begin{minipage}{0.65\columnwidth}\includegraphics{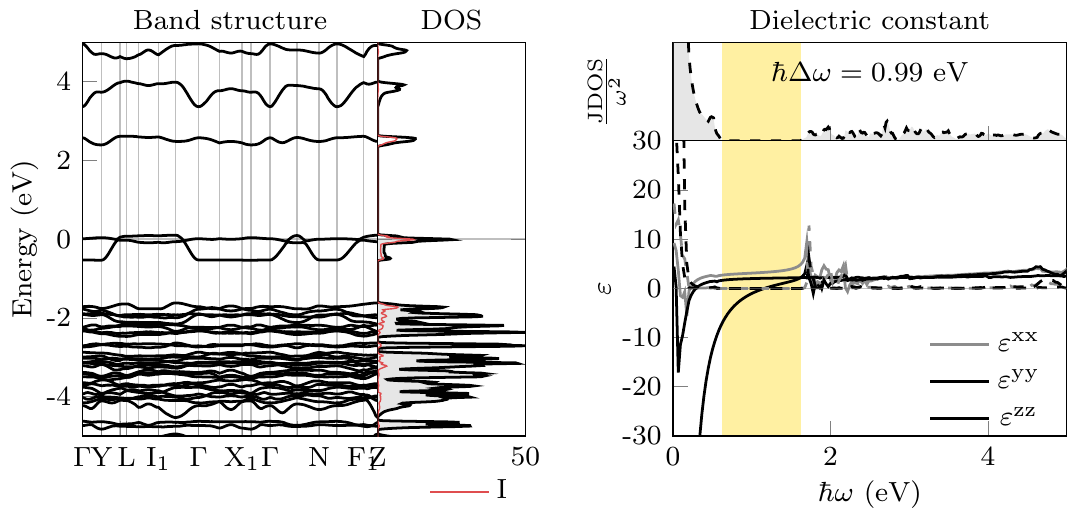}\end{minipage}}\\
		\cline{1-2}Formula&$\rm I_2Sb_2F_{11}$&\\
		\cline{1-2}Space group&5~(C2)&\\
		\cline{1-2}\emph{W}&0.63~eV&\\
		\cline{1-2}$\hbar\Delta\omega$&0.99~eV&\\\hline
	\end{tabular}
\end{table}

\begin{table}
	\centering
	\renewcommand\arraystretch{2.6}
	\begin{tabular}{|C{0.13\columnwidth}|C{0.15\columnwidth}|C{0.67\columnwidth}|}
		\hline
		Class&\romanNum{3}&\multirow{4}[10]*{\begin{minipage}{0.65\columnwidth}\includegraphics{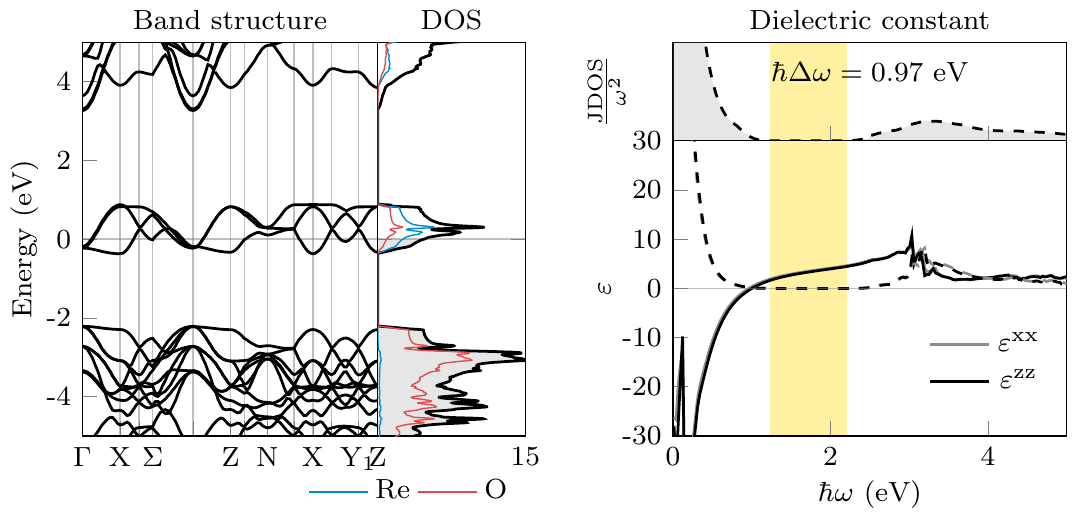}\end{minipage}}\\
		\cline{1-2}Formula&$\rm Sr_2MgReO_6$&\\
		\cline{1-2}Space group&139~(I4/mmm)&\\
		\cline{1-2}\emph{W}&1.24~eV&\\
		\cline{1-2}$\hbar\Delta\omega$&0.97~eV&\\\hline
	\end{tabular}
\end{table}

\clearpage
\begin{table}
	\centering
	\renewcommand\arraystretch{2.6}
	\begin{tabular}{|C{0.13\columnwidth}|C{0.15\columnwidth}|C{0.67\columnwidth}|}
		\hline
		Class&\romanNum{3}&\multirow{4}[10]*{\begin{minipage}{0.65\columnwidth}\includegraphics{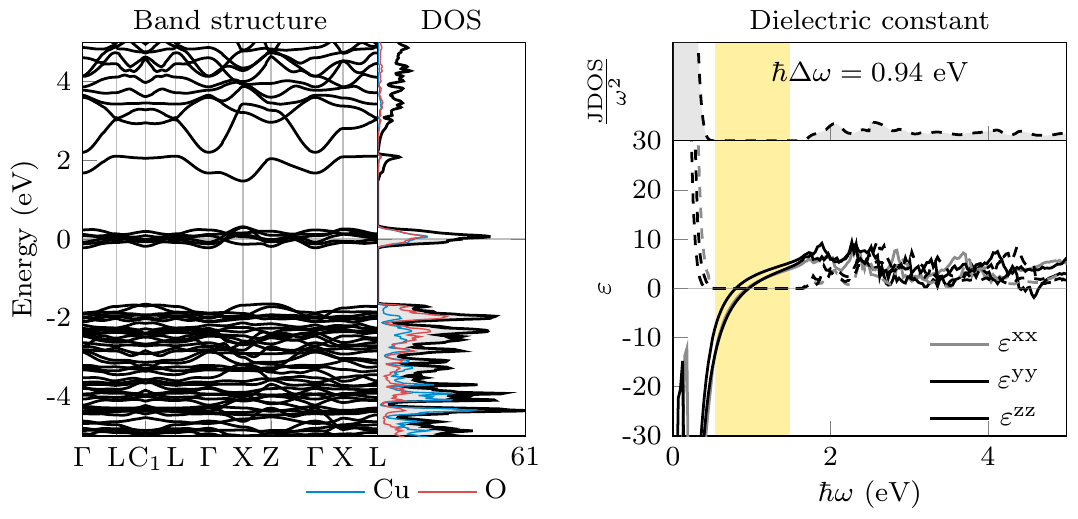}\end{minipage}}\\
		\cline{1-2}Formula&$\rm LiBa_2Cu_3O_6$&\\
		\cline{1-2}Space group&69~(Fmmm)&\\
		\cline{1-2}\emph{W}&0.54~eV&\\
		\cline{1-2}$\hbar\Delta\omega$&0.94~eV&\\\hline
	\end{tabular}
\end{table}

\begin{table}
	\centering
	\renewcommand\arraystretch{2.6}
	\begin{tabular}{|C{0.13\columnwidth}|C{0.15\columnwidth}|C{0.67\columnwidth}|}
		\hline
		Class&\romanNum{3}&\multirow{4}[10]*{\begin{minipage}{0.65\columnwidth}\includegraphics{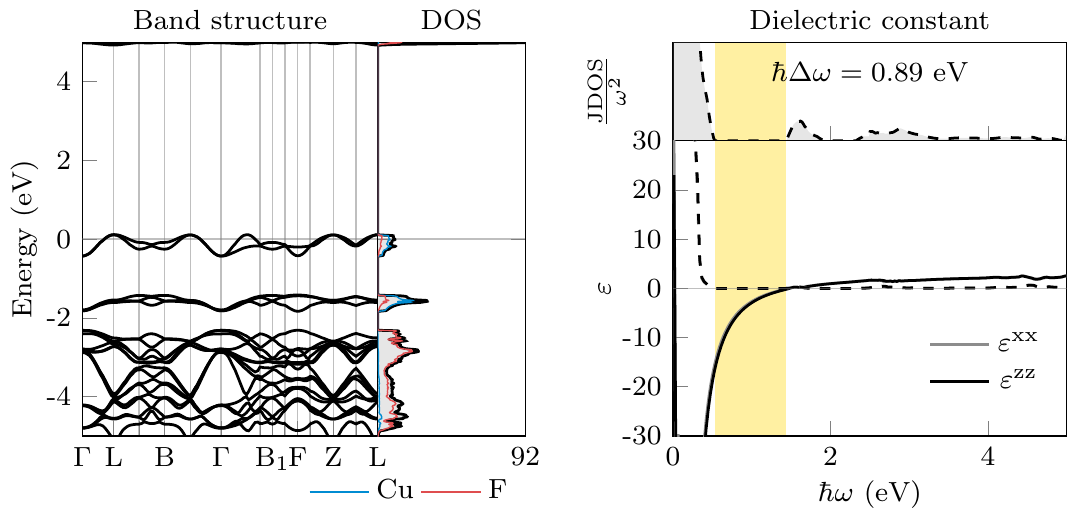}\end{minipage}}\\
		\cline{1-2}Formula&$\rm CuZrF_6$&\\
		\cline{1-2}Space group&148~(R-3)&\\
		\cline{1-2}\emph{W}&0.54~eV&\\
		\cline{1-2}$\hbar\Delta\omega$&0.89~eV&\\\hline
	\end{tabular}
\end{table}

\begin{table}
	\centering
	\renewcommand\arraystretch{2.6}
	\begin{tabular}{|C{0.13\columnwidth}|C{0.15\columnwidth}|C{0.67\columnwidth}|}
		\hline
		Class&\romanNum{3}&\multirow{4}[10]*{\begin{minipage}{0.65\columnwidth}\includegraphics{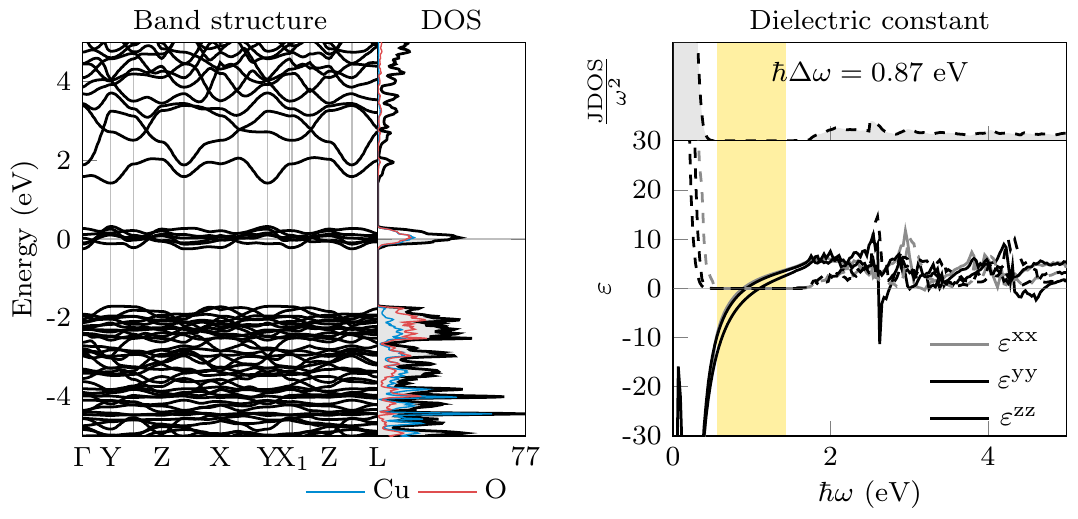}\end{minipage}}\\
		\cline{1-2}Formula&$\rm NaBa_2Cu_3O_6$&\\
		\cline{1-2}Space group&69~(Fmmm)&\\
		\cline{1-2}\emph{W}&0.56~eV&\\
		\cline{1-2}$\hbar\Delta\omega$&0.87~eV&\\\hline
	\end{tabular}
\end{table}

\begin{table}
	\centering
	\renewcommand\arraystretch{2.6}
	\begin{tabular}{|C{0.13\columnwidth}|C{0.15\columnwidth}|C{0.67\columnwidth}|}
		\hline
		Class&\romanNum{3}&\multirow{4}[10]*{\begin{minipage}{0.65\columnwidth}\includegraphics{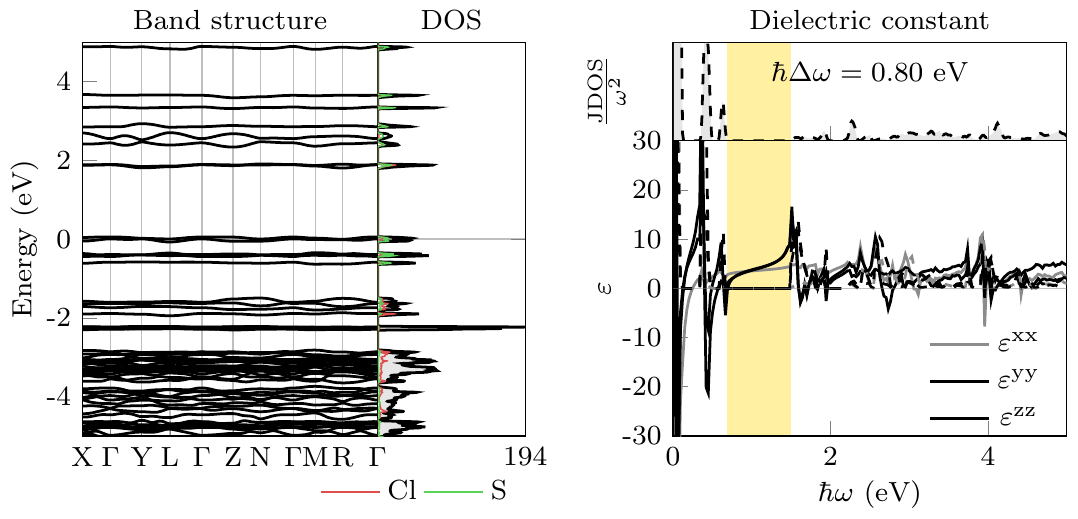}\end{minipage}}\\
		\cline{1-2}Formula&$\rm S_3Cl_3AsF_6$&\\
		\cline{1-2}Space group&2~(P-1)&\\
		\cline{1-2}\emph{W}&0.69~eV&\\
		\cline{1-2}$\hbar\Delta\omega$&0.80~eV&\\\hline
	\end{tabular}
\end{table}

\clearpage
\begin{table}
	\centering
	\renewcommand\arraystretch{2.6}
	\begin{tabular}{|C{0.13\columnwidth}|C{0.15\columnwidth}|C{0.67\columnwidth}|}
		\hline
		Class&\romanNum{3}&\multirow{4}[10]*{\begin{minipage}{0.65\columnwidth}\includegraphics{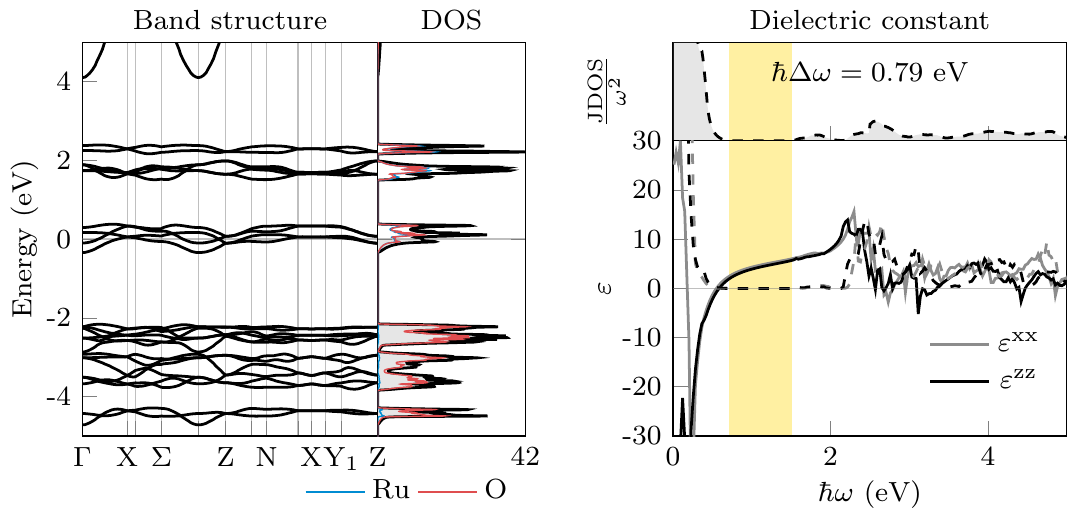}\end{minipage}}\\
		\cline{1-2}Formula&$\rm KRuO_4$&\\
		\cline{1-2}Space group&88~(I41/a)&\\
		\cline{1-2}\emph{W}&0.72~eV&\\
		\cline{1-2}$\hbar\Delta\omega$&0.79~eV&\\\hline
	\end{tabular}
\end{table}

\begin{table}
	\centering
	\renewcommand\arraystretch{2.6}
	\begin{tabular}{|C{0.13\columnwidth}|C{0.15\columnwidth}|C{0.67\columnwidth}|}
		\hline
		Class&\romanNum{3}&\multirow{4}[10]*{\begin{minipage}{0.65\columnwidth}\includegraphics{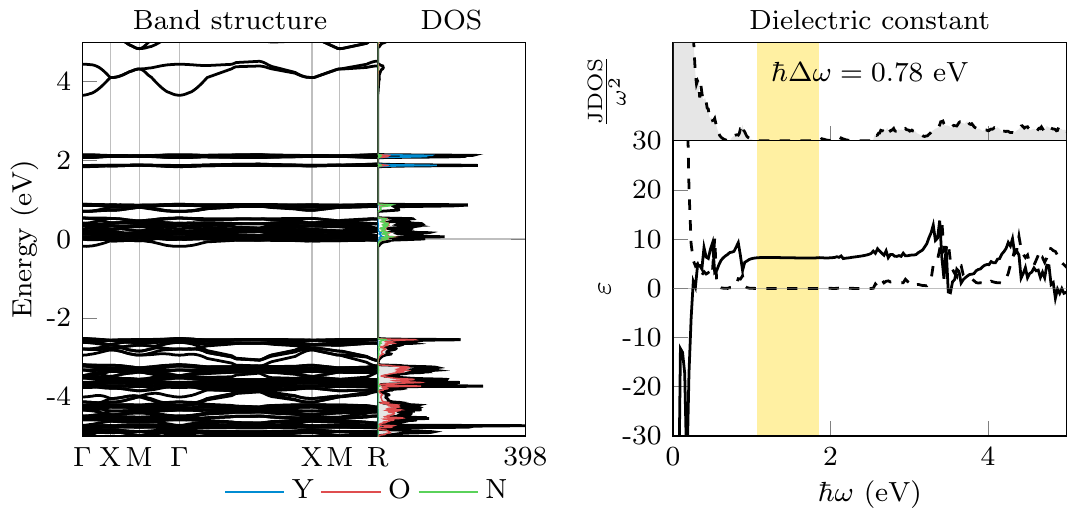}\end{minipage}}\\
		\cline{1-2}Formula&$\rm K_5YCo_2{(NO_2)}_{12}$&\\
		\cline{1-2}Space group&201~(Pn-3)&\\
		\cline{1-2}\emph{W}&1.07~eV&\\
		\cline{1-2}$\hbar\Delta\omega$&0.78~eV&\\\hline
	\end{tabular}
\end{table}

\begin{table}
	\centering
	\renewcommand\arraystretch{2.6}
	\begin{tabular}{|C{0.13\columnwidth}|C{0.15\columnwidth}|C{0.67\columnwidth}|}
		\hline
		Class&\romanNum{3}&\multirow{4}[10]*{\begin{minipage}{0.65\columnwidth}\includegraphics{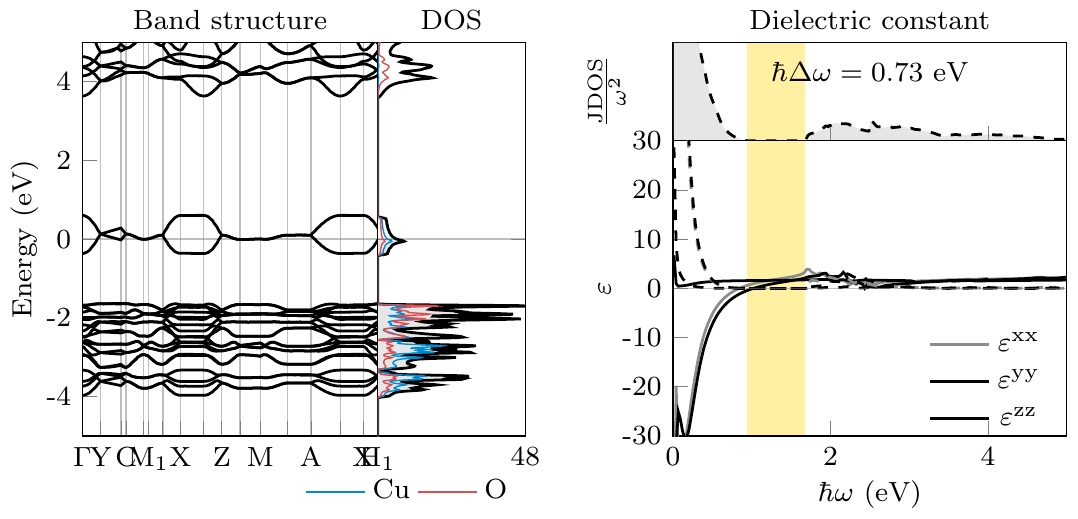}\end{minipage}}\\
		\cline{1-2}Formula&$\rm Cu{(HCOO)}_2$&\\
		\cline{1-2}Space group&14~(P21/c)&\\
		\cline{1-2}\emph{W}&0.95~eV&\\
		\cline{1-2}$\hbar\Delta\omega$&0.73~eV&\\\hline
	\end{tabular}
\end{table}

\begin{table}
	\centering
	\renewcommand\arraystretch{2.6}
	\begin{tabular}{|C{0.13\columnwidth}|C{0.15\columnwidth}|C{0.67\columnwidth}|}
		\hline
		Class&\romanNum{3}&\multirow{4}[10]*{\begin{minipage}{0.65\columnwidth}\includegraphics{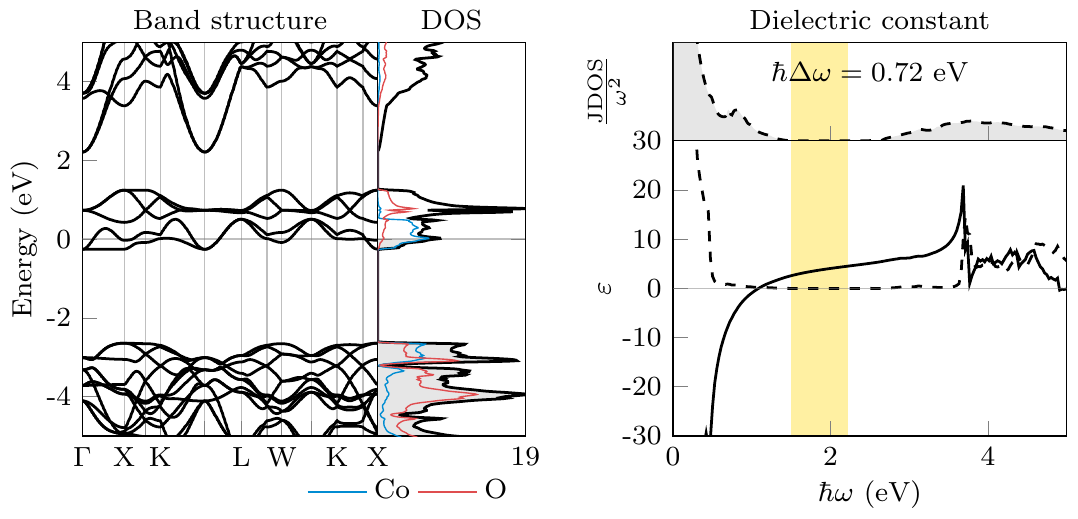}\end{minipage}}\\
		\cline{1-2}Formula&$\rm Ba_2CoWO_6$&\\
		\cline{1-2}Space group&225~(Fm-3m)&\\
		\cline{1-2}\emph{W}&1.50~eV&\\
		\cline{1-2}$\hbar\Delta\omega$&0.72~eV&\\\hline
	\end{tabular}
\end{table}

\clearpage
\begin{table}
	\centering
	\renewcommand\arraystretch{2.6}
	\begin{tabular}{|C{0.13\columnwidth}|C{0.15\columnwidth}|C{0.67\columnwidth}|}
		\hline
		Class&\romanNum{3}&\multirow{4}[10]*{\begin{minipage}{0.65\columnwidth}\includegraphics{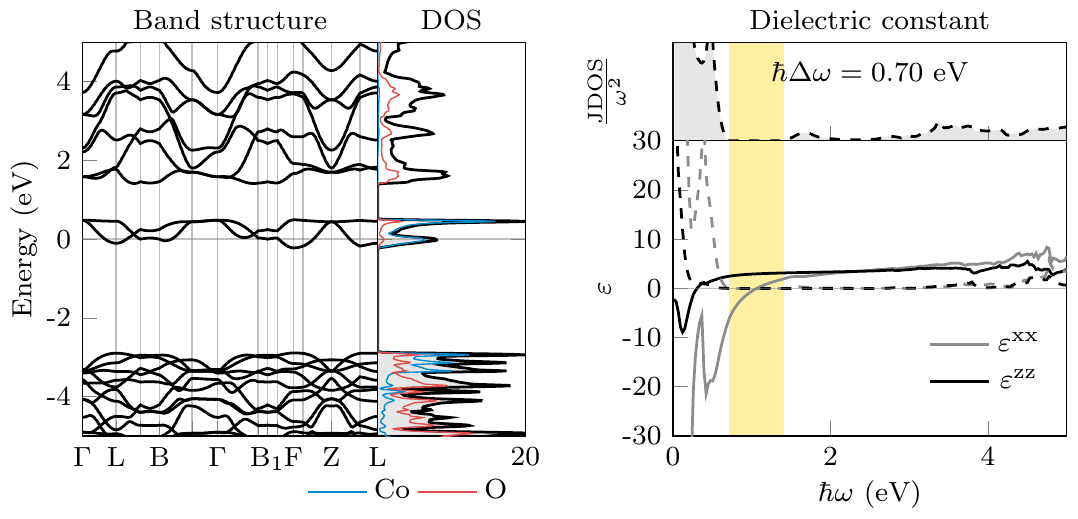}\end{minipage}}\\
		\cline{1-2}Formula&$\rm K_2Co{(SeO_3)}_2$&\\
		\cline{1-2}Space group&166~(R-3m)&\\
		\cline{1-2}\emph{W}&0.71~eV&\\
		\cline{1-2}$\hbar\Delta\omega$&0.70~eV&\\\hline
	\end{tabular}
\end{table}

\begin{table}
	\centering
	\renewcommand\arraystretch{2.6}
	\begin{tabular}{|C{0.13\columnwidth}|C{0.15\columnwidth}|C{0.67\columnwidth}|}
		\hline
		Class&\romanNum{3}&\multirow{4}[10]*{\begin{minipage}{0.65\columnwidth}\includegraphics{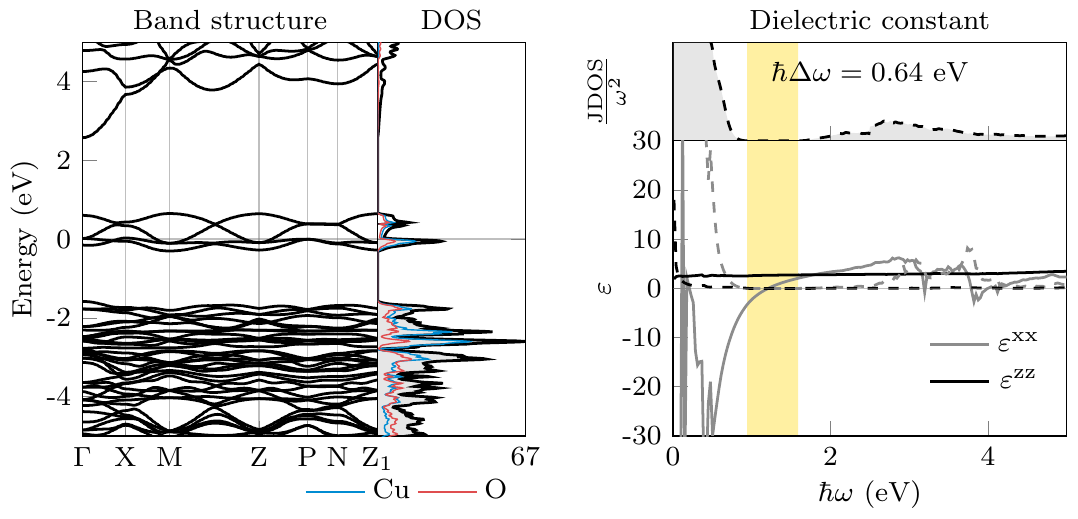}\end{minipage}}\\
		\cline{1-2}Formula&$\rm SrCu_2{(BO_3)}_2$&\\
		\cline{1-2}Space group&121~(I-42m)&\\
		\cline{1-2}\emph{W}&0.95~eV&\\
		\cline{1-2}$\hbar\Delta\omega$&0.64~eV&\\\hline
	\end{tabular}
\end{table}

\begin{table}
	\centering
	\renewcommand\arraystretch{2.6}
	\begin{tabular}{|C{0.13\columnwidth}|C{0.15\columnwidth}|C{0.67\columnwidth}|}
		\hline
		Class&\romanNum{3}&\multirow{4}[10]*{\begin{minipage}{0.65\columnwidth}\includegraphics{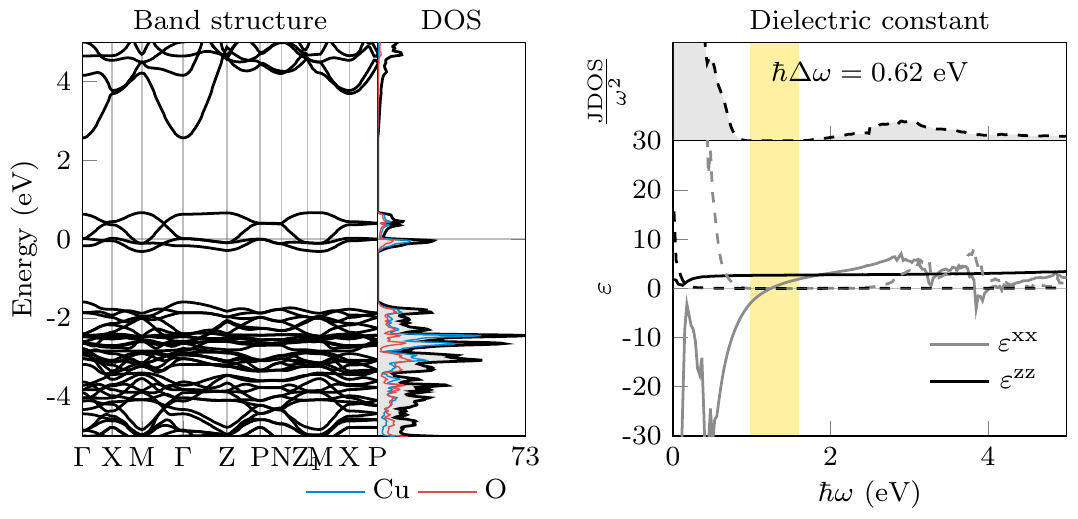}\end{minipage}}\\
		\cline{1-2}Formula&$\rm SrCu_2{(BO_3)}_2$&\\
		\cline{1-2}Space group&140~(I4/mcm)&\\
		\cline{1-2}\emph{W}&0.98~eV&\\
		\cline{1-2}$\hbar\Delta\omega$&0.62~eV&\\\hline
	\end{tabular}
\end{table}

\begin{table}
	\centering
	\renewcommand\arraystretch{2.6}
	\begin{tabular}{|C{0.13\columnwidth}|C{0.15\columnwidth}|C{0.67\columnwidth}|}
		\hline
		Class&\romanNum{3}&\multirow{4}[10]*{\begin{minipage}{0.65\columnwidth}\includegraphics{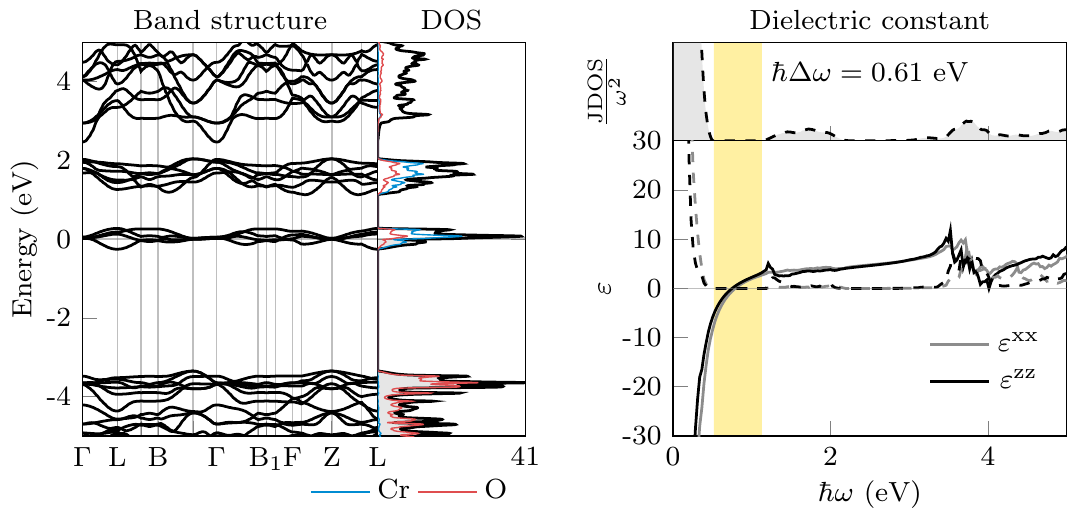}\end{minipage}}\\
		\cline{1-2}Formula&$\rm Ba_3Cr_2O_8$&\\
		\cline{1-2}Space group&166~(R-3m)&\\
		\cline{1-2}\emph{W}&0.52~eV&\\
		\cline{1-2}$\hbar\Delta\omega$&0.61~eV&\\\hline
	\end{tabular}
\end{table}

\clearpage
\begin{table}
	\centering
	\renewcommand\arraystretch{2.6}
	\begin{tabular}{|C{0.13\columnwidth}|C{0.15\columnwidth}|C{0.67\columnwidth}|}
		\hline
		Class&\romanNum{3}&\multirow{4}[10]*{\begin{minipage}{0.65\columnwidth}\includegraphics{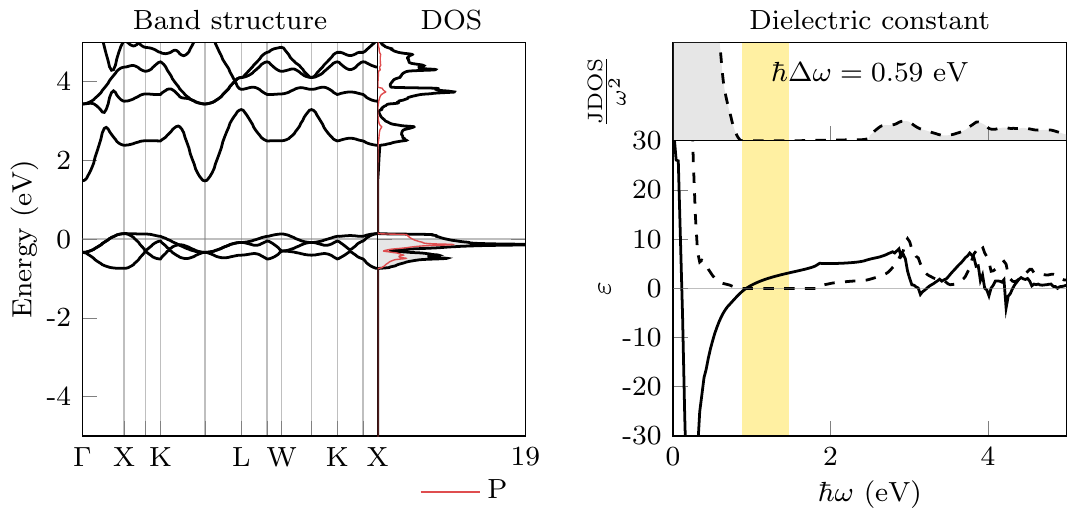}\end{minipage}}\\
		\cline{1-2}Formula&$\rm CsRbP$&\\
		\cline{1-2}Space group&216~(F-43m)&\\
		\cline{1-2}\emph{W}&0.88~eV&\\
		\cline{1-2}$\hbar\Delta\omega$&0.59~eV&\\\hline
	\end{tabular}
\end{table}

\begin{table}
	\centering
	\renewcommand\arraystretch{2.6}
	\begin{tabular}{|C{0.13\columnwidth}|C{0.15\columnwidth}|C{0.67\columnwidth}|}
		\hline
		Class&\romanNum{3}&\multirow{4}[10]*{\begin{minipage}{0.65\columnwidth}\includegraphics{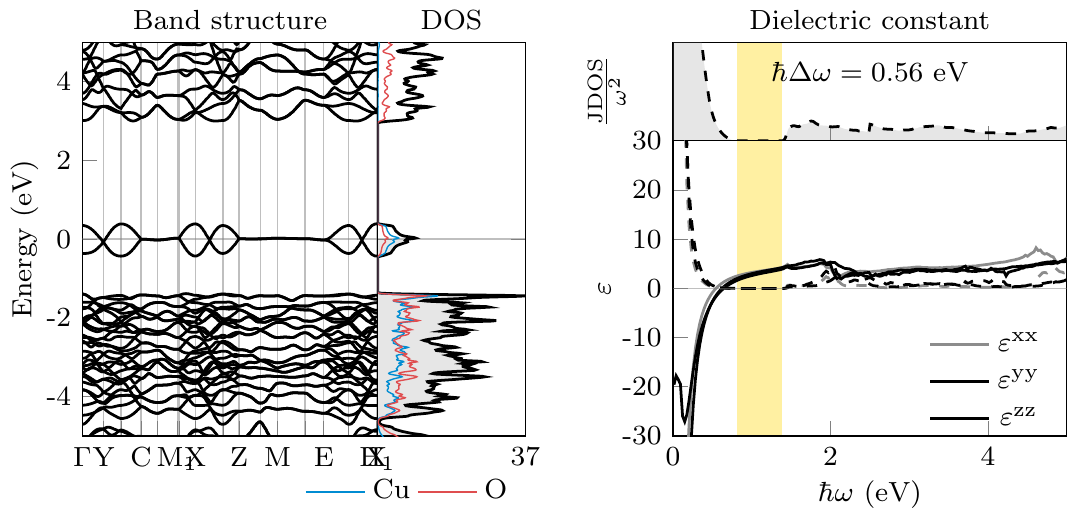}\end{minipage}}\\
		\cline{1-2}Formula&$\rm Cu{(HSeO3)}_2$&\\
		\cline{1-2}Space group&14~(P21/c)&\\
		\cline{1-2}\emph{W}&0.82~eV&\\
		\cline{1-2}$\hbar\Delta\omega$&0.56~eV&\\\hline
	\end{tabular}
\end{table}

\begin{table}
	\centering
	\renewcommand\arraystretch{2.6}
	\begin{tabular}{|C{0.13\columnwidth}|C{0.15\columnwidth}|C{0.67\columnwidth}|}
		\hline
		Class&\romanNum{3}&\multirow{4}[10]*{\begin{minipage}{0.65\columnwidth}\includegraphics{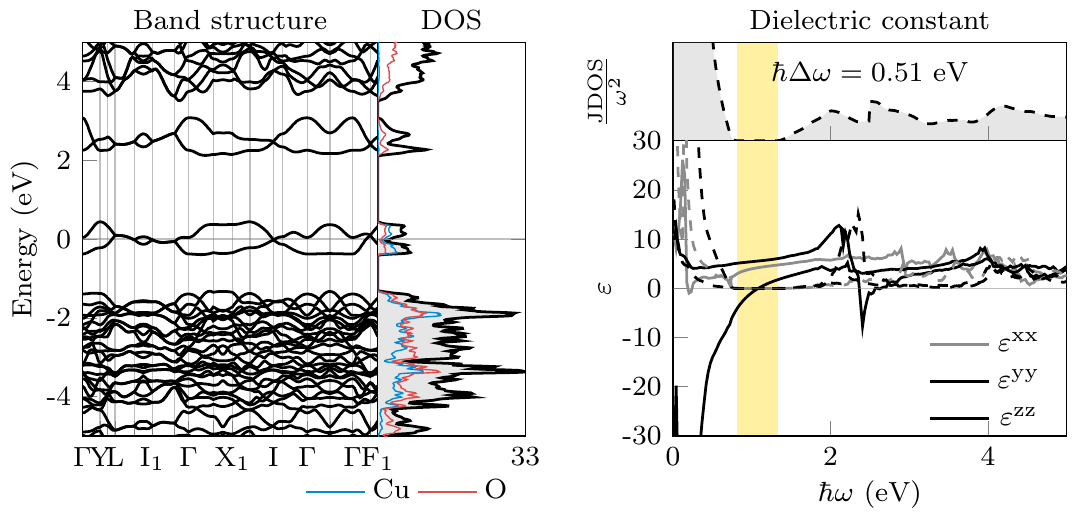}\end{minipage}}\\
		\cline{1-2}Formula&$\rm CuSe_2O_5$&\\
		\cline{1-2}Space group&15~(C2/c)&\\
		\cline{1-2}\emph{W}&0.82~eV&\\
		\cline{1-2}$\hbar\Delta\omega$&0.51~eV&\\\hline
	\end{tabular}
\end{table}

\begin{table}
	\centering
	\renewcommand\arraystretch{2.6}
	\begin{tabular}{|C{0.13\columnwidth}|C{0.15\columnwidth}|C{0.67\columnwidth}|}
		\hline
		Class&\romanNum{3}&\multirow{4}[10]*{\begin{minipage}{0.65\columnwidth}\includegraphics{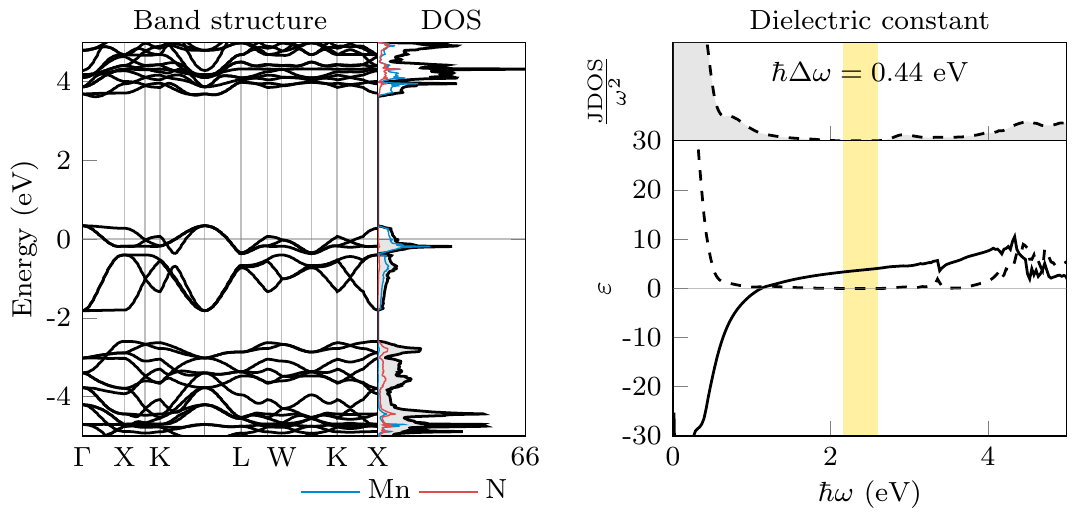}\end{minipage}}\\
		\cline{1-2}Formula&$\rm CsMn{(CN)}_3$&\\
		\cline{1-2}Space group&225~(Fm-3m)&\\
		\cline{1-2}\emph{W}&2.16~eV&\\
		\cline{1-2}$\hbar\Delta\omega$&0.44~eV&\\\hline
	\end{tabular}
\end{table}

\clearpage
\begin{table}
	\centering
	\renewcommand\arraystretch{2.6}
	\begin{tabular}{|C{0.13\columnwidth}|C{0.15\columnwidth}|C{0.67\columnwidth}|}
		\hline
		Class&\romanNum{3}&\multirow{4}[10]*{\begin{minipage}{0.65\columnwidth}\includegraphics{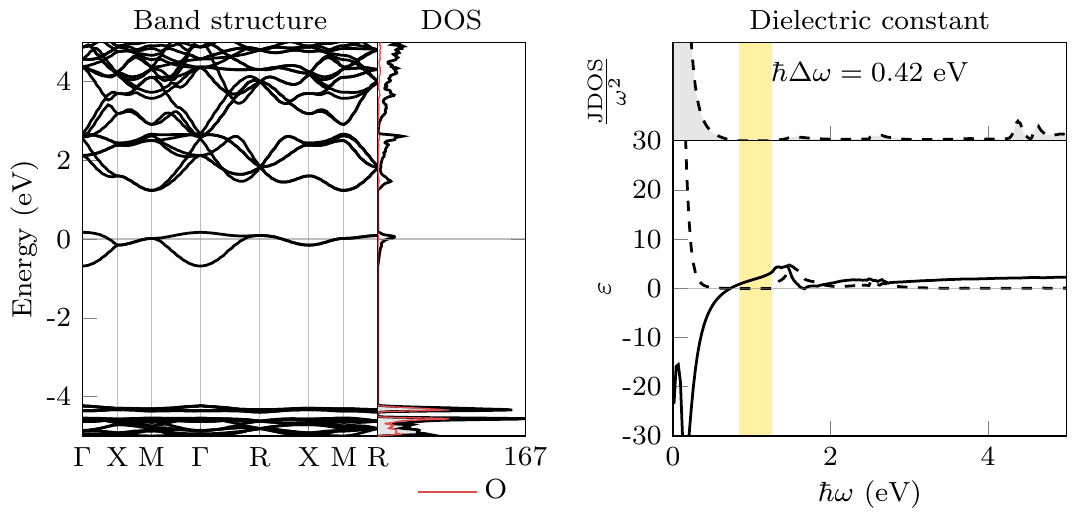}\end{minipage}}\\
		\cline{1-2}Formula&$\rm Na_4Al_3{(SiO_4)}_3$&\\
		\cline{1-2}Space group&218~(P-43n)&\\
		\cline{1-2}\emph{W}&0.84~eV&\\
		\cline{1-2}$\hbar\Delta\omega$&0.42~eV&\\\hline
	\end{tabular}
\end{table}

\begin{table}
	\centering
	\renewcommand\arraystretch{2.6}
	\begin{tabular}{|C{0.13\columnwidth}|C{0.15\columnwidth}|C{0.67\columnwidth}|}
		\hline
		Class&\romanNum{3}&\multirow{4}[10]*{\begin{minipage}{0.65\columnwidth}\includegraphics{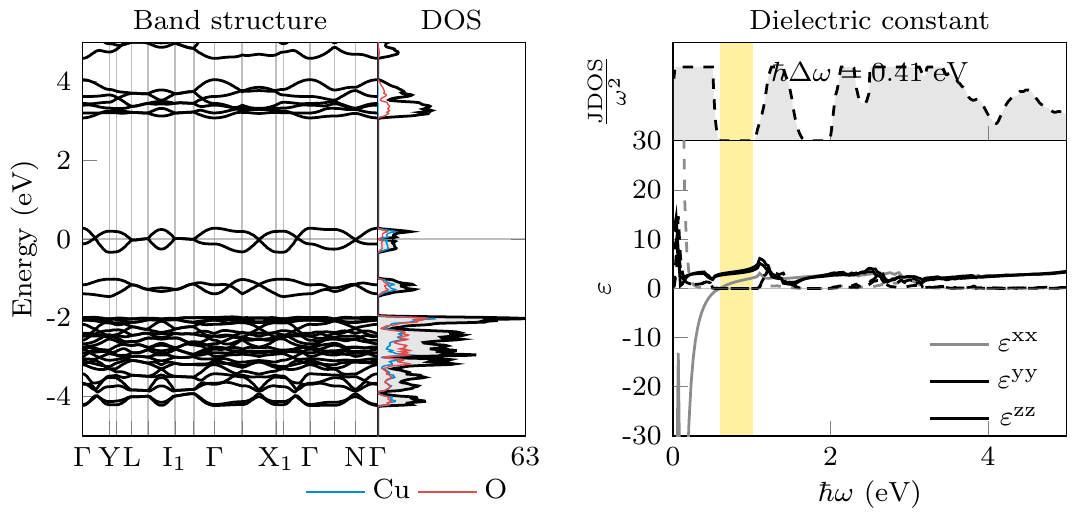}\end{minipage}}\\
		\cline{1-2}Formula&$\rm CuC_5H_7NO_4\cdot 2H_2O$&\\
		\cline{1-2}Space group&15~(C2/c)&\\
		\cline{1-2}\emph{W}&0.60~eV&\\
		\cline{1-2}$\hbar\Delta\omega$&0.41~eV&\\\hline
	\end{tabular}
\end{table}

\begin{table}
	\centering
	\renewcommand\arraystretch{2.6}
	\begin{tabular}{|C{0.13\columnwidth}|C{0.15\columnwidth}|C{0.67\columnwidth}|}
		\hline
		Class&\romanNum{3}&\multirow{4}[10]*{\begin{minipage}{0.65\columnwidth}\includegraphics{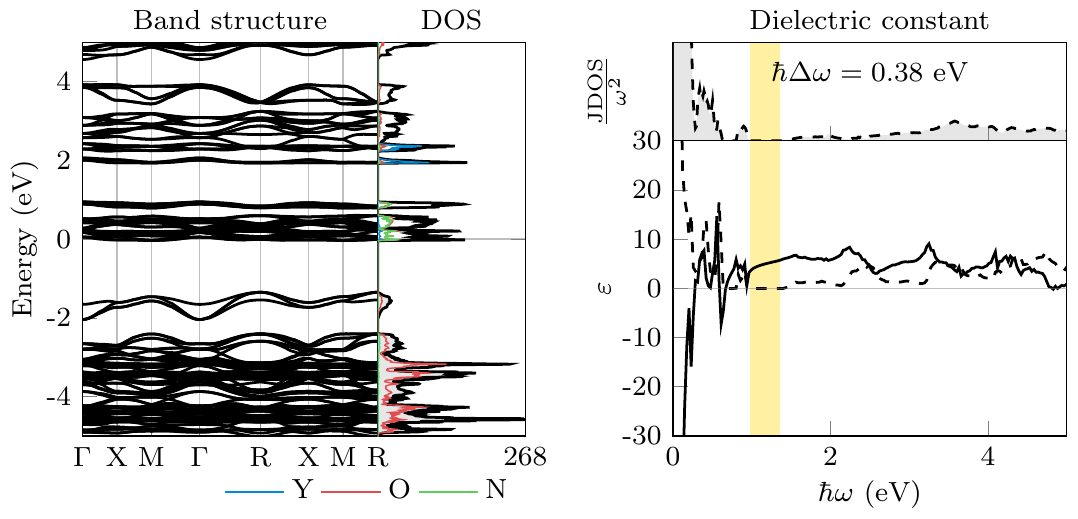}\end{minipage}}\\
		\cline{1-2}Formula&$\rm Tl_5YCo_2{(NO_2)}_{12}$&\\
		\cline{1-2}Space group&201~(Pn-3)&\\
		\cline{1-2}\emph{W}&0.98~eV&\\
		\cline{1-2}$\hbar\Delta\omega$&0.38~eV&\\\hline
	\end{tabular}
\end{table}

\begin{table}
	\centering
	\renewcommand\arraystretch{2.6}
	\begin{tabular}{|C{0.13\columnwidth}|C{0.15\columnwidth}|C{0.67\columnwidth}|}
		\hline
		Class&\romanNum{3}&\multirow{4}[10]*{\begin{minipage}{0.65\columnwidth}\includegraphics{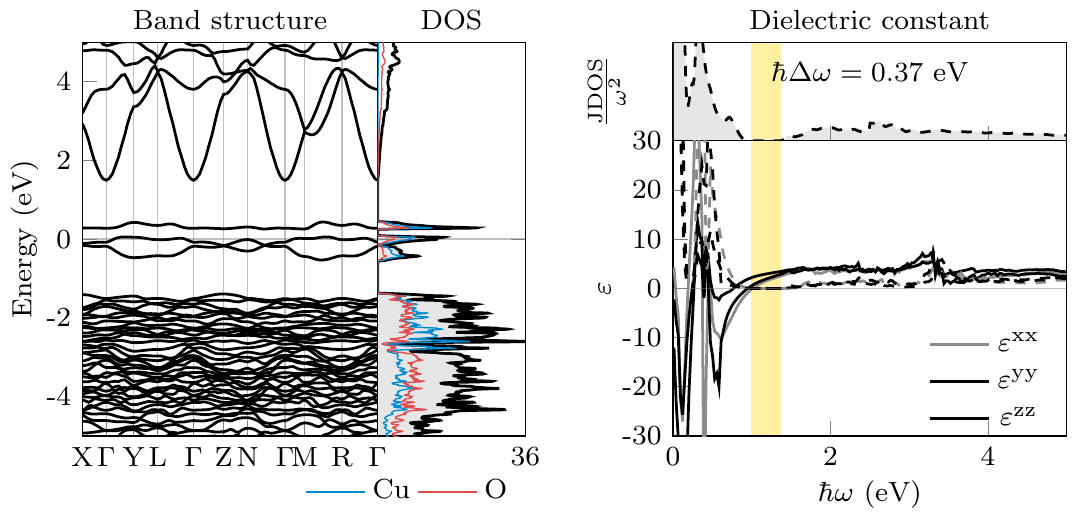}\end{minipage}}\\
		\cline{1-2}Formula&$\rm Na_2Cu_3Ge_4O_{12}$&\\
		\cline{1-2}Space group&2~(P-1)&\\
		\cline{1-2}\emph{W}&1.00~eV&\\
		\cline{1-2}$\hbar\Delta\omega$&0.37~eV&\\\hline
	\end{tabular}
\end{table}

\clearpage
\begin{table}
	\centering
	\renewcommand\arraystretch{2.6}
	\begin{tabular}{|C{0.13\columnwidth}|C{0.15\columnwidth}|C{0.67\columnwidth}|}
		\hline
		Class&\romanNum{3}&\multirow{4}[10]*{\begin{minipage}{0.65\columnwidth}\includegraphics{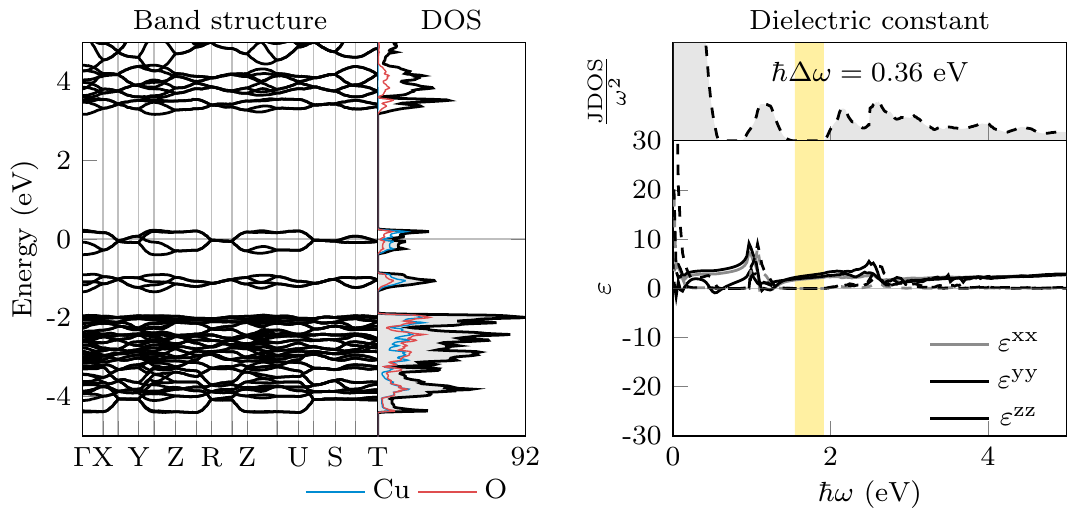}\end{minipage}}\\
		\cline{1-2}Formula&$\rm CH_3NH_3Cu(HCOO)_3$&\\
		\cline{1-2}Space group&62~(Pnma)&\\
		\cline{1-2}\emph{W}&1.55~eV&\\
		\cline{1-2}$\hbar\Delta\omega$&0.36~eV&\\\hline
	\end{tabular}
\end{table}

\begin{table}
	\centering
	\renewcommand\arraystretch{2.6}
	\begin{tabular}{|C{0.13\columnwidth}|C{0.15\columnwidth}|C{0.67\columnwidth}|}
		\hline
		Class&\romanNum{3}&\multirow{4}[10]*{\begin{minipage}{0.65\columnwidth}\includegraphics{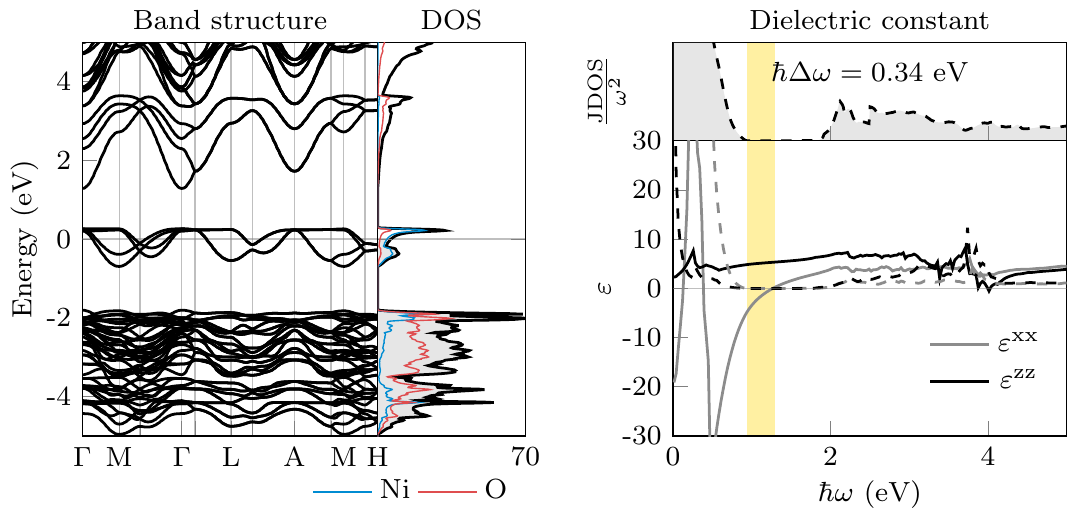}\end{minipage}}\\
		\cline{1-2}Formula&$\rm Ba_3NiSb_2O_9$&\\
		\cline{1-2}Space group&194~(P63/mmc)&\\
		\cline{1-2}\emph{W}&0.95~eV&\\
		\cline{1-2}$\hbar\Delta\omega$&0.34~eV&\\\hline
	\end{tabular}
\end{table}

\clearpage

\end{document}